\documentclass{aa}

\usepackage{graphicx}
\usepackage{txfonts}
\usepackage{placeins}
\newcommand{\asec}{$^{\prime\prime}$}

\def\SigmaH2{$\Sigma $(${\rm H_2}$)}
\def\r1415{$^{14}$N/$^{15}$N}

\def\H{N$_{2}$H$^{+}$}

\def\15N{$^{15}$NNH$^+$}
\def\N15{N$^{15}$NH$^+$}

\def\H13CN{\mbox{H$^{13}$CN}}

\def\kms{\mbox{km~s$^{-1}$}}
\def\cmc{cm$^{-3}$}
\def\cmq{cm$^{-2}$}

\def\soll{\mbox{L$_\odot$}}

\def\Ntot{\mbox{$N_{\rm tot}$}}
\def\Tex{\mbox{$T_{\rm ex}$}}

\def\tau17{\mbox{$\tau_{\rm 17}$}}
\def\tau18{\mbox{$\tau_{\rm 18}$}}

\def\Td{\mbox{$T_{\rm d}$}}

\def\Tk{\mbox{$T_{\rm k}$}}

\def\kms{km\,s$^{-1}$}

\def\Td{$T_{\rm dust}$}

\begin{document}

   \title{The evolution of sulphur-bearing molecules in high-mass star-forming cores}

%   \subtitle{I. Overviewing the $\kappa$-mechanism}

   \author{F. Fontani
          \inst{1,2,3}
          \and
          E. Roueff\inst{2}
          \and
          L. Colzi\inst{4}
          \and
    %      L. Bizzocchi\inst{4}
    %      \and
          P. Caselli\inst{3}
          %\fnmsep\thanks{Just to show the usage          of the elements in the author field}
          }

\institute{INAF - Osservatorio Astrofisico di Arcetri,
              Largo E. Fermi 5,
              I-50125, Florence (Italy)\\
            \email{francesco.fontani@inaf.it}
         \and
         LERMA, Observatoire de Paris, PSL Research University, CNRS, Sorbonne
         Universit\'e, F-92190 Meudon (France)
   %     \and
   %       Dipartimento di Chimica “Giacomo Ciamician”, Universit\'a di Bologna, Via F. Selmi 2, I-40126 Bologna (Italy)
          \and
          Centre for Astrochemical Studies, Max-Planck-Institute for Extraterrestrial Physics, Giessenbachstrasse 1, 85748 Garching, Germany
          \and
        Centro de Astrobiolog\'ia (CSIC-INTA), Ctra Ajalvir km 4, 28850, Torrej\'on de Ardoz, Madrid (Spain)
             }

   \date{Received xxx; accepted yyy}

% \abstract{}{}{}{}{} 
% 5 {} token are mandatory
 
  \abstract
   {To understand the chemistry of sulphur (S) in the interstellar medium, models need to be tested by observations of S-bearing molecules in different physical conditions.}
   {We aim to derive column densities and abundances of S-bearing molecules in high-mass dense cores in different evolutionary stages and with different physical properties.}
   {We analyse observations obtained with the Institut de RadioAstronomie Millim\'etrique (IRAM) 30m telescope towards 15 well-known cores classified in the three main evolutionary stages of the high-mass star-formation process: high-mass starless cores, high-mass protostellar objects, and ultracompact HII regions.}
   {We have detected rotational lines of SO, SO$^+$, NS, C$^{34}$S, $^{13}$CS,
   SO$_2$, CCS, H$_2$S, HCS$^+$, OCS, 
   H$_2$CS, and CCCS. 
   We also analyse for the first time lines of the NO molecule to complement the analysis.
   From a local thermodynamic equilibrium approach, we have derived column densities of each species and excitation temperatures for those detected in multiple lines with different excitation.
   Based on a statistical analysis on the line widths and the excitation temperatures, we find that: NS, C$^{34}$S, $^{13}$CS, CCS, and HCS$^+$ trace cold, quiescent, and likely extended material; OCS, and SO$_2$ trace warmer, more turbulent, and likely denser and more compact material; SO and perhaps SO$^+$ trace both quiescent and turbulent material depending on the target.
   The nature of the emission of H$_2$S, H$_2$CS and CCCS is less clear.
%   H$_2$CS and CCCS show linewidths smaller than $\sim 5$\kms, but do not seem correlated with those of other quiescent gas tracers.
   The molecular abundances of SO, SO$_2$, and H$_2$S show the strongest positive correlations with the kinetic temperature, believed to be an evolutionary indicator. 
Moreover, the sum of all molecular abundances show an enhancement of gaseous S from the less evolved to the more evolved stages.
   These trends could be due to the increasing amount of S sputtered from dust grains owing to the increasing protostellar activity with evolution.
   The average abundances in each evolutionary group increase especially in the oxygen-bearing molecules, perhaps due to the increasing abundance of atomic oxygen with evolution owing to photodissociation of water in gas phase.}
   {Our observational work represents a test-bed for theoretical studies aimed at modelling the chemistry of sulphur during the evolution of high-mass star-forming cores.}
%   {The increasing trend of the abundances of SO, SO$_2$, H$_2$S with the evolutionary parameters is likely due to the increasing amount of sulphur sputtered from dust grains, owing to the increasing protostellar activity with evolution.}

   \keywords{astrochemistry – line:
          identification – 
          ISM: molecules – 
          stars: formation 
               }

   \maketitle
%
%-------------------------------------------------------------------

\section{Introduction}
\label{intro}

Sulphur (S) is one of the most mysterious elements from an astrochemical point of view.
It has a cosmic abundance relative to hydrogen of $1.73\times 10^{-5}$ (Lodders~\citeyear{lodders03}), and its ionised form has a relative abundance of $1.66\times 10^{-5}$ (Esteban et al.~\citeyear{esteban04}), in close agreement with previous derivation in the Ionised Orion Nebula.
%({Dom{\'\i}nguez-Guzm{\'a}n} et al.~\citeyear{esteban22})
This makes it the tenth most abundant element in the Universe.
Sulphur is also the sixth most important biogenic element after hydrogen, oxygen, carbon, nitrogen, and phosphorus. 
It is a key component in most proteins, since it is contained in the methionine (C$_5$H$_{11}$NO$_2$S) and cysteine (C$_3$H$_7$NO$_2$S) amino acids, and a (minor) constituent of fats, biological fluids, and skeletal minerals. Also, in the form of H$_2$S, it can replace water in the photosynthesis of some bacteria.
Nevertheless, its chemical behaviour in the interstellar medium (ISM), and in particular in star-forming molecular clouds, is still not well-understood.

The chemistry of sulphur in the ISM has been studied for a long time (e.g.~Oppenheimer \& Dalgarno~\citeyear{oed74}; Pineau des For\^{e}ts et al.~\citeyear{pineau86}; Drdla et al.~\citeyear{drdla89}; Turner~\citeyear{turner95}; Goicoechea et al.~\citeyear{goicoechea06}; Fuente et al.~\citeyear{fuente19}; Navarro-Almaida et al.~\citeyear{navarro20}; Esplugues et al.~\citeyear{esplugues22}), but still a major problem remains unsolved: what is the main reservoir %volatile 
(i.e. both in gas and in ice mantles of dust grains) %reservoir 
of sulphur in the ISM?
The problem arises from the fact that, despite observations of the diffuse interstellar medium suggest a negligible elemental depletion of S in solid phase 
(Howk, Sembach \& Savage~\citeyear{howk06}; Jenkins~\citeyear{jenkins09}), its abundance in the gas-phase seen through molecular emission is at most a few percent of the cosmic value (McGuire~\citeyear{mcguire18}; see also references in Table~1 of Woods et al.~\citeyear{woods15}).
A natural explanation would be severe freeze-out of S-bearing species on ice mantles of dust grains. However, measured abundances of S-bearing molecules on ices are too low to solve the problem. 
OCS and SO$_2$ (e.g.~Boogert et al.~\citeyear{boogert15}; McClure et al.~\citeyear{mcclure23}) were detected on ice mantles, but their abundances can account for only $\sim 4\%$ of the cosmic sulphur abundance.
Theoretical models predict that H$_2$S should be one of the most important sulphur reservoirs on ice mantles owing to hydrogenation of atomic S (Vidal et al.~\citeyear{vidal17}), 
but observations of H$_2$S on ices around high-mass young stellar objects again provide upper limit abundances much lower than the S cosmic abundance (Jim\'enez-Escobar \& Mu\~{n}oz-Caro~\citeyear{jem11}). 
Recently Heyl et al~(\citeyear{heyl22}) have constrained the binding energies of various species on dust grains from available observations but acknowledge the uncertainties linked to sulphur containing species.

In addition to the problem of the missing volatile sulphur, even the formation of some S-bearing molecules detected in the gas-phase is unclear. 
For example, not only S$^+$ + H$_2$ $\rightarrow$ SH$^+$ + H, which  should be one of the most important reactions initiating sulphur chemistry, 
 is strongly endothermic 
(by 9860~K,  ~Millar et al.~\citeyear{millar:86}), but the following SH$^+$ + H$_2$ $\rightarrow$ H$_2$S$^+$ + H and H$_2$S$^+$ + H$_2$ $\rightarrow$ H$_3$S$^+$ + H reactions are also endothermic by 6380 and 2900~K, respectively, due to the much higher binding energy of the reactants with respect to the products. 
Even in shocked material such as L1157 B1, S-bearing species can account for only small fractions of the cosmic sulphur abundance (e.g.~Holdship et al.~\citeyear{holdship19}).

To progress in this 
compelling but controversial aspect of astrochemistry 
and put stringent constraints on models, (accurate) abundance measurements of S-bearing species must be provided. The approach adopted in previous studies was
twofold: either from line surveys of targeted objects with known physical structure (e.g.~Esplugues et al.~\citeyear{esplugues14}; Fuente et al.~\citeyear{fuente16}; Vastel et al.~\citeyear{vastel18}; Cernicharo et al.~\citeyear{cernicharo21}; Fuente et al.~\citeyear{fuente23}), 
or from selected atomic or molecular lines in source surveys (e.g.~Hatchell et al.~\citeyear{hatchell98}; Herpin et al.~\citeyear{herpin09}; Anderson et al.~\citeyear{anderson13}; Hily-Blant et al.~\citeyear{hily-blant22}).
Direct (Anderson et al.~\citeyear{anderson13}) %and indirect (Hily-Blant et al.~\citeyear{hily-blant22}) 
observational studies of atomic S in shocked regions or indirect (Hily-Blant et al.~\citeyear{hily-blant22}) studies of young starless or protostellar cores suggest that 
an important component of volatile sulphur in the ISM may be in gaseous atomic form. 
In comet 67P/Churyumov-Gerasimenko, the abundance of atomic S in the coma is very high, and the total elemental abundance does not show depletion with respect to the Solar photospheric abundance \citep{calmonte16,altwegg19}.
Model predictions indicate that H$_2$S, as well as organosulphur species and allotropes of S such as S$_8$, can be important sulphur reservoirs in ices \citep{lec19,shingledecker20}. These predictions are corroborated by laboratory experiments. 
For example, \citet{cazaux22} proposed that S$^+$ in translucent clouds can favour the formation of long S chains, contributing significantly to S depletion in dense regions.
But further, accurate abundance measurements of S-bearing molecules both in the gas-phase and on ices are required to better constrain  the abundances of sulphur reservoirs and their variation with evolution. 

In this framework, high-mass star-forming cores can play a relevant role. 
First, they have molecular spectra richer of lines than their low-mass counterparts, in particular in the stage of "hot molecular cores" (HMC, e.g.~Kurtz et al.~\citeyear{kurtz00}; Fontani et al.~\citeyear{fontani07}; Rivilla et al.~\citeyear{rivilla17}), but also in earlier evolutionary stages (e.g.~Vasyunina et al.~\citeyear{vasyunina11}; Taniguchi et al.~\citeyear{taniguchi18}; 
Coletta et al.~\citeyear{coletta20}; 
Mininni et al.~\citeyear{mininni21}).
Second, they are likely the birthplace of most stars in the Galaxy (e.g. Carpenter~\citeyear{carpenter00}; Evans et al.~\citeyear{evans09}), 
probably including the Sun (Adams~\citeyear{adams10}; Pfalzner~\citeyear{pfalzner13}), 
and hence the chemical evolution of these regions can give us relevant constraints on the environmental conditions of the birthplace of the Solar system.

In this work, we present observations of S-bearing molecules towards 15 high-mass star forming regions, equally divided in three evolutionary classes:
high-mass starless cores (HMSCs), which are infrared-dark, dense ($n\geq 10^{3}-10^{5}$)~\cmc, and cold (\Tk $\sim 10-20$~K) molecular clouds in an evolutionary stage immediately before (or at the very beginning of) the gravitational collapse; high-mass protostellar objects (HMPOs), which are collapsing cores with evidence of at least one deeply embedded infrared-bright protostar,
characterised typically by densities and temperatures higher than in the previous stage
($n\simeq 10^6$~\cmc, $T\geq 20$~K);
ultra-compact HII regions (UCHII), which are cores containing at least one embedded Zero-Age-Main-Sequence star associated with an expanding HII region, whose 
surrounding molecular cocoon ($n\geq 10^5$~\cmc, 
\Tk $\sim 20 - 100$~K) can be affected by its progressive expansion and by heating and irradiation from the central star.

The paper is organised as follows: the source sample, the observations, and the data reduction are described in Sect.~\ref{sampleobsred}; the results are shown in Sect.~\ref{res}, and discussed in Sect.~\ref{discu}. Conclusions and future perspectives are given in Sect.~\ref{conc}

%--------------------------------------------------------------------
\section{Sample, observations, and data reduction}
\label{sampleobsred}

\subsection{Sample}
\label{sample}

\begin{table}
\setlength\tabcolsep{1 pt}
\begin{center}
\caption[] {List of the observed sources. }
\label{tab:sources}
\normalsize
\begin{tabular}{lllll}
\hline \hline
source & RA(J2000) & Dec(J2000) & $V_{\rm LSR}$\tablefootmark{a} & acronym\tablefootmark{b} \\ %& $d$ & $L_{\rm bol}$ & Ref. \\
    & h:m:s & $\circ$:$\prime$:$\prime\prime$ & (\kms) &  \\ % & kpc & \soll\ & \\
\cline{1-5}
\multicolumn{5}{c}{HMSC} \\
\cline{1-5}
I00117--MM2   & 00:14:26.3     & +64:28:28 & $-36.3$ & 00117b \\ %& 1.8  & $10^{3.1}$ & (1) \\
AFGL5142--EC \tablefootmark{w}  & 05:30:48.7	&  +33:47:53 & $-3.9$ & AFGLa\\ %& 1.8 & $10^{3.6}$ & (2) \\
05358--mm3    \tablefootmark{w}  & 05:39:12.5 & +35:45:55 & $-17.6$ & 05358a \\ %& 1.8 & $10^{3.8}$ & (3,11) \\ 
%G034--G2(MM2) & 18:56:50.0 & +01:23:08 &  $+43.6$ \\ %& 2.9 & $10^{1.6}$\tablefootmark{r} & (4)  \\
%G034--F2(MM7) & 18:53:19.1  & +01:26:53 & $+57.7$ \\ % & 3.7 & $10^{1.9}$\tablefootmark{r}   & (4)  \\
%G034--F1(MM8) & 18:53:16.5  & +01:26:10 & $+57.7$ \\ % & 3.7 & -- & (4) \\
%G028--C1(MM9) & 18:42:46.9  & $-$04:04:08 & $+78.3$ \\ %  & 5.0  & -- & (4) \\
%G028--C3(MM11)\tablefootmark{a}   & 18:42:44  & $-$04:01:54 & $+78.3$ \\
I20293--WC    & 20:31:10.7  &   +40:03:28 & $+6.3$ & 20293a \\ %& 2.0 & $10^{3.6}$ & (5,6) \\
%I22134--G     \tablefootmark{w}  & 22:15:10.5  &	+58:48:59 & $-18.3$ \\ % & 2.6 & $10^{4.1}$ & (7) \\
I22134--B     & 22:15:05.8 &  +58:48:59 & $-18.3$ & 22134b \\ %& 2.6  & $10^{4.1}$ & (7) \\
% 22134+5834-3 (HMSC) & 22:15:06.77   &  +58:48:49.30 & \\
\cline{1-5}
\multicolumn{5}{c}{HMPO}   \\
\cline{1-5}
I00117--MM1   & 00:14:26.1      & +64:28:44 & $-36.3$ & 00117a \\ %& 1.8 & $10^{3.1}$ & (1) \\
%I04579--VLA1  & 05:01:39.9 &  +47:07:21 & $-17.0$ \\ %  & 2.5 & $10^{3.6}$ &  (8) \\
%AFGL5142--MM  & 05:30:48.0     & +33:47:54 &  $-3.9$ \\ %& 1.8  & $10^{3.6}$ & (2) \\
05358--mm1    & 05:39:13.1 & +35:45:51 & $-17.6$ & 05358b \\ %& 1.8  & $10^{3.8}$ & (3) \\
%18089--1732   & 18:11:51.4 & $-$17:31:28 & $+32.7$ \\ % & 3.6 & $10^{4.5}$ & (9) \\
18517+0437    & 18:54:14.2 & +04:41:41 & $+43.7$ & 18517 \\ %  & 2.9 & $10^{4.1}$ & (10) \\
%G75--core     & 20:21:44.0 &    +37:26:38 & $+0.2$  \\ %& 3.8 & $10^{4.8}$ & (11,12) \\
%I20293--MM1   & 20:31:12.8 &	+40:03:23 & $+6.3$ \\ %& 2.0 & $10^{3.6}$ & (5) \\
I21307        & 21:32:30.6  &    +51:02:16  & $-46.7$ & 21307 \\ %& 3.2 & $10^{3.6}$ & (13) \\ 
I23385        & 23:40:54.5 &      +61:10:28 & $-50.5$ & 23385 \\ % & 4.9 & $10^{4.2}$ & (14) \\
\cline{1-5}
\multicolumn{5}{c}{UCHII }   \\
\cline{1-5}
G75--core\tablefootmark{c}     & 20:21:44.0 &    +37:26:38 & $+0.2$ & G75 \\ %& 3.8 & $10^{4.8}$ & (11,12) \\
%G75.78+0.74 (HC HII) & 20:21:44.01 &	+37:26:37.6 & \\
%05137(UC)\tablefootmark{c}  &  05:17:13.3 &  +39:22:14 & $-25.4$  & \\
%G5.89--0.39   & 18:00:30.5    &    $-$24:04:01 & $+9.0$ \\ %& 1.28 & $10^{5.1}$ & (15,16) \\
%I19035--VLA1  & 19:06:01.5 &   +06:46:35 & $+32.4$ \\ % & 2.2 & $10^{3.9}$ & (11) \\
19410+2336    & 19:43:11.4 &    +23:44:06 & $+22.4$ & 19410 \\ % & 2.1 & $10^{4.0}$ & (17) \\
%ON1           & 20:10:09.1  &	 +31:31:36 & $+12.0$  \\ %& 2.5 & $10^{4.3}$ & (18,19) \\
I22134--VLA1  & 22:15:09.2 &    +58:49:08 & $-18.3$ & 22134 \\ %& 2.6 & $10^{4.1}$ & (11) \\
23033+5951    & 23:05:24.6 & +60:08:09 & $-53.0$  & 23033 \\ %& 3.5 & $10^{4.0}$ & (17) \\
NGC7538-IRS9  & 23:14:01.8   &   +61:27:20 & $-57.0$ & NGC7538 \\ % & 2.8 & $10^{4.6}$ & (8) \\
\hline
\end{tabular}
\end{center}
\tablefoot{
\tablefoottext{a}{velocity at which
we centred the spectra, corresponding to the systemic velocity;}
\tablefoottext{b}{acronym used in this paper;}
\tablefoottext{c}{Hyper-compact HII region.}
%\tablefoottext{a}{Source not included in paperI,
%selected from Butler \& Tan~(\citeyear{bet}). See also Butler et al.~(\citeyear{butler14});}
%\tablefoottext{b}{Observed in \H\ (1--0), \H\ (3--2), and \D\ (2--1);}
%\tablefoottext{c}{Observed in \H\ (1--0) and \D\ (2--1);}
\tablefoottext{w}{"warm" ($T \geq 20$~K) HMSCs externally heated \citep{fontani11}.}
%\tablefoottext{r}{Luminosity of the core and not of the whole associated star-forming region (Rathborne et al.~\citeyear{rathborne});}
%%\tablefoottext{n}{non-perturbed;} 
%\tablefoottext{1}{Palau et al.~(\citeyear{palau10})}
%\tablefoottext{2}{Busquet et al.~(\citeyear{busquet11})}
%\tablefoottext{3}{Beuther et al.~(\citeyear{beuther07b})}
%\tablefoottext{4}{Butler \& Tan~(\citeyear{bet})}
%\tablefoottext{5}{Palau et al.~(\citeyear{palau07})}
%\tablefoottext{6}{Busquet et al.~(\citeyear{busquet})}
%\tablefoottext{7}{Busquet~(\citeyear{busquetphd})}
%\tablefoottext{8}{S\'anchez-Monge et al.~(\citeyear{sanchez})}
%\tablefoottext{9}{Beuther et al.~(\citeyear{beuther04})}
%\tablefoottext{10}{Schnee \& Carpenter~(\citeyear{schnee})}
%\tablefoottext{11}{S\'anchez-Monge~(\citeyear{sanchez11})}
%\tablefoottext{12}{Ando et al.~(\citeyear{ando})}
%\tablefoottext{13}{Fontani et al.~(\citeyear{fonta04a})}
%%\tablefoottext{14}{S\'anchez-Monge et al.~(\citeyear{sanchez10})}
%\tablefoottext{14}{Fontani et al.~(\citeyear{fonta04b})}
%\tablefoottext{15}{Hunter et al.~(\citeyear{hunter})}
%\tablefoottext{16}{Motogi et al.~(\citeyear{motogi})}
%\tablefoottext{17}{Beuther et al.~(\citeyear{beuther02})}
%\tablefoottext{18}{Su et al.~(\citeyear{su})}
%\tablefoottext{19}{Nagayama et al.~(\citeyear{nagayama})}
}
\end{table}

We have studied 15 high-mass star-forming cores selected from the sample of \citet{fontani11}, and
extensively used to investigate specific aspects of (astro-)chemical evolution 
(e.g.~Fontani et al.~\citeyear{fontani14};~\citeyear{fontani15a};\citeyear{fontani15b};~\citeyear{fontani18};~\citeyear{fontani21}; Colzi et al.~\citeyear{colzi18}; Mininni et al.~\citeyear{mininni18}; Coletta et al.~\citeyear{coletta20}; Rivilla et al.~\citeyear{rivilla20}). 
We have selected a comparable number of sources belonging to the three evolutionary groups described in Sect.~\ref{intro} 
(i.e. HMSCs, HMPOs, and UCHIIs) without applying any specific selection criterion to avoid selection biases due to a specific physical parameter.
In \citet{fontani11}, five of these cores are classified as HMSCs, six as HMPOs, and four as UCHIIs. One of the HMPOs, G75, contains a hyper-compact HII region, and in this paper we have considered it as belonging to the UCHII group, bearing in mind that this object is in between the HMPO and the UCHII groups from the evolutionary point of view. 
The HMSCs AFGLa and 05358a have kinetic temperature, \Tk, higher than 20~K. 
Both cores are likely externally heated by a nearby protostellar core \citep{colzi19, rivilla20b} whose emission at 3mm could be partly included in the telescope beam, and are classified as "warm" cores in \citet{fontani11}.
In the case of 05358a, we observed also the nearby protostar 05358b.
The other HMSCs have \Tk$< 20$~K and are labelled as "quiescent" in \citet{fontani11}.
The targets are listed in Table~\ref{tab:sources}.
Additional information (e.g.~source distances, bolometric luminosities of the associated star 
forming regions, reference papers) are given in Table~1 of \citet{fontani11}.

\subsection{Observations}
\label{obs}

The observed transitions of S-bearing molecules are listed in Table~\ref{tab:lines}.
Rest frequencies, quantum numbers, energy of the lower and upper energy level, and Einstein coefficients are taken from the Cologne Database for Molecular Spectroscopy (CDMS, Endres et al.~\citeyear{endres16}) and the Jet Propulsion Laboratory (JPL, Pickett et al.~\citeyear{pickett98}).
Some transitions have been detected in the datasets described in \citet{fontani15b} and \citet{colzi18}. They are labelled in Table~\ref{tab:lines}. The others have been observed in January and March 2017 (IRAM 30m project 116-16, PI: Fontani). For the latter observing run,
observations were performed in band E2 of the EMIR receiver, covering the frequency ranges 250.540-258.320~GHz and 266.220-274.000~GHz with the Fast Fourier Transform spectrometer at 200 kHz resolution (FTS200).
We observed each source in Wobbler-switching mode, with a Wobbler throw of $\pm 120$\asec. The observations were calibrated with the Chopper-Wheel technique \citep{keu81}.
Pointing was checked at the beginning of each observing day, and every hour towards nearby quasars. Focus was checked on planets at the beginning of each observing run, and after sunset and sunrise.
System temperature was variable source by source, ranging from $\sim 170$~K to $\sim 700$~K, with an average value of $\sim 300$~K.

The spectra were obtained in antenna temperature units, $T_{\rm A}^*$ and converted in main beam temperature units, $T_{\rm MB}$, from the formula $T_{\rm A}^*=T_{\rm MB}(B_{\rm eff}/F_{\rm eff})=T_{\rm MB}\eta_{\rm MB}$, where $\eta_{\rm MB}=B_{\rm eff}/F_{\rm eff}$ is the ratio between the main beam efficiency and  the  forward  efficiency of the telescope.
At the frequencies observed in this work, $B_{\rm eff}\sim 0.46$ and $F_{\rm eff}\sim 0.88$.
For completeness, in Table~\ref{tab:lines-2} we show the spectral parameters of NO, a species that does not contain S but that we analyse in this paper for the first time to help the overall interpretation of the results (Sections~\ref{res} and~\ref{discu}).
In particular, the comparison between NS and NO is useful to explore the relative S/O ratio in the various sources, and investigate trends as a function of the evolutionary stage.

\begin{table*}
\caption{Spectral parameters of the observed lines of the analysed S-bearing species}    
\label{tab:lines}      
\centering          
\begin{tabular}{l c c c c c c}     % 7 columns 
\hline\hline 
   Line\tablefootmark{a}       & $\nu_0$         &   $E_{\rm l}$  &   $E_{\rm u}$ & $A_{\rm ul}$ &  Ref.\tablefootmark{b} & $\theta_{\rm MB}$\tablefootmark{c} \\
              & (GHz)          &   (K)     &  (K)   &  (s$^{-1}$)  &                  & (\arcsec) \\
\hline   
$^{13}$CS          &              &         &      &              &   &             \\
$J=2-1$           &  92.494308   &    2    &  7   &  $1.41\times 10^{-5}$   & \citet{fontani15b}  & $\sim 27$ \\
\hline
C$^{34}$S          &              &         &      &                &         &   \\
$J=2-1$           &  96.41294    &    2    &  6   &  $1.60\times 10^{-5}$  & \citet{fontani15b}  & $\sim 26$  \\
\hline        
SO            &              &         &      &                 &          &  \\
$J(K)=2(2)-1(1)$     &  86.09395    &    15   &  19  &  $0.50\times 10^{-5}$  & \citet{colzi18}  & $\sim 29$  \\
$J(K)=6(5)-5(4)$     &  219.949442  &    24   &  35  &  $1.34\times 10^{-4}$  & \citet{fontani15b}  & $\sim 11$  \\
$J(K)=5(6)-4(5)$     &  251.82577   &    39   &  51  &  $1.93\times 10^{-4}$   & this work & $\sim 10$  \\
$J(K)=8(9)-8(8)$     &  254.573628  &    87   &  100 &  $0.42\times 10^{-5}$   & this work & $\sim 10$  \\
$J(K)=6(6)-5(5)$     &  258.255826  &    44   &  56  &  $2.12\times 10^{-4}$   & this work & $\sim 10$  \\
$J(K)=3(4)-4(3)$     &  267.197746  &    16   &  29  &  $7.1\times 10^{-7}$   & this work  & $\sim 9$ \\
\hline        
SO$^+$     &              &         &      &          &                  \\
$J=11/2-9/2\;,\Omega=1/2\;,l=e$ & 254.977935   &   27    & 39   & $8.63\times 10^{-5}$       &  this work  & $\sim 10$ \\
$J=11/2-9/2\;,\Omega=1/2\;,l=f$ & 255.353237   &   27    & 39   & $8.67\times 10^{-5}$        & this work & $\sim 10$ \\
\hline        
NS\tablefootmark{d}            &              &         &      &                   &        &  \\
$J=11/2-9/2\;,F=13/2-11/2$   &              &         &      &      &                    &   \\
$\Omega=1/2\;,l=e$ &  253.570476  &    27   &  39 & $2.83\times 10^{-4}$   & this work   & $\sim 10$  \\
$J=11/2-9/2\;,F=13/2-11/2$   &              &         &      &         &                  & \\
$\Omega=1/2\;,l=f$ &  253.968393  &    28   &  40  & $2.84\times 10^{-4}$    & this work  & $\sim 10$  \\
\hline        
CCS           &              &         &      &                    &       & \\
$N(J)=7(6)-6(5)$    &  86.181391   &    19   &  23  &  $2.78\times 10^{-5}$   & \citet{colzi18}  & $\sim 29$  \\
$N(J)=7(7)-6(6)$     &  90.686381   &    21   &  26  &  $3.29\times 10^{-5}$  & \citet{fontani15b}  & $\sim 27$  \\
$N(J)=7(8)-6(7)$     &  93.870107   &    15   &  20  &  $3.80\times 10^{-5}$   & \citet{fontani15b} & $\sim 26$ \\
$N(J)=13(12)-12(12)$ & 94.9394789 & 57 & 62 &  $3.9 \times 10^{-7}$  & \citet{fontani15b} & $\sim 26$ \\
$N(J)=12(11)-11(10)$ &  153.4497738  &   46   &  54  &  $1.66\times 10^{-4}$   & \citet{colzi18}  & $\sim 16$ \\
$N(J)=17(18)-16(17)$ & 221.0711222 &  88 &  98 &  $5.07 \times 10^{-4}$  & \citet{fontani15b} & $\sim 11$ \\
$N(J)=7(6)-7(7)$ & 223.5624057 &  15 &  26 &  $3.06 \times 10^{-6}$  & \citet{fontani15b} & $\sim 11$ \\
$N(J)=10(10)-9(10)$ & 254.2139143 &  31 &  43 &  $3.21 \times 10^{-6}$  & \citet{fontani15b} & $\sim 10$ \\
\hline        
HCS$^+$          &              &         &      &                  &      &    \\
$J=2-1$       &  85.34789    &    2    &  6   &  $1.11\times 10^{-5}$   & \citet{colzi18}  & $\sim 29$ \\
$J=6-5$       &  256.0271    &    31   &  43  &  $3.46\times 10^{-4}$   & this work  & $\sim 10$ \\
\hline        
p-H$_2$S           &              &         &      &                 &           &  \\
$J(K_a,K_b)=2(2,0)-2(1,1)$ &  216.710437  &    74   &   84 &   $4.87\times 10^{-5}$   & \citet{fontani15b} & $\sim 11$ \\     
\hline        
OCS           &              &         &      &                   &      &   \\
$J=18-17$         &  218.903356  &    89   &  100 &   $3.04\times 10^{-5}$  & \citet{fontani15b}  & $\sim 11$ \\
$J=21-20$         &  255.374456  &    123  &  135 &   $4.84\times 10^{-5}$  & this work   & $\sim 10$ \\
$J=22-21$         &  267.530219  &    135  &  148 &   $5.57\times 10^{-5}$  & this work  & $\sim 9$ \\
\hline
SO$_2$     &            &         &      &                   &      &   \\
$J(K_a,K_b)=8(3,5)-9(2,8)$  &  86.6390877  &  51  & 55 &  $1.34 \times 10^{-6}$  &   \citet{colzi18} & $\sim 28$ \\
$J(K_a,K_b)=11(1,11)-10(0,10)$  &  221.9652196  &  50  & 60 & $1.14 \times 10^{-4}$ &  \citet{fontani15b} & $\sim 11$\\
$J(K_a,K_b)=13(1,13)-12(0,12)$  &  251.199675  &  70  & 82  &  $1.76 \times 10^{-4}$ & this work & $\sim 10$ \\
$J(K_a,K_b)=8(3,5)-8(2,6)$  &  251.2105851  &  43  & 55  &  $1.20 \times 10^{-4}$ & this work & $\sim 10$ \\
$J(K_a,K_b)=6(3,3)-6(2,4)$  &  254.2805358  &  29  & 41  &  $1.14 \times 10^{-4}$ & this work & $\sim 10$ \\
$J(K_a,K_b)=4(3,1)-4(2,2)$  &  255.5533022 &  19 &  31  &  $9.28 \times 10^{-5}$  & this work & $\sim 10$ \\
$J(K_a,K_b)=3(3,1)-3(2,2)$  &  255.9580440 &  15 &  28  &  $6.63 \times 10^{-5}$  & this work & $\sim 10$ \\
$J(K_a,K_b)=5(3,3)-5(2,4)$  &  256.2469451 &  24 &  36  &  $1.07 \times 10^{-4}$  & this work & $\sim 10$ \\
$J(K_a,K_b)=7(3,5)-7(2,6)$ & 257.0999657 &  35 &  48  &  $1.22 \times 10^{-4}$  & this work & $\sim 10$ \\
$J(K_a,K_b)=9(3,7)-9(2,8)$  &  258.9421992 &  51 &  63  &  $1.32 \times 10^{-4}$  & this work & $\sim 9$ \\
$J(K_a,K_b)=7(2,6)-6(1,5)$  &  271.5290139 &  22 &  35  &  $1.11 \times 10^{-4}$  & this work & $\sim 9$ \\
\hline        
CCCS          &              &         &      &                &         &   \\
$J=15-14$         &  86.708379   &    29   &   33 &   $5.04\times 10^{-5}$   & \citet{colzi18} & $\sim 29$ \\
$J=16-15$         &  92.488490   &    33   &   38 &   $6.13\times 10^{-5}$  & \citet{fontani15b}  & $\sim 27$ \\
\hline        
o-H$_2$CS          &              &         &      &                 &    &       \\
$J(K_a,K_b)=8(1,8)-7(1,7)$ &  270.521931  &    59   &  72  &  $2.90\times 10^{-4}$  & this work  & $\sim 9$  \\
\hline              
\end{tabular}
\tablefoot{For CCS, H$_2$S, CCCS, and SO$_2$, only the transitions with $E_{\rm u}\leq 100$K are listed;
\tablefoottext{a}{All parameters are taken from the CDMS catalogue \citep{endres16}, except that for SO$^+$, for which we use the Jet Propulsory Laboratory (JPL) catalogue \citep{pickett98};}
\tablefoottext{b}{Reference paper where the observations are presented;}
\tablefoottext{c}{Beam size;}
\tablefoottext{d}{we list only the two strongest hyperfine components.}
}
\end{table*}

\begin{table*}
\caption{Spectral line parameters of the NO molecule}    
\label{tab:lines-2}      
\centering          
\begin{tabular}{l c c c c c c}     % 6 columns 
\hline\hline 
   line       & $\nu_0$         &   $E_{\rm l}$  &   $E_{\rm u}$ & $A_{\rm ul}$ &  Ref.\tablefootmark{a} & $\theta_{\rm MB}$\tablefootmark{b}\\
              & GHz          &   K     &  K   &  s$^{-1}$  &                  \\
\hline 
 NO         &    & & & & \\
 $J = 5/2 - 3/2$, $\Omega = 1/2$, $F = 7/2-5/2$  & 250.796436  & 7.2  & 19.3  &  $1.85\times 10^{-6}$  & this work & $\sim 10$\\
$J = 5/2 - 3/2$, $\Omega = 1/2$, $F = 5/2-3/2$  & 250.815594  & 7  & 19  &  $1.55\times 10^{-6}$  &  this work & $\sim 10$ \\
\hline
\end{tabular}
\tablefoot{All parameters are taken from the CDMS catalogue \citep{endres16}.
\tablefoottext{a}{Reference paper where the observations are presented;}
\tablefoottext{b}{Beam size;}
}
\end{table*}

\subsection{Data reduction}
\label{reduction}

The first steps of the data reduction (e.g. average of the scans, baseline removal, flag of 
bad scans and channels) were made with the {\sc class} package of the 
{\sc gildas}\footnote{https://www.iram.fr/IRAMFR/GILDAS/} software using standard procedures.
Then, the baseline-subtracted spectra in main beam temperature ($T_{\rm MB}$) units were fitted with the MAdrid Data CUBe Analysis 
({\sc madcuba}\footnote{{\sc madcuba} is a software developed in the Madrid Center of Astro-biology (INTA-CSIC) which enables to visualise and analyse single spectra and data cubes: https://cab.inta-csic.es/madcuba/.}, Mart\'in et al.~\citeyear{martin19}) software.

The transitions of S-bearing species listed in Table~\ref{tab:lines}
were identified via the Spectral Line Identification and 
Local Thermodymanic Equilibrium Modelling (SLIM) tool of {\sc madcuba},
which assumes Local Thermodynamic Equilibrium (LTE) conditions. 
The lines were fitted with the AUTOFIT function of SLIM. 
This function produces the synthetic spectrum that best matches the data assuming as input parameters: total molecular column density 
($N_{\rm tot}$), radial systemic velocity of the source ($V$), line full-width at half-maximum (FWHM), excitation temperature (\Tex),
and angular size of the emission ($\theta_{\rm S}$). AUTOFIT assumes that $V$, FWHM, $\theta_{\rm S}$, and \Tex\
are the same for all transitions fitted simultaneously. 
In each source, all transitions of a given molecule have been fit simultaneously.

The input parameters have all been left free except $\theta_{\rm S}$, for which we assumed that the emission fills the telescope beam.
This assumption could not be appropriate in some sources and some lines. However, we do not have interferometer observations of the observed lines from which the angular emitting size can be estimated. Some targets have been observed at high-angular resolution in S-bearing molecules. 
For example, \citet{beuther09} observed 05358-mm3 and derived angular emitting sizes of a few arcseconds in C$^{34}$S and SO$_2$ but in lines at much higher excitation (i.e. above 100~K) than ours. 
\citet{wang16} observed the SO $J(K)=6(5)-5(4)$ transition towards I22134-VLA1, and found several clumps of a few arcseconds within the beam of our data, which is unlikely to be representative of all transitions we have observed. 
Therefore, we decided to assume a unity filling factor, bearing in mind that this can introduce large uncertainties in the individual column densities. 
However, in computing the fractional abundances, the neglected dilution factor in the molecular column density is compensated (at least partially) by the comparable neglected dilution factor on the H$_2$ column density.

Some spectra show deviations from the LTE approximation, especially in the SO and SO$_2$ lines. 
Figs.~\ref{fig:spec-so-hmsc}, \ref{fig:spec-so-hmpo} and \ref{fig:spec-so-uchii} show that in some sources (e.g. AFGLa, G75, 05358b, NGC7538), some lines of SO and SO$_2$ are underestimated by the best fit, while others are overestimated. 
The lines that are underestimated by the fit could be either blended with nearby transitions of other species, and/or be in non-LTE conditions. 
The lines that are overestimated are likely in non-LTE conditions, for example due to multiple components with different temperatures that make the approximation of a single \Tex, inappropriate.

In this respect, some sources require a comment:

(i) the warm HMSC AFGLa (Fig.~\ref{fig:spec-so-hmsc}): the SO $J(K)=6(5)-5(4)$ line is underestimated by a factor three by AUTOFIT. We have used RADEX on-line\footnote{https://var.sron.nl/radex/radex.php} to investigate the effect of non-LTE conditions.
Using the line width estimated in the LTE approach, that is 4.4~\kms, assuming \Ntot~$5\times 10^{14}$\cmq, \Tex = 100~K, and a H$_2$ volume density of $\sim 10^6$\cmc, the intensity of the $J(K)=6(5)-5(4)$ transition is $\sim 6$~K and that of the $J(K)=2(2)-1(1)$ line is $\sim 1$~K, as observed. 
However, the other two lines should be stronger than observed, maybe because of a different beam dilution not taken into account by RADEX.
Hence, even a non-LTE approach does not allow to fit all lines properly. 
The OCS lines also show deviations from the LTE approach (Fig.~\ref{fig:spec-ocs-hmsc}), but even in this case neither a non-LTE approach significantly improves the predicted intensity of the three lines, nor a fit with two velocity features with different line width. 

(ii) the HMPO 23385 (Fig.~\ref{fig:spec-so-hmpo}): the presence of two Gaussian features separated in radial velocity by $\sim +3.5$\kms\ is clear in SO $J(K)=2(1)-1(1)$ and $J(K)=6(5)-5(4)$, and maybe also in the CCCS line (Fig.~\ref{fig:spec-cccs}).
This second velocity feature was already revealed in previous observations of HCN and HNC isotopologues \citep{colzi18}, and it is likely due to a cloud south of 23385 detected in interferometer images \citep{fontani04}. Only the main component at the systemic velocity of $-49.5$~\kms\ has been analysed in the following.

\begin{figure*}
   \centering
   \includegraphics[width=15cm]{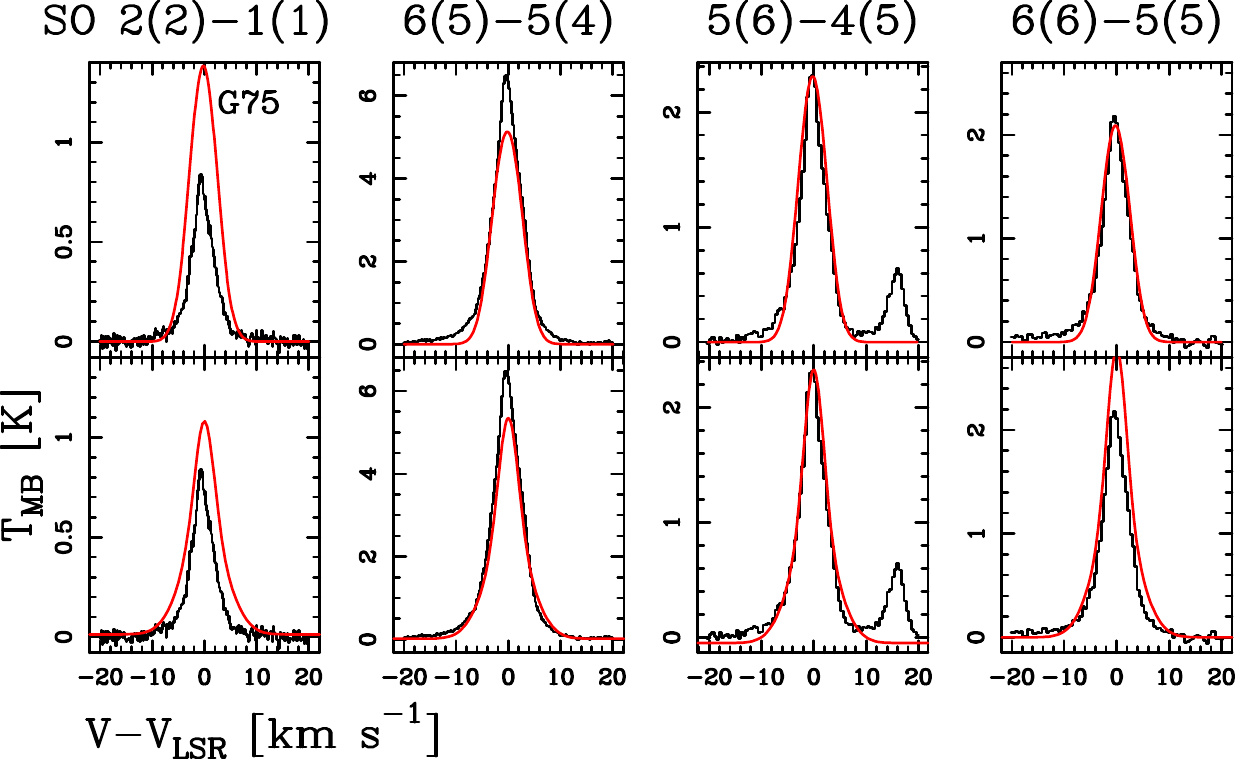}
      \caption{Best fit to the SO lines towards the UCHII G75. On the x-axis we show the velocity shift from the systemic velocity $V_{\rm LSR}$ listed in Table~\ref{tab:sources}.
      The red curves superimposed on the observed spectrum (in black) represent the best fit with one Gaussian component in the upper panels, and with two Gaussian components in the lower panels.}
         \label{fig:G75-2gauss}
   \end{figure*}

(ii) {\it the UCHII G75:} the SO $J(K)=2(2)-1(1)$ line is overestimated by the LTE fit. 
%In this case, assuming two Gaussian velocity components (a narrow one and a broad one at the same radial velocity), improves the best fit of the $J(K)=2(2)-1(1)$ line, but worsens that of the other lines.
In Fig.~\ref{fig:G75-2gauss} we show the best fit to the SO lines considering one and two Gaussian components.
The fit with only one Gaussian provides a FWHM of $\sim 6.2$~\kms, and there is a clear residual in the high velocity wings.
The fit with two Gaussian components provides FWHM of $\sim 4.2$~\kms\ and $\sim 9.6$~\kms, and the profile is better fitted in the wings.
Similarly, assuming two Gaussian components improves the fit of the SO$_2$ $J(K_a,K_b)=11(1,11)-10(0,10)$ line in G75 and also in NGC7538, but worsens that of the other lines.
However, the column density of the most intense component of the 2-Gaussian method is smaller but comparable within the errors as the one obtained with one component only.
Therefore, in summary, in the sources in which some lines are under- or overestimated by the fit in LTE with a single Gaussian component there is likely a mix of non-LTE effects and multiple velocity features. 
Because the relative effect of both features is not easy to estimate, we will adopt the results obtained from the best-fit LTE approach using a single Gaussian component, except when an alternative approach improves significantly (i.e. difference larger than the uncertainties) the residuals.

Finally, the OCS $J=22-21$ line is often contaminated by the broad high-velocity red wing of HCO$^+$ $J=3-2$ at 267557.6259~MHz. 
Therefore, we have simultaneously fit OCS and HCO$^+$ to properly derive the OCS best fit in the sources with clear contamination.

\section{Results}
\label{res}

The spectra of the lines listed in Tables~\ref{tab:lines} and \ref{tab:lines-2}, as well as their best fit obtained with {\sc madcuba} (Sect.~\ref{reduction}), are shown in Figures~\ref{fig:spec-13cs} - \ref{fig:spec-no} in Appendix~\ref{app:spectra}. 
For the species with more than three lines detected, e.g. SO, CCS, and SO$_2$, we show only the most intense ones.

\subsection{Line profiles}
\label{profiles}

The best fit results for all sources and all molecular species are reported in Appendix~\ref{app:tables}.
Tables~\ref{tab:velocities} and \ref{tab:velocities2} show the best fit radial centroid velocities, and
tables~\ref{tab:linewidths} and \ref{tab:linewidths2} list the best fit line widths at half maximum.
Both parameters have been obtained with {\sc madcuba} as described in Sect.~\ref{reduction}.

In most cases, the lines are very well fit with a single Gaussian. This is especially apparent in $^{13}$CS, C$^{34}$S, SO$^+$, NS, CCS, HCS$^+$, CCCS, and NO.
The transitions detected in this work have upper level energies up to $\sim 40$~K (Table~\ref{tab:lines}), suggesting that they likely trace relatively cold material. 
Their spectral shapes confirm, overall, this interpretation, even though in some sources clear high-velocity wings are present: the warm HMSC AFGLa, and the two HII regions G75 and NGC7538.
The transitions that are likely associated with shocked or warmer gas are those of SO, OCS, p-H$_2$S, and SO$_2$. In fact, their upper level energies are higher than 40~K, and they show often high-velocity wings.
These wings are clearly present in the three sources already mentioned, and also towards: the warm HMSC 05358a, the HMPOs 05358b, 18517, and 23385, and all UCHII regions except 22134.
The SO$^+$, NS, and NO transitions do not show high-velocity wings, however their signal-to-noise ratios are on average much lower than those of the transitions discussed above, and hence the lack of these wings can be due to the limited sensitivity.

\begin{figure*}
   \centering
   \includegraphics[width=16cm]{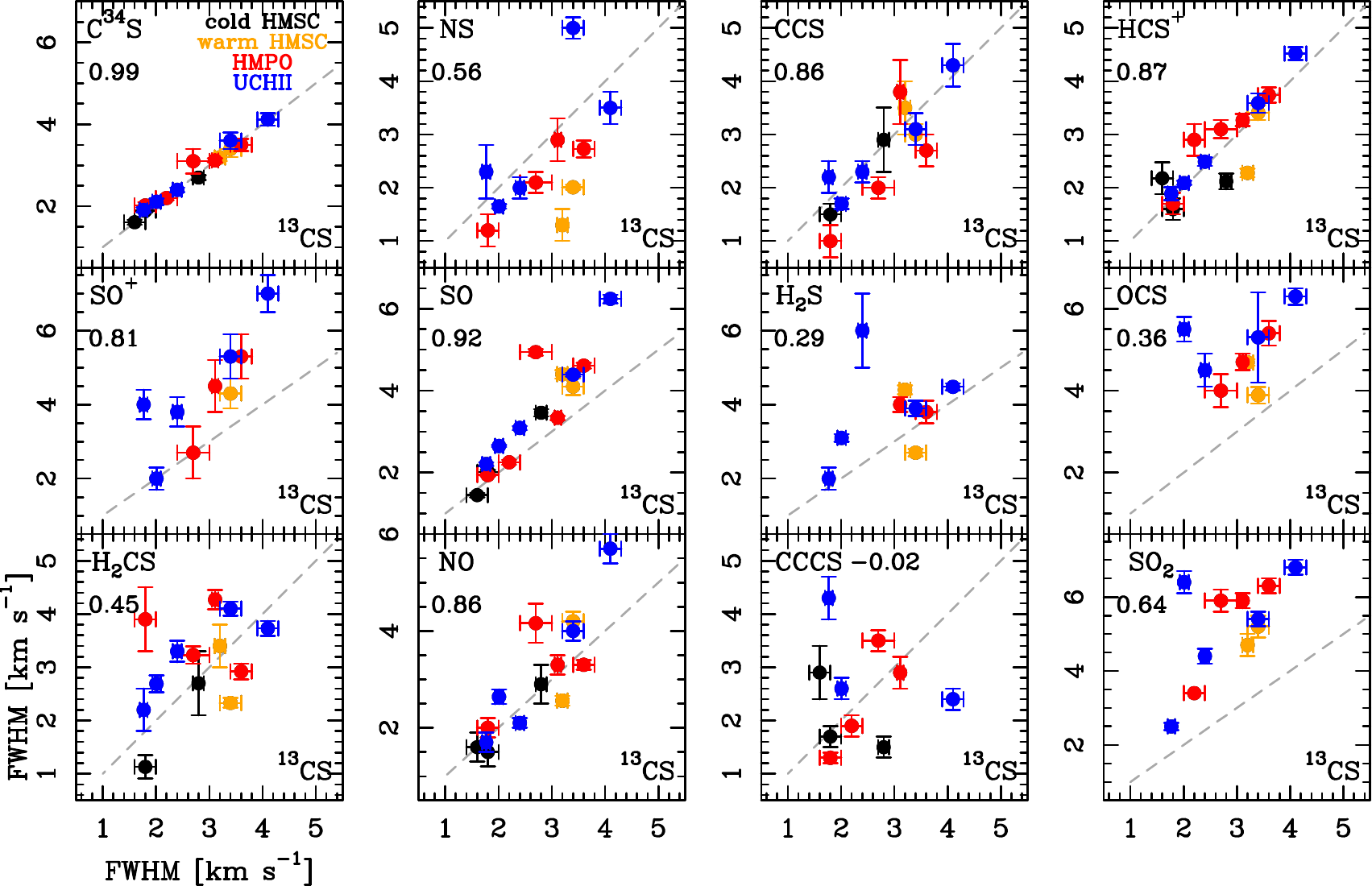}
      \caption{Comparison between the lines FWHMs of the observed species and those of $^{13}$CS. 
      In all plots, the black points indicate the cold HMSCs, the orange ones the two warm HMSCs, the red ones the HMPOs, and the blue ones the UCHII regions.
      The dashed grey line indicates the locus y=x.
      The number in the upper left corner of each panel is the Pearson's $\rho$ correlation coefficient.}
         \label{fig:fwhm}
   \end{figure*}

\begin{figure*}
   \centering
   \includegraphics[width=16cm]{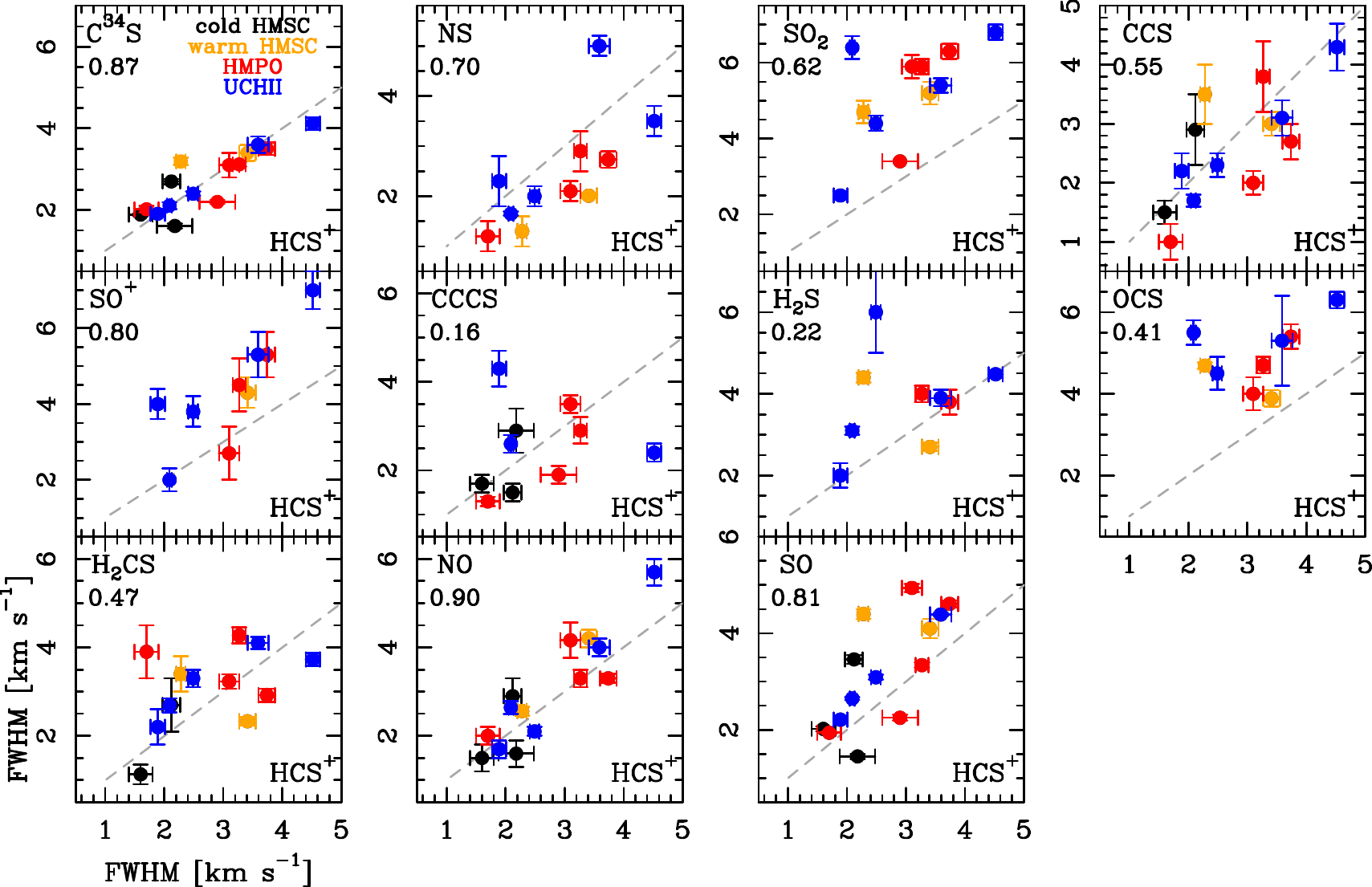}
      \caption{Same as Fig.~\ref{fig:fwhm}, but using HCS$^+$ as reference species. The comparison with $^{13}$CS is shown already in Fig.~\ref{fig:fwhm}.}
         \label{fig:fwhm-HCSp}
   \end{figure*}

\begin{figure*}
   \centering
   \includegraphics[width=16cm]{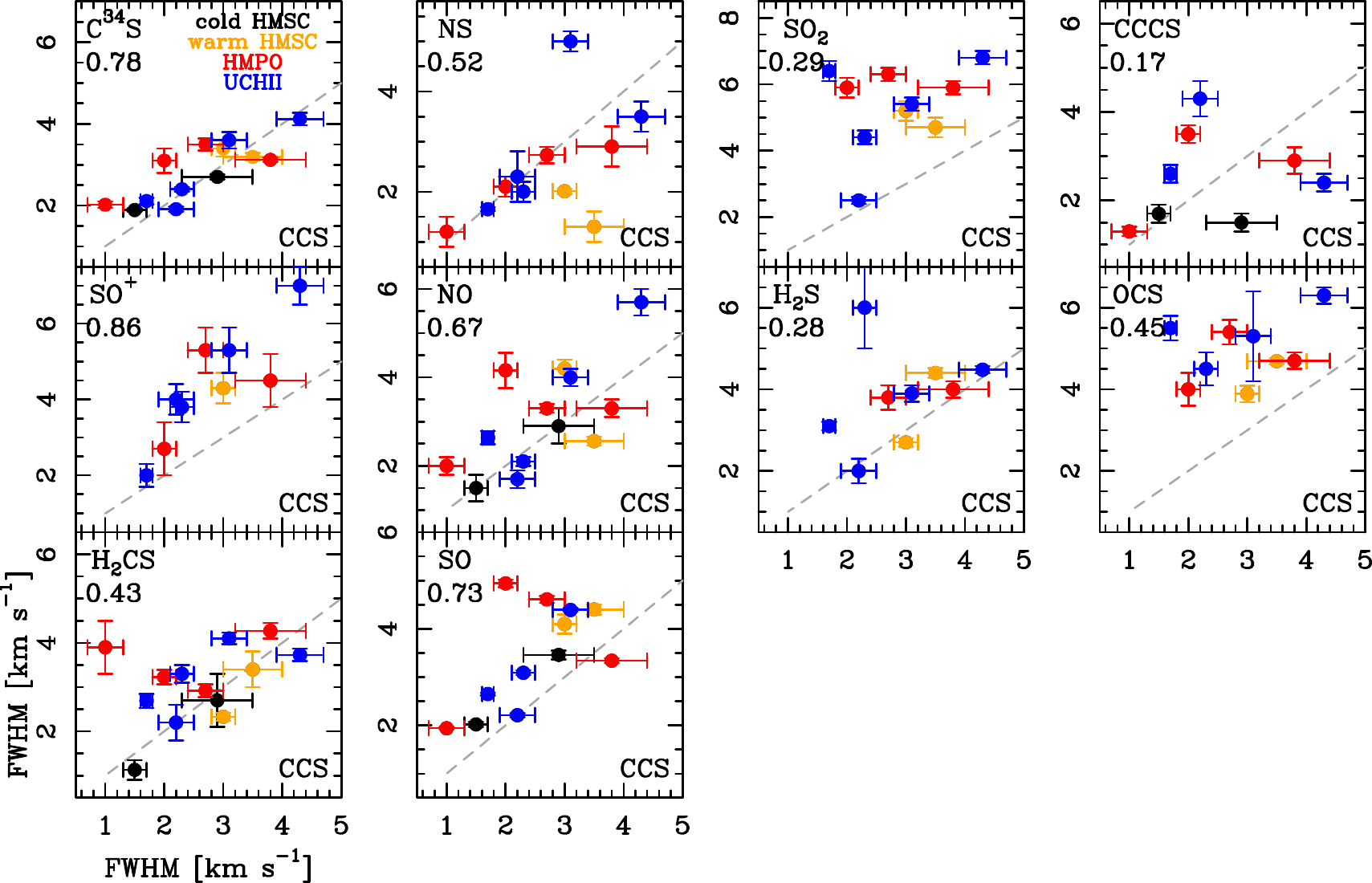}
      \caption{Same as Fig.~\ref{fig:fwhm}, but using CCS as reference species. The comparison with $^{13}$CS and HCS$^+$ is shown already in Figs.~\ref{fig:fwhm} and \ref{fig:fwhm-HCSp}, respectively.}
         \label{fig:fwhm-CCS}
   \end{figure*}

Figures~\ref{fig:fwhm} -- \ref{fig:fwhm-CCS} show the comparison between the FWHM of the lines obtained with {\sc madcuba} as explained in Sect.~\ref{reduction}.
Among all possible combinations, we have chosen $^{13}$CS as reference species being the best tracer of quiescent gas (Fig.~\ref{fig:fwhm}).
We also investigate correlations with HCS$^+$ and CCS, which both have a temperature estimate. 
Such comparisons allow us to understand the species that are more likely associated with similar gas, and to check for significant variations among the different lines/species.

Figure~\ref{fig:fwhm} indicates that the FWHMs of the $^{13}$CS lines are positively correlated with those of almost all species, except CCCS. The correlation is perfect with C$^{34}$S (Pearson's $\rho$ correlation coefficient 0.99), good with HCS$^+$, SO, NO, CCS, and SO$^+$ ($\rho \sim 0.81-0.92$), and faint with H$_2$CS, NS, H$_2$S, OCS, and SO$_2$ ($\rho\sim 0.29-0.64$).
Clearly, the $^{13}$CS, C$^{34}$S, and HCS$^+$ observed transitions trace the same gas, based on their correlations and on the almost identical range of FWHMs measured.
This is consistent with the fact that these species are all strictly chemically related to CS, and the observed transitions have the same quantum numbers (except for HCS$^+$ $J=6-5$). 
CCS is also tightly correlated with $^{13}$CS, and in
all these species (C$^{34}$S, $^{13}$CS, HCS$^+$, and CCS), the FWHM never exceeds $\sim 5$~\kms, indicating that they are all associated with relatively quiescent material. 

Interestingly, the FWHMs derived from NS are narrower than those of $^{13}$CS except than in two UCHII regions, perhaps indicating that NS tends to traces the most quiescent and extended material in the less evolved stages.
There is a very good positive correlation also with SO ($\rho=0.92$), but SO lines have systematically larger FWHM. 
This suggests that the SO emission is affected by a warmer, denser, and more turbulent component, for example associated with outflow cavities, which likely adds to the quiescent component.
Indeed, high-velocity non-Gaussian wings in SO are found towards HMPOs and UCHIIs, while in cold HMSCs only the quiescent component is detected.
The FWHMs of the other O-bearing molecules, namely SO$^+$, OCS, and SO$_2$, are systematically larger than those of $^{13}$CS as well. This, and the additional evidence of hints of non-Gaussian wings in the spectra of OCS and SO$_2$ (see Figs.~\ref{fig:spec-ocs-hmpo},~\ref{fig:spec-ocs-uchii},~\ref{fig:spec-so2-hmpo},~\ref{fig:spec-so2-uchii}), indicate that these species trace more turbulent material. 
The nature of the SO$^+$ emission is not easy to determine: its FWHMs are well correlated with those of $^{13}$CS, but, like SO, are systematically larger. 
Hence, they could contain both a quiescent and a turbulent component like SO. 
However, the lines do not show clear hints of non-Gaussian wings. 
This could be due to the fact that the SO$^+$ lines are fainter, and hence the lower signal-to-noise ratio in the spectra prevents us to detect the high-velocity emission in the wings
%Inspection of Fig.~\ref{fig:fwhm-SO} indicates that the FWHM of SO is positively correlated with almost all tracers except NS, OCS, CCCS, and o-H$_2$CS. 
%The smallest dispersions are with C$^{34}$S, $^{13}$CS, and NO, and in all cases the SO lines are broader. 
%Therefore, the SO emission that we detect is likely made of a quiescent envelope component, and a warmer, denser, and more turbulent component (e.g. associated with outflow cavities).
%Which of the two components is dominant depends on the target. 
%Indeed, FWHMs larger than $\sim 3$~\kms\ and high-velocity non-Gaussian wings are found only towards HMPOs and UCHIIs, with the exception of the warm HMSC AFGLa, while in cold HMSCs only the quiescent component is detected.

About NO, this molecule is found to be associated with both envelope material and outflow cavities in low-mass star-forming cores \citep{codella18}. 
In Fig.~\ref{fig:fwhm} (see also Table~\ref{tab:linewidths2}) one can see that the FWHMs of NO are smaller than those of SO, indicating that the emission from the envelope component is dominant in our NO lines. 
This is futher supported by the lack of non-Gaussian wings in the spectra (see Fig.~\ref{fig:spec-no}), and it is
consistent with the larger range in $E_{\rm u}$ of the SO lines with respect to the NO ones (19--100~K against 19~K, Tables~\ref{tab:lines} and \ref{tab:lines-2}).

Figures~\ref{fig:fwhm-HCSp} and \ref{fig:fwhm-CCS} show the comparison between the lines FWHM of HCS$^+$ and CCS and those of the other tracers. 
The FWHMs of HCS$^+$ correlate very well with those of C$^{34}$S, SO, SO$^+$, and NO ($\rho \geq 0.8$).
However, the relations closest to a y=x relation are found with C$^{34}$S and NO, while SO and SO$^+$ have larger FWHMs. Those of CCS (Fig.~\ref{fig:fwhm-CCS}) correlates well ($\rho \geq 0.8$) only with SO$^+$ and C$^{34}$S, but the range of values indicate that SO$^+$ is associated with more turbulent material, as for HCS$^+$.
%Because for HCS$^+$ and CCS we can estimate the excitation temperature, the plots shown in Figs.~\ref{fig:fwhm-HCSp}, and \ref{fig:fwhm-CCS} are useful to check the species that are most likely associated with similar gas.
In summary, the lines of the C- and N-bearing species tend to trace the most quiescent, and probably extended, envelope of the sources. 
In particular NS seems associated with the most quiescent gas in early sources, and then $^{13}$CS, C$^{34}$S, HCS$^+$, and CCS trace a similar, but maybe slightly inner, envelope.
The O-bearing species are associated with more turbulent, and thus likely inner, regions of the cores based both on the larger FWHMs and on the presence of non-Gaussian wings in evolved objects. 
Among these molecules, NO seems to trace the outermost layers, then SO and SO$^+$ arise from slightly more turbulent inner regions, and finally OCS and SO$_2$ from the most turbulent inner regions.
The origin of o-H$_2$CS, p-H$_2$S, and CCCS is very source-dependent, but while the o-H$_2$CS and p-H$_2$S lines show hints of non-Gaussian high velocity wings in some targets (05358a, 05358b, 23385, 18517, G75, NGC7538),
the CCCS lines are always Gaussian and hence likely associated uniquely with an envelope component.
%The larger dispersions are with SO$^+$, CCS, p-H$_2$S, and SO$^2$. 

%$We have checked also the differences between the centroid line velocities and the $V_{\rm LSR}$.
%Again, we use as reference values the difference $V-V_{\rm LSR}$ of $^{13}$CS as template for C-bearing species, and SO for O-bearing species. 
%This comparison is illustrated in Fig.~\ref{fig:velocities}.
%The plots confirm a correlation close to the $y=x$ line between $^{13}$CS, C$^{34}$S, and HCS$^+$. 
%The other comparisons show also positive correlations, but are associated with large dispersions and are, hence, less compelling than for the line widths at half maximum.
%should we keep this figure or not?}

%        \begin{figure*}
%   \centering
%   \includegraphics[width=16cm]{correlations-fwhm-SO.pdf}
%      \caption{Same as Fig.~\ref{fig:fwhm}, but using SO as reference species.}
%         \label{fig:fwhm-SO}
%   \end{figure*}

%     \begin{figure*}
%   \centering
%   \includegraphics[width=16cm]{correlations-velocities.pdf}
%      \caption{Comparison between the centroid velocities of the lines with respect to \VLSR. In all plots, the black points indicate the cold HMSCs, the orange ones the two warm HMSCs, the red squares the HMPOs, and the blue penthagons the UCHII regions.
%      The dashed grey line indicates the locus y=x.}
%         \label{fig:velocities}
%   \end{figure*}

\subsection{Excitation temperatures and total column densities}
\label{column}

The excitation temperatures, \Tex, of the molecules listed in Tables~\ref{tab:lines} and \ref{tab:lines-2} are shown in Table~\ref{tab:temperatures}. The molecular total column densities are listed in Tables~\ref{tab:column1} and \ref{tab:column2}. They have been derived assuming LTE conditions as described in Sect.~\ref{reduction}.
\Tex\ were derived from {\sc madcuba} as explained in Sect.~\ref{reduction} for the five species for which we detected at least two transitions with different energies, namely SO, CCS, HCS$^+$, OCS, and SO$_2$.
For the species for which we have only one transition, that is $^{13}$CS, C$^{34}$S, p-H$_2$S, o-H$_2$CS, and CCCS, or multiple transitions with too similar energies of the levels (i.e. SO$^+$, NS, and NO), we had to fix \Tex\ to compute \Ntot. The \Tex\ adopted is discussed below.

\begin{figure*}
   \centering
   \includegraphics[width=16cm]{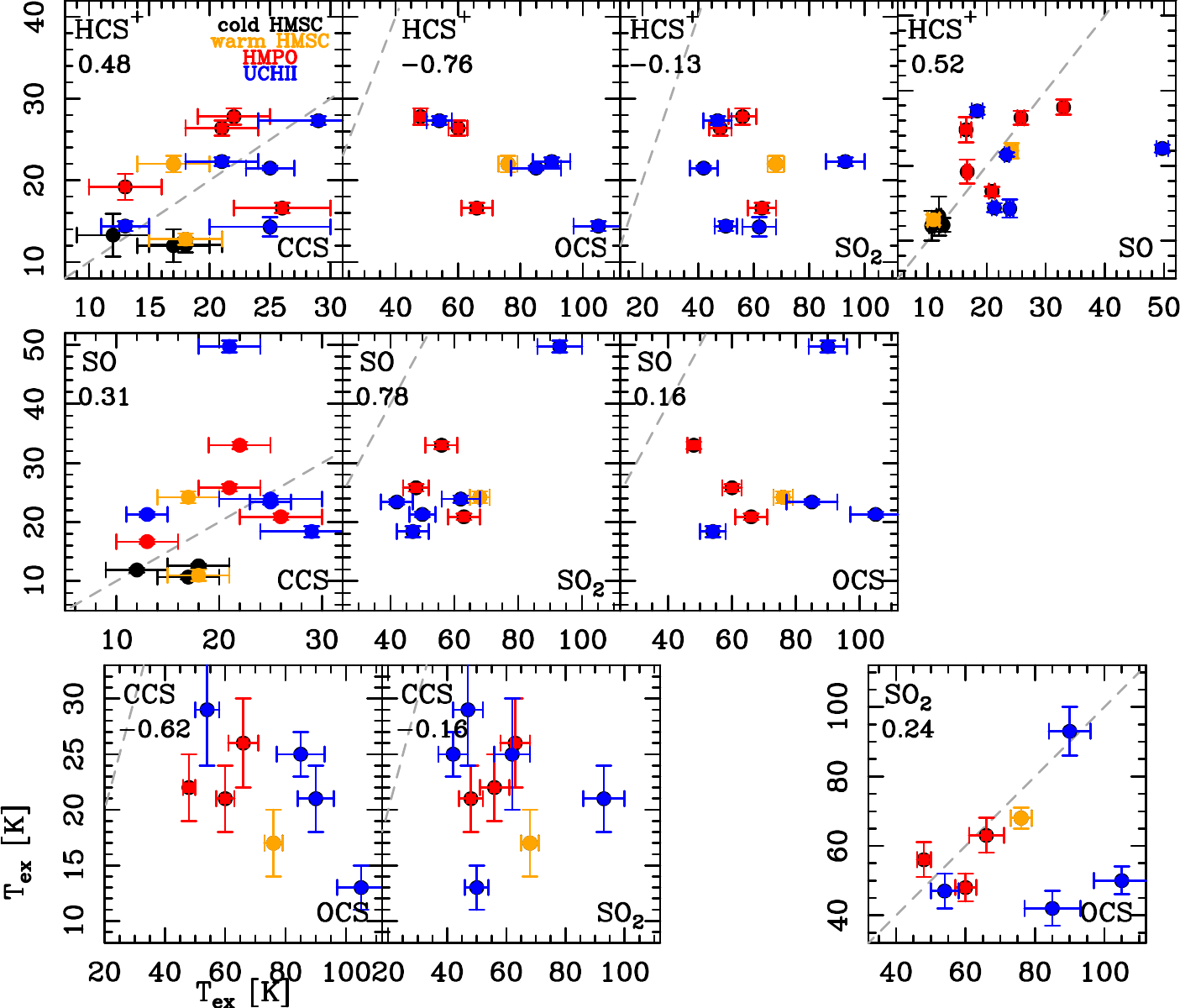}
      \caption{Comparison between the excitation temperatures of the various molecules. In all plots, the black points indicate the cold HMSCs, the orange ones the two warm HMSCs, the red ones the HMPOs, and the blue ones the UCHII regions.
      The dashed grey line indicates the locus y=x.}
         \label{fig:temperatures}
   \end{figure*}

In Figure~\ref{fig:temperatures} we investigate correlations between each pair of measured excitation temperatures.  
As done for the FWHMs, we make a quantitative analysis computing the Pearson's $\rho$ correlation coefficients.
A clear positive correlation is found only between SO and SO$_2$ ($\rho=0.78$), even though the trend is strongly influenced by one source only. Fainter, still positive correlations are found between HCS$^+$ and CCS, HCS$^+$ and SO, SO and CCS. Negative correlations are found between HCS$^+$ and OCS, and CCS and OCS.
The tracers associated with gas having temperature lower than $\sim 30$~K are HCS$^+$ and CCS. 
The molecule SO also traces relatively cold gas because its \Tex\ is between $\sim 10-33$~K, except for the UCHII region NGC7538, for which \Tex\ from SO is $\sim 50$~K. This source is also responsible for the positive correlation between SO and SO$_2$, which would be not significant without it ($\rho\sim 0.14$), and it is hence likely an outlier.
The tracers that are clearly associated with material warmer than $\sim 40$~K are SO$_2$ and OCS, both having temperature above 45~K. 
All this is consistent with the previous suggestion based on the lines FWHMs that the lines of the C-bearing species trace a colder extended envelope, SO maybe a denser and more compact but still relatively cold region, and OCS and SO$_2$ the warmer and likely innermost regions.

%Figure~\ref{fig:temperatures} does not show any clear correlation, except for SO and HCS$^+$.
%However, an anticorrelation between HCS$^+$ and OCS, as well as CCS and OCS, may be present.

The correlations between the FWHM of the lines shown in Figs.~\ref{fig:fwhm} -- \ref{fig:fwhm-CCS} allow us to determine which species are most likely arising from similar gas, and hence to assign to the molecules for which we do not have a direct \Tex\ estimate the best representative one. 
Fig.~\ref{fig:fwhm-HCSp} shows that $^{13}$CS, C$^{34}$S, NS, and o-H$_2$CS, have a good positive correlation with the FWHM of HCS$^+$, and a similar velocity range.
Hence, we have fixed \Tex\ from that of HCS$^+$ to derive the column density of these species.
Among the O-bearing species, the molecule NO shows the best positive correlation with HCS$^+$ ($\rho=0.90$). However, we have computed the correlation also with SO, and found a similar coefficient ($\rho=0.91$). 
Therefore, due to the fact that NO and SO are oxygen-containing species while HCS$^+$ is not, we have computed the NO column density fixing \Tex\ to that of SO. 
The FWHM of SO$^+$ also shows the best positive correlation with HCS$^+$ ($\rho=0.8$), but because of the fair correlation coefficient with SO ($\rho=0.65$), the similar range of FWHMs measured, and their chemical relation, we adopted \Tex\ from SO to compute the SO$^+$ column density.
The molecule p-H$_2$S does not show a compelling correlation with any of the tracers inspected.
We thus decided to adopt \Tex\ from SO because it is the most accurate estimate, owing to the large number of transitions with relatively wide energy range ($E_{\rm u}=19-56$~K) on which it is based.
For CCCS we do not have correlation with any tracer, and the dispersions are similar. We have used \Tex\ from HCS$^+$ to be consistent with other C-only-bearing species.

For undetected species, an upper limit on \Ntot\ has been evaluated fixing FWHM and \Tex\ to those of the species that likely trace similar gas. For example, for OCS and SO$_2$ we have used FWHM and \Tex\ from SO. 
Fixing FWHM and \Tex\ in this way, we have computed the upper limit on \Ntot\ simulating Gaussian lines below an intensity peak of $\sim 3\sigma$~rms.

\subsection{Molecular fractional abundances}
\label{abundances}

From the molecular column densities in Tables~\ref{tab:column1} and ~\ref{tab:column2}, 
we computed the fractional abundances, X, dividing them 
by the H$_2$ total column densities, $N({\rm H_2})$, given in \citet{fontani18}. 
These are average values within an angular beam of $28$\asec.
Because the total column densities have been estimated assuming that the emission is more extended than the beam, the abundances are hence average values within $28$\asec.
They are given in Tables~\ref{tab:abundances-1} and \ref{tab:abundances-2}.

\begin{figure*}
   \centering
   \includegraphics[width=16cm]{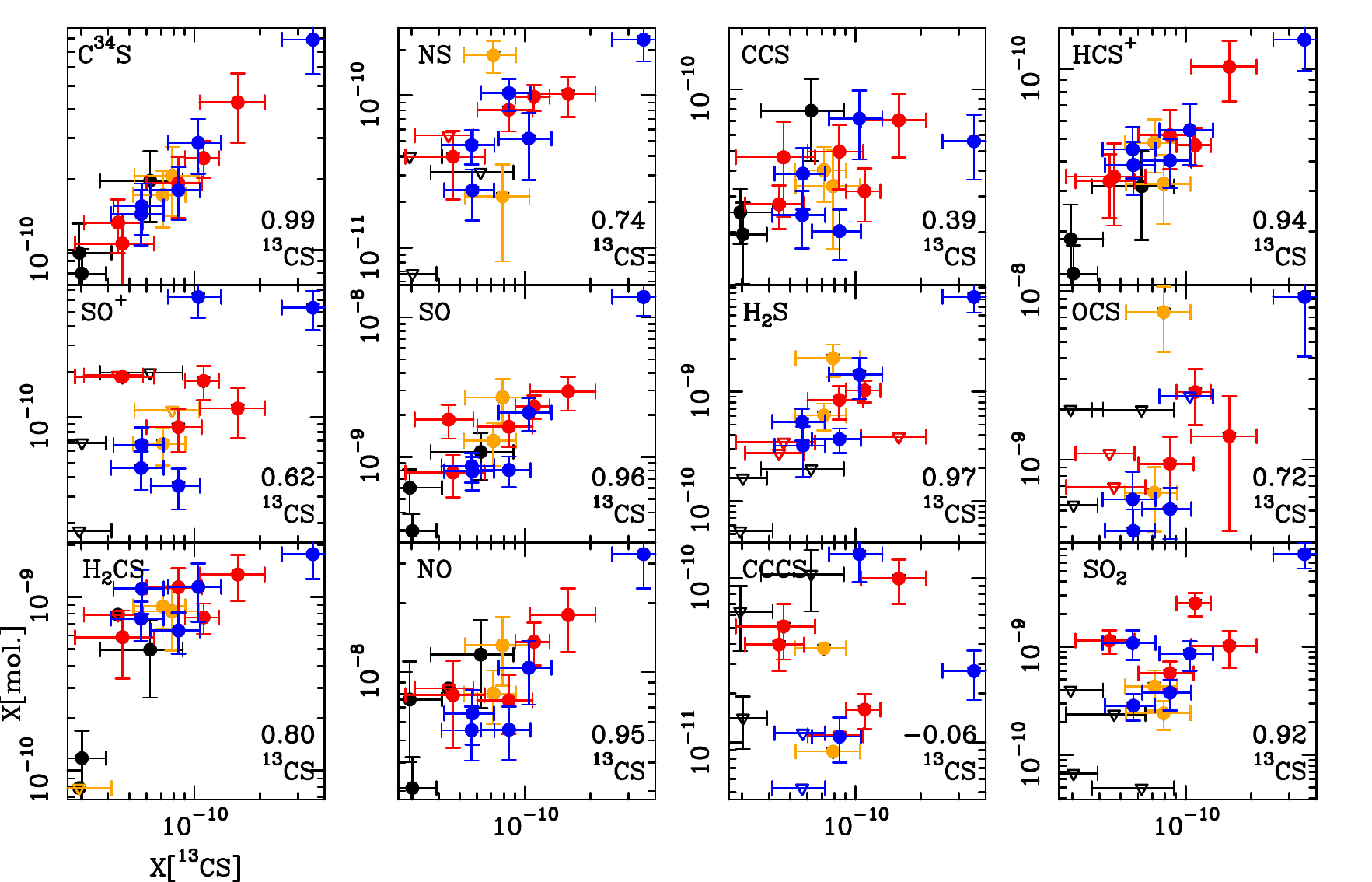}
      \caption{Comparison between the fractional abundances of the observed species and that of $^{13}$CS. In all plots, the black points indicate the cold HMSCs, the orange ones the two warm HMSCs, the red ones the HMPOs, and the blue ones the UCHII regions. 
      The empty triangles are upper limits.
      The number in the lower-right corner of each panel is the Pearson's $\rho$ correlation coefficient.}
         \label{fig:abb-corr}
   \end{figure*}

\begin{figure*}
   \centering
   \includegraphics[width=16cm]{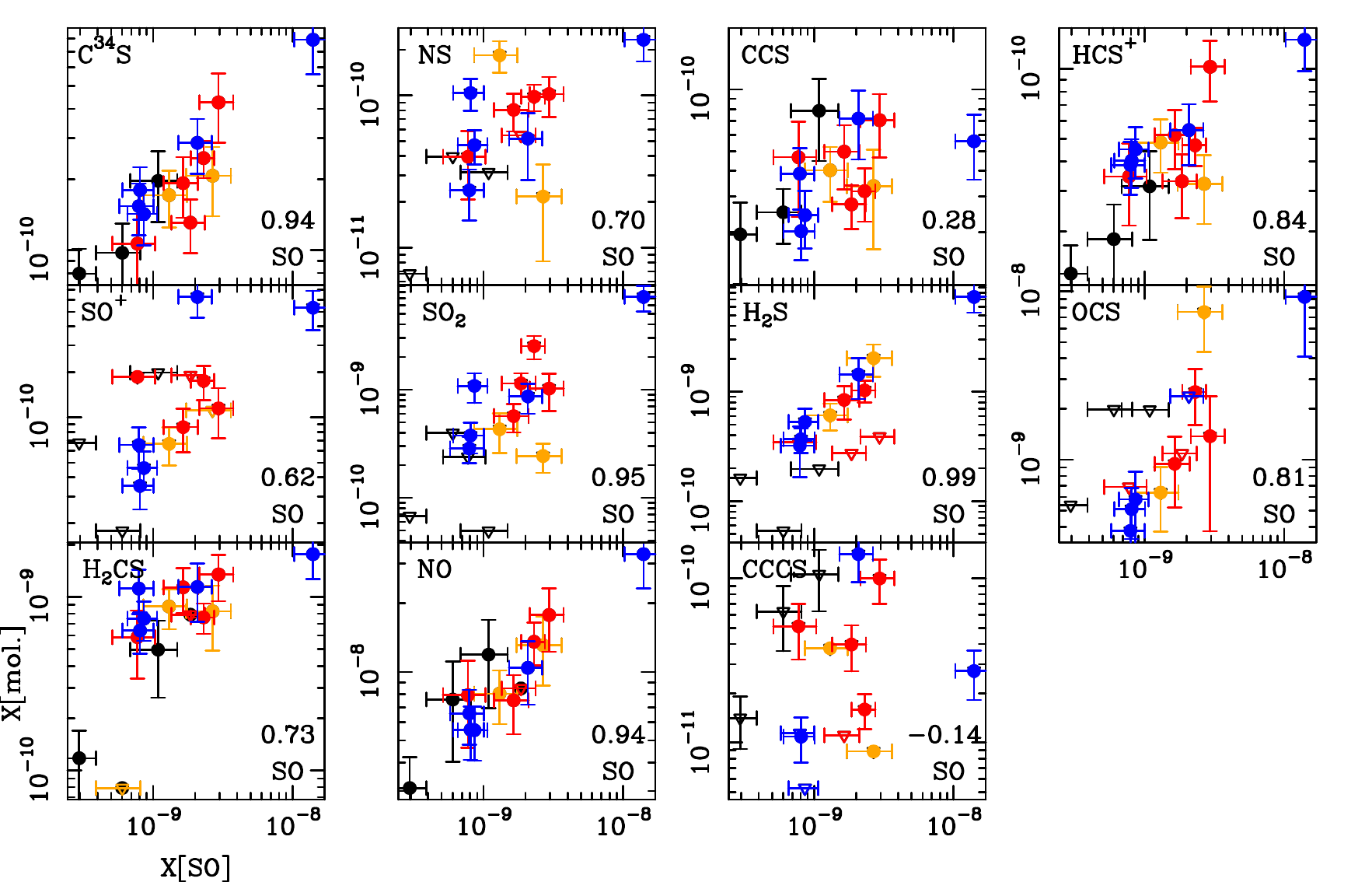}
      \caption{Same as Fig.~\ref{fig:abb-corr}, but using SO as reference species. The comparison between SO and $^{13}$CS is already shown in Fig.~\ref{fig:abb-corr}.}
         \label{fig:abb-corr-SO}
   \end{figure*}

We have checked if the correlations, and tentative correlations, found based on the FWHMs with $^{13}$CS (Fig.~\ref{fig:fwhm}) are also apparent in the fractional abundances. 
We performed the same check with the abundances of SO as reference species for the abundances of O-bearing molecules.
The comparison with $^{13}$CS and SO are shown in Figures~\ref{fig:abb-corr} and \ref{fig:abb-corr-SO}, respectively.
Let us consider first $^{13}$CS (Fig.~\ref{fig:abb-corr}). The clear correlations already found with C$^{34}$S, SO, HCS$^+$, and NO based on the FWHMs are confirmed between the abundances of such species and that of $^{13}$CS.
Among the other species, we find a positive correlation of X[$^{13}$CS] with X[NS], X[SO$^+$], X[p-H$_2$S], X[SO$_2$], X[OCS], and X[o-H$_2$CS]. 
We stress, however, that these trends consider only the detected sources and do not include the upper limits. 
Hence, they must be taken with caution.
Moreover, the strong correlations with OCS, p-H$_2$S, and SO$_2$, are influenced mostly by the UCHII G75.
These correlations are less strong ($\rho\leq 0.5$) excluding this target, while those with C$^{34}$S, SO, HCS$^+$, NO, and o-H$_2$CS remain strong even excluding G75 from the statistical analysis.
A very poor correlation ($\rho\leq 0.4$) is found with CCS, and no correlation is found with CCCS, like in the FWHMs.

As for the correlations with the SO fractional abundances (Fig.~\ref{fig:abb-corr-SO}), we find positive correlations with $^{13}$CS, C$^{34}$S, HCS$^+$, NO, but also with o-H$_2$CS, p-H$_2$S, SO$^+$, and OCS, that confirms the double origin of SO in both quiescent and turbulent material.
SO is also positively correlated with NS and SO$_2$, but the trend is determined mostly by one source only, that is again the UCHII region G75, as for $^{13}$CS. No correlation between SO and CCS or CCCS is found.
As for $^{13}$CS, we stress that for some molecules the correlation coefficient is calculated without the upper limits (i.e. NS, SO$^+$, p-H$_2$S, OCS, SO$_2$, and o-H$_2$CS), and hence they should be considered with caution.

\section{Discussion}
\label{discu}

\begin{table*}
\label{tab:abundances-1}
\caption{Abundances with respect to H$_2$ of the analysed diatomic molecules}
    \begin{tabular}{ccccccc}
\hline
\hline
      source &  $^{13}$CS              &          C$^{34}$S  &       SO                    &      SO$^+$               &      NS                        &    NO   \\
\hline
\multicolumn{7}{c}{HMSCs} \\
\hline
   00117b  &  6(3)$\times 10^{-11}$ & 2.0(0.7)$\times 10^{-10}$  &  1.1(0.4)$\times 10^{-9}$ &  2.0$\times 10^{-10}$  & 3.0(1.0)$\times 10^{-11}$  & 1.2(0.5)$\times 10^{-8}$  \\
    AFGLa  &   8(3)$\times 10^{-11}$     & 2.1(0.7)$\times 10^{-10}$   & 2.7(0.9)$\times 10^{-9}$     &  1.1$\times 10^{-10}$  & 2.1(1.4)$\times 10^{-11}$  & 1.3(0.4)$\times 10^{-8}$  \\
   05358a  &   7(2)$\times 10^{-11}$     & 1.7(0.5)$\times 10^{-10}$   & 1.3(4)$\times 10^{-9}$       &  7(2)$\times 10^{-11}$      & 1.8(0.4)$\times 10^{-10}$  &  8(2)$\times 10^{-9}$     \\
   20293a  &   3.0(0.9)$\times 10^{-11}$  & 8(2)$\times 10^{-11}$      & 3.0(0.9)$\times 10^{-10}$     &  7$\times 10^{-11}$     & 7$\times 10^{-12}$  & 3(1)$\times 10^{-9}$   \\
   22134b  &   3(2)$\times 10^{-11}$      & 1.0(0.3)$\times 10^{-10}$   & 6(2)$\times 10^{-10}$    &  1.8$\times 10^{-11}$    & 4$\times 10^{-11}$   & 8(4)$\times 10^{-9}$   \\
\hline
\multicolumn{7}{c}{HMPOs} \\
\hline
   00117a  &  4.7(1.9)$\times 10^{-11}$  & 1.1(0.4)$\times 10^{-10}$   & 8(3)$\times 10^{-10}$  & 1.9$\times 10^{-10}$  & 4.0$\times 10^{-11}$  & 8(3)$\times 10^{-9}$  \\
   05358b  &   8(2)$\times 10^{-11}$    & 1.9(0.6)$\times 10^{-10}$   & 1.6(0.5)$\times 10^{-9}$   &  9(3)$\times 10^{-11}$  & 8(2)$\times 10^{-11}$  & 8(2)$\times 10^{-9}$  \\
    18517  &   1.1(0.2)$\times 10^{-10}$  & 2.5(0.4)$\times 10^{-10}$  & 2.3(0.4)$\times 10^{-9}$   &  1.7(0.5)$\times 10^{-10}$  & 1.0(0.2)$\times 10^{-10}$  & 1.4(0.3)$\times 10^{-8}$  \\
    21307  &   4.4(1.3)$\times 10^{-11}$  & 1.3(0.3)$\times 10^{-10}$  & 1.9(0.5)$\times 10^{-9}$   &  1.9$\times 10^{-10}$  & 5.5$\times 10^{-11}$  & 8$\times 10^{-9}$   \\
    23385  &   1.6(0.5)$\times 10^{-10}$  & 4.3(1.4)$\times 10^{-10}$  & 3.0(0.8)$\times 10^{-9}$   &  1.2(0.4)$\times 10^{-10}$  & 1.0(0.3)$\times 10^{-10}$  & 1.8(0.6)$\times 10^{-8}$  \\
\hline
\multicolumn{7}{c}{UCHIIs} \\
\hline
       G75  &   3.5(1.0)$\times 10^{-10}$  & 8.0(2.0)$\times 10^{-10}$  & 1.4(0.4)$\times 10^{-8}$   &  5.3(1.6)$\times 10^{-10}$  & 2.3(0.7)$\times 10^{-10}$  & 3.3(1.0)$\times 10^{-8}$  \\
    19410  &   8(2)$\times 10^{-11}$     & 1.8(0.5)$\times 10^{-10}$   & 8(2)$\times 10^{-10}$      &  3.5(1.1)$\times 10^{-11}$   & 1.0(0.2)$\times 10^{-10}$  & 5.6(1.5)$\times 10^{-9}$  \\
    22134  &   1.0(0.3)$\times 10^{-10}$  & 2.9(0.8)$\times 10^{-10}$  & 2.1(0.6)$\times 10^{-9}$   &  6.3(1.7)$\times 10^{-10}$   & 5(2)$\times 10^{-11}$   & 1.0(0.3)$\times 10^{-8}$    \\
    23033  &   5.7(1.5)$\times 10^{-11}$  & 1.5(0.4)$\times 10^{-10}$  & 8(2)$\times 10^{-10}$     &  7(2)$\times 10^{-11}$   & 2.4(0.9)$\times 10^{-11}$     & 6.6(1.8)$\times 10^{-9}$    \\
     NGC7538  &   5.7(1.6)$\times 10^{-11}$  & 1.4(0.4)$\times 10^{-10}$  & 9(2)$\times 10^{-10}$     &  4.6(1.3)$\times 10^{-11}$  & 4.7(1.2)$\times 10^{-11}$  & 5.6(1.5)$\times 10^{-9}$    \\
\hline
\end{tabular}
\tablefoot{The numbers is brackets are the uncertainties calculated according to the propagation of errors.}
\end{table*}

\subsection{Chemistry of the tracers with correlated physical parameters}
\label{chemistry}

Based on the FWHMs, the \Tex, and the molecular fractional abundances we have found correlations between some tracers.
Let us discuss here the possible chemical reasons of these correlations.
We base our discussion on the chemical network of \citet{shingledecker20}.

First, we note the perfect correlation in both FWHMs and abundances between $^{13}$CS and C$^{34}$S that shows a common origin for the two isotopologues, independent of either evolution and source physical properties.
In particular, the correlation of the abundances also implies that fractionation effects of both carbon and sulphur are irrelevant. We will discuss better this point in Sect.~\ref{isotopic-stage}.
The relation between both CS isotopologues and HCS$^+$ 
points to a ion-molecule schema for the formation of CS where dissociative recombination of HCS$^+$ leads to CS. 
Formation of HCS$^+$ may occur via both: 
\begin{equation}
{\rm  CS + H_3^+ \rightarrow HCS^+ + H_2}\;,
\end{equation}
and 
\begin{equation}
{\rm  CS^+ + H_2 \rightarrow HCS^+ + H}\;,
\end{equation}
where CS$^+$ results from ${\rm SO + C^+}$, and/or ${\rm CH + S^+}$ reactions, and charge transfer between CS and H$^+$. 
CS may also be formed through neutral-neutral reaction pathways from the reactions ${\rm O + CCS}$ and ${\rm C + SO}$ that may stand for the FWHM correlations observed in Fig.~\ref{fig:fwhm} between $^{13}$CS and CCS, and $^{13}$CS and SO. 
The latter correlation is also observed between their abundances (Fig.~\ref{fig:abb-corr}).
%The relation between SO and CS could be attributed to a common origin from gas-phase reactions involving atomic S in the extended envelope of the cores:
%\begin{equation}
%{\rm  S + C_2 \rightarrow CS + C}\;
%\end{equation}
%\begin{equation}
%{\rm  S + OH,O_2 \rightarrow SO + H,O}\;,
%\end{equation}
%or via the direct link:
%\begin{equation}
%{\rm  C + SO \rightarrow CS + O}\;.
%\end{equation}
Also, because both CS isotopologues and SO have non-Gaussian high-velocity wings in the HMPO and UCHII stages
(Figs.~\ref{fig:spec-13cs},~\ref{fig:spec-c34s},~\ref{fig:spec-so-hmpo}, and~\ref{fig:spec-so-uchii}), 
they are likely both partially produced from grain sputtering in protostellar outflows in evolved sources.

NS is the molecule that is associated with the smallest FWHMs, except for the UCHII NGC7538. 
NS is produced starting from atomic S either on grain surfaces or in the gas via:
\begin{equation}
{\rm  S + NH \rightarrow NS + H}\;,
\end{equation}
\begin{equation}
{\rm  N + SH \rightarrow NS + H}\;,
\end{equation}
and destroyed mainly via:
\begin{equation}
{\rm NS + O \rightarrow NO + N}\;.
\label{eq:NS}
%{\rm NS + O \rightarrow NO + S}\;.
\end{equation}
Therefore, the fact that NS is associated with very quiescent gas could be due to the destruction of NS by atomic oxygen in evolved stages.
Atomic oxygen could be more abundant in these stages owing to photodissociation of water (see e.g.~van Dishoeck et al.~\citeyear{vandishoeck21}).
If so, a positive correlation between the abundances of NS and NO should be expected in evolved stages. 
In fact, we find a positive correlation between the two abundances ($\rho\sim 0.65$).
We will discuss better this point in Sect.~\ref{abundances-stage}.

\subsection{Abundances as a function of the evolutionary stage}
\label{abundances-stage}

%Overall, the column densities listed in Tables~\ref{tab:abundances-1} and \ref{tab:abundances-2} have a large dispersion that is usually of one order of magnitude.
We have searched for relations between the fractional abundances and the source parameters supposed to be evolutionary indicators, such as the gas kinetic temperature (\Tk), the dust temperature (\Td), the luminosity ($L$), and the luminosity-to-mass ratio ($L/M$). 
All parameters were derived from {\it Herschel} dust thermal emission measurements \citep{mininni21}, except for \Tk\ that was derived from ammonia observations \citep{fontani11}.

\begin{figure*}
       \centering
   \includegraphics[width=16cm]{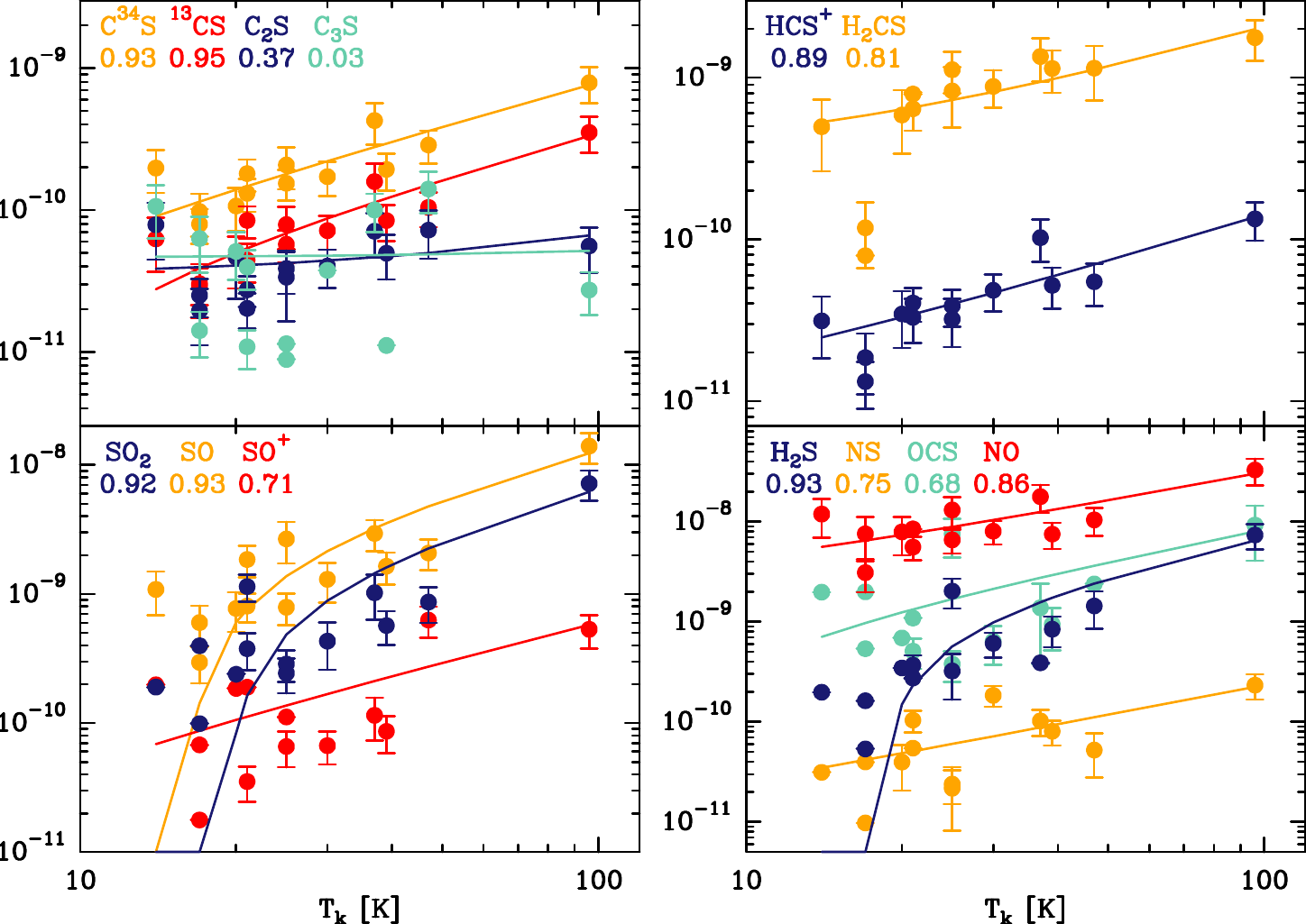}
      \caption{Fractional abundances as a function of the kinetic temperature. \Tk\ is derived from ammonia \citep{fontani11}.
      The four panels show sulphur-bearing molecules which contain: only carbon (top-left panel); carbon and hydrogen (top-right panel); only oxygen (bottom-left panel); anything else (bottom-right panel). 
      The symbols without uncertainty are upper limits. 
      The numbers below the label of each molecule is the Pearson's $\rho$ correlation coefficient.
      The curves represent linear fits to the data including upper limits plotted in logaritmic scale.
      }
         \label{fig:evol-Tk}
\end{figure*}

 Molecular fractional abundances as a function of the gas kinetic temperature are shown in Fig.~\ref{fig:evol-Tk}. 
 All except those of CCS and CCCS are positively correlated with \Tk. 
 For each molecule, we computed the Pearson's $\rho$ correlation coefficient and performed a linear regression fit to the data (plotted in logaritmic scale in Fig.~\ref{fig:evol-Tk}).
 The strongest positive correlations are found for C$^{34}$S, $^{13}$CS, HCS$^+$, SO, SO$_2$, and H$_2$S, having all $\rho \geq 0.9$, while the other tracers show a
 lower correlation. 
No significant correlation is found for CCS and CCCS.
Plots of the abundances as a function of the other evolutionary indicators (\Td, $L$, $M/L$) are shown in appendix~\ref{app:plots-abb}. We have grouped the sulphuretted molecules in C-only-bearing (Fig.~\ref{fig:abundances-parameters-C}), CH-bearing (Fig.~\ref{fig:abundances-parameters-CH}), O-only-bearing (Fig.~\ref{fig:abundances-parameters-O}), and others (Fig.~\ref{fig:abundances-parameters-others}).
The C-only-bearing species, namely C$^{34}$S, $^{13}$CS, CCS, and CCCS, and the CH-bearing species HCS$^+$ and o-H$_2$CS, overall do not show obvious trends with the four evolutionary parameters except that with \Tk.
However, while C$^{34}$S, $^{13}$CS, HCS$^+$, and o-H$_2$CS show hints of an increasing trend with \Td, $L$, and $M/L$, especially for \Td $\geq 25$~K and $L\geq 10^3$~\soll, CCS and CCCS do not show any trend over the full temperature ($\sim 10-100$~K) and luminosity ($\sim 10^2-10^5$~\soll) ranges.
Furthermore, CCCS is the only species detected in all the cold HMSCs but not in the warm HMSCs (see Fig.~\ref{fig:spec-cccs}).

Both CCS and CCCS are thought to be gas-phase products mainly via:
\begin{equation}
{\rm  HC_2S^+ + e^- \rightarrow CCS + H}\;,
\label{eq:diss-1}
\end{equation}
\begin{equation}
{\rm  HC_3S^+ + e^- \rightarrow CCS + CH}\;,
\label{eq:diss-2}
\end{equation}
\begin{equation}
{\rm  S + C_2H \rightarrow CCS + H}\;,
\end{equation}
\begin{equation}
{\rm  HCS + C \rightarrow CCS + H}\;,
\end{equation}
\begin{equation}
{\rm  HC_3S^+ + e^- \rightarrow CCCS + H}\;,
\label{eq:diss-3}
\end{equation}
\begin{equation}
{\rm  S + C_3H \rightarrow CCCS + H}\;,
\end{equation}
\begin{equation}
{\rm  CS + C_2H \rightarrow CCCS + H}\;,
\end{equation}
and the dissociative recombination reactions (Eq.~\ref{eq:diss-1}, \ref{eq:diss-2}, and \ref{eq:diss-3}) are the dominant formation pathways for both CCS and CCCS.
%All these reactions need significant C not locked in CO, and hence occur efficiently only in very extended core envelopes.
%If this envelope is not affected significantly by the inner protostellar activity, its chemical composition would reflect that of the natal cloud.
%This would explain the constant abundances of CCS and CCCS with evolution,
%but not the fact that CCCS is detected in the cold HMSCs and not in the warm HMSCs.
The main destruction channel of CCS is via reactions with atomic oxygen and nitrogen that give back CS, whereas CCCS does not react with atomic oxygen nor nitrogen as a consequence of the presence of a (small) barrier, as discussed in \citet{vidal17}.
%Instead, CCCS undergoes different ion-molecule reactions giving back HC$_3$S$^+$, C$_3$S$^+$, and other ions.
If this barrier is overtaken in warm(er) environments, CCCS starts to be destroyed. This would explain the lower abundance in warm HMSCs than in cold HMSCs.
Then, the fact that CCCS is present again at later stages could be due to a renewed efficient production, even though the non-detection in two UCHIIs again demonstrates an efficient destruction also in these later stages.

The abundances of the O-bearing species SO, SO$^+$ and SO$_2$ are overall positively correlated with \Td, \Tk, $L$, and $M/L$. The clearest trends are for SO and especially SO$_2$, as both molecules show the best correlation with \Tk\ (Fig.~\ref{fig:abundances-parameters-O}). 
Finally, among the remaining species that are NS, OCS, and p-H$_2$S (Fig.~\ref{fig:abundances-parameters-others}), the latter shows a clear increasing trend with all evolutionary parameters, in particular with \Tk, while NS and OCS show less obvious trends.
Both SO$_2$ and H$_2$S are well-known shock tracers, and our study indicates that they are evolutionary indicators for high-mass star-forming regions. 

\begin{figure*}
       \centering
     \includegraphics[clip,trim=3cm 1cm 3cm 1cm,width=9cm]{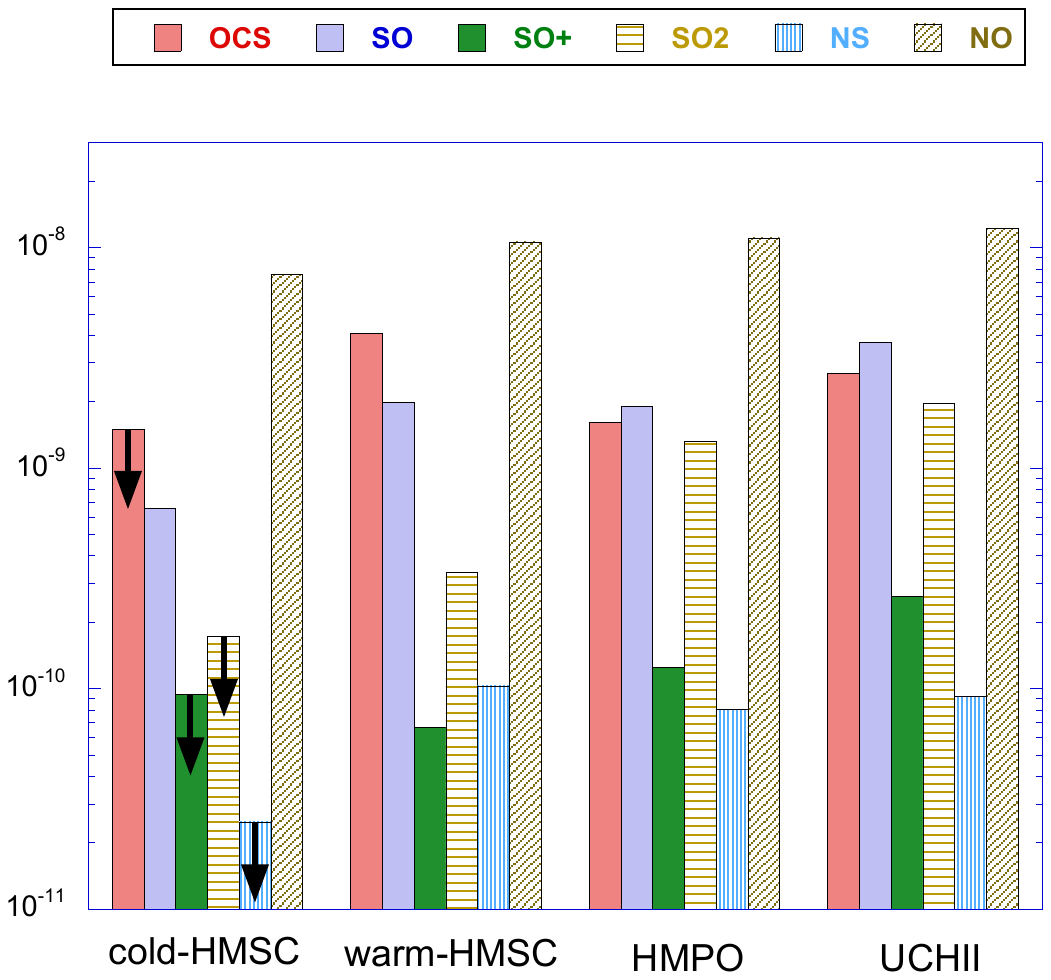}
     \includegraphics[clip,trim=3cm 1cm 3cm 1cm,width=9cm]{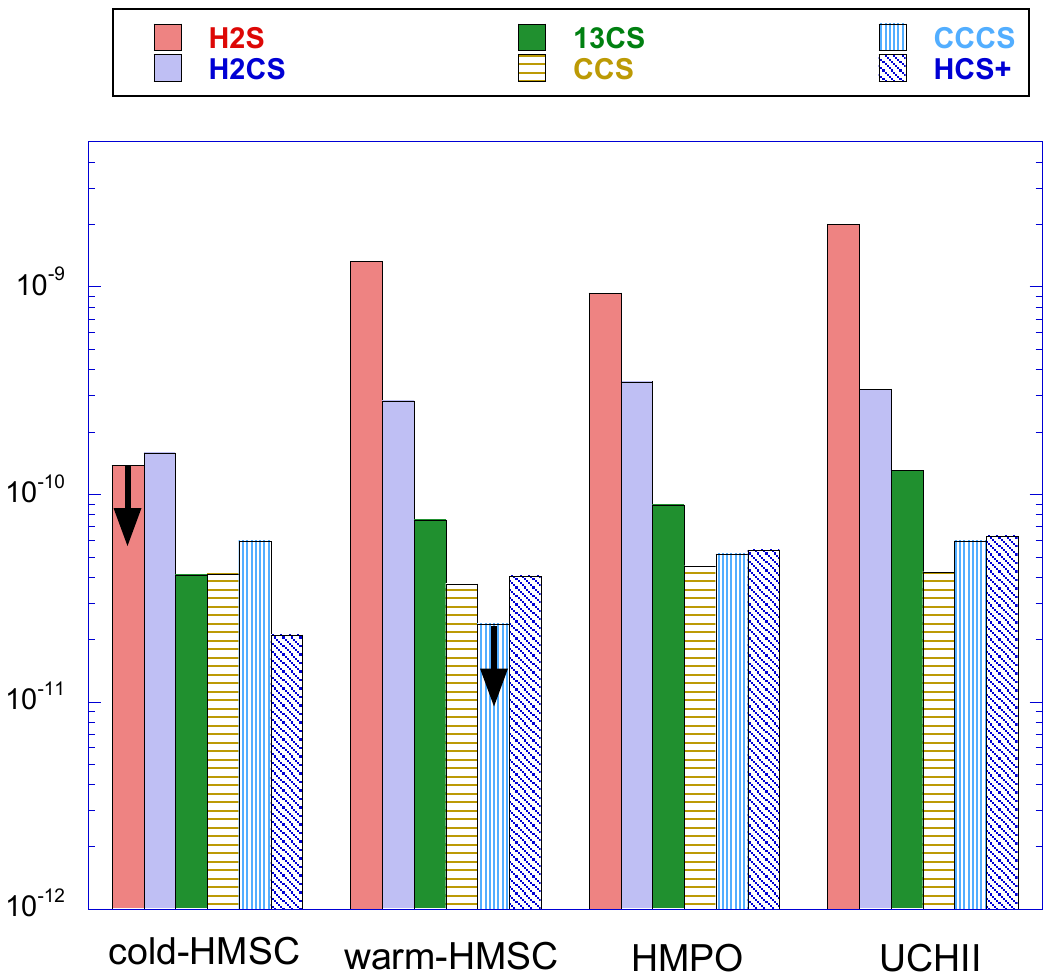}
      \caption{Mean fractional abundances as a function of the evolutionary stage. Pointing-down arrows indicate mean upper limits.
      {\it Left panel:} measured mean abundances of S-bearing species containing oxygen and/or nitrogen. The sources are ordered from left to right according to their evolutionary stage (from HMSCs to UCHIIs).
%      {\it Central panel:} same as top panel for OCS, CCS, and CCCS.
      {\it Right panel:} same as the left panel for S-bearing species containing only carbon and/or hydrogen.}
         \label{fig:abundances-stage-mean}
\end{figure*}

We have also computed the mean fractional abundances for each evolutionary stage. The results are shown in Fig.~\ref{fig:abundances-stage-mean}). 
In particular, for the HMSC group the mean has been computed separately for the cold sources (00117b, 20293a and 22134b) and the warm sources (AFGLa, 05358a).
In some tracers, the cold HMSCs were both undetected: SO$^+$, NS, p-H$_2$S, OCS, and SO$_2$. For these, we have computed the mean upper limit.
The plot shows an increase of the mean abundances with the evolutionary stage in almost all molecules.
However, the increase is different: (1) it jumps by about an order of magnitude from the cold HMSC group to the other groups for NS, p-H$_2$S, and o-H$_2$CS, and then stays almost constant; 
(2) it gradually increases through all stages in SO, SO$^+$, and SO$_2$;
(3) it increases by less than an order of magnitude from the cold HMSC group to the other groups for $^{13}$CS and HCS$^+$, and then stays almost constant; (4) it is (almost) constant in all groups for NO, OCS, and CCS (but OCS contains several upper limits).

Let us start with the molecules that show marginal or no enhancement.
The negligible increase for $^{13}$CS, NO, CCS, and HCS$^+$ is consistent with the fact that they are associated with extended material likely not affected, or marginally affected, by inner protostellar activity. 
However, OCS is thought to be produced in protostellar shocks, but because we only have upper limits in the cold HMSC group, the negligible enhancement could simply be due to the fact that these upper limits are too high with respect to the real abundances.
About the species that show significant enhancement:
the clear increase observed for SO, SO$^+$, SO$_2$, p-H$_2$S, and o-H$_2$CS indicates a favoured production of these species in warm gas and/or protostellar outflows. 
In particular, the gradual increase up to the UCHII phase of SO, SO$^+$, and SO$_2$, could be due to the increasingly larger abundance of atomic oxygen with evolution, owing to photodissociation of water in gas phase.
In fact, low water abundances of $\sim 10^{-4}$ measured in protostellar environments (much lower than $\sim 4\times 10^{-4}$ expected if all volatile oxygen is driven into water) were explained by strong UV radiation in shocks that dissociates water into O and OH \citep{vandishoeck21,tabone21}. 

\begin{figure}
       \centering
   \includegraphics[width=9cm]{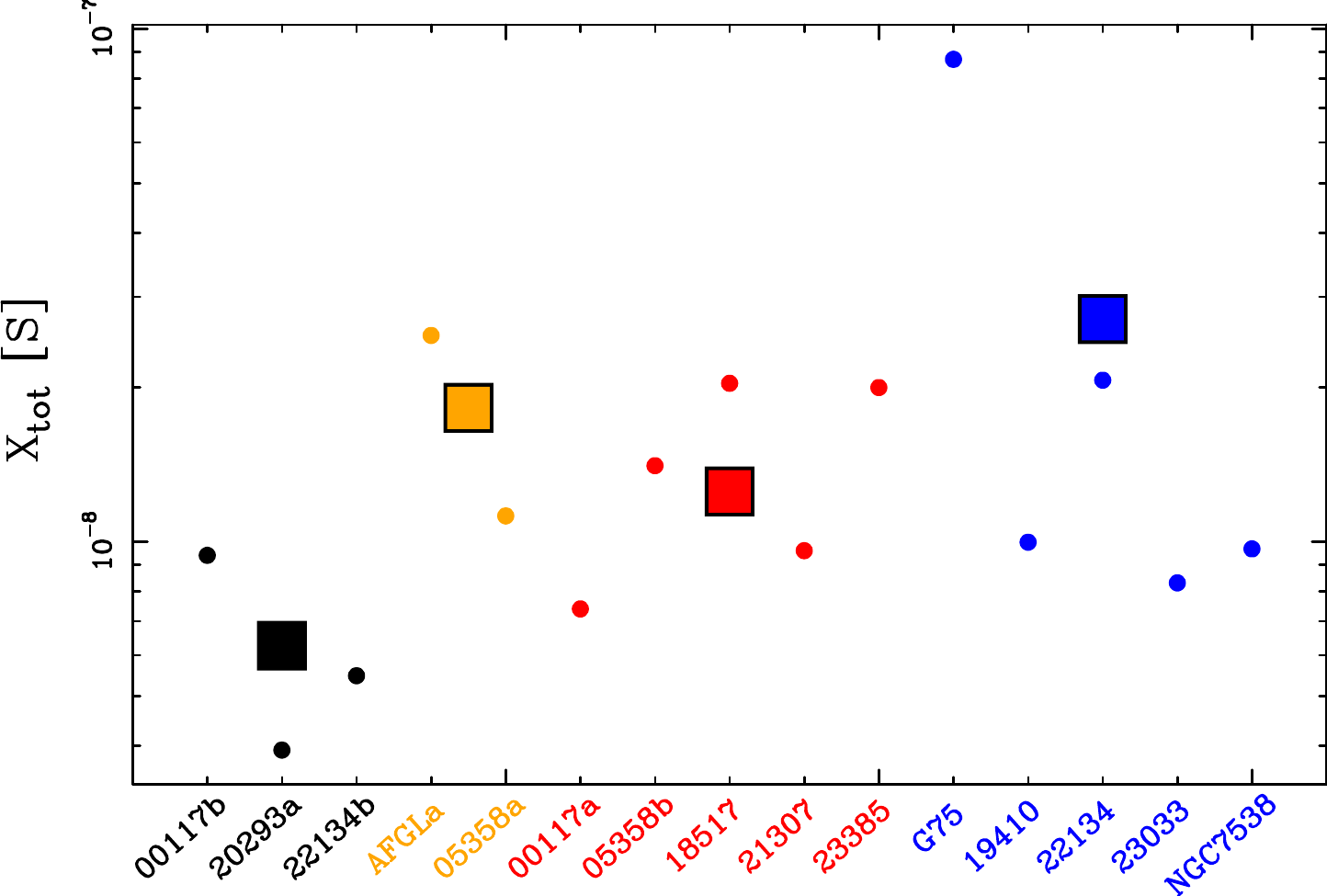}
      \caption{Sum of the molecular fractional abundances in each source.
      The sources are indicated on the x-axis. The small symbols represent the total molecular abundances in each source, while the big symbols are the average values calculated in each group: cold HMSCs (in black); warm HMSCs (in orange); HMPOs (in red); UCHIIs (in blue).
      }
         \label{fig:abb-tot-S}
\end{figure}

We summed the abundances of all S-bearing molecules to estimate the total budget of gaseous sulphur in each source.
This sum, that we call $X_{\rm tot}[S]$, is shown in Fig.~\ref{fig:abb-tot-S}. The {\it o-}H$_2$CS and {\it p-}H$_2$S abundances have been converted into the total ones assuming their LTE relative ratio of $3/4$ and $1/4$, respectively. 
The total abundance of CS has been computed multiplying the $^{13}$CS abundance for the average local ISM carbon isotope ratio of 68 \citep{milam05}.
We considered also the upper limits in the sum, and hence in each source $X_{\rm tot}[S]$ is an upper limit to the total budget of gaseous sulphur, especially for the cold HMSCs that are associated with the highest number of upper limits.
Despite this, the minimum average value is found towards the cold HMSCs, and the maximum enhancement of sulphur is found towards the UCHIIs, that are the most evolved objects.
The global increasing trend is likely due to the increasing amount of S sputtered from dust grains owing to the increasing protostellar activity with evolution.
The largest $X_{\rm tot}[S]$ is $\sim 10^{-7}$ found towards G75.
Because the reference elemental abundance of atomic sulphur is $1.73\times 10^{-5}$ (Lodders~\citeyear{lodders03}), still this maximum enhancement is lower by more than a factor 100 than the cosmic elemental abundance, in agreement with previous observational results and theoretical models (e.g.~Esplugues et al.~\citeyear{esplugues14}, Fuente et al.~\citeyear{fuente16}).

\subsection{Abundance ratios as a function of the evolutionary stage}
\label{abundances-ratios-stage}

\begin{figure}
       \centering
       \includegraphics[clip,trim=3cm 1cm 3cm 1cm,width=8cm]{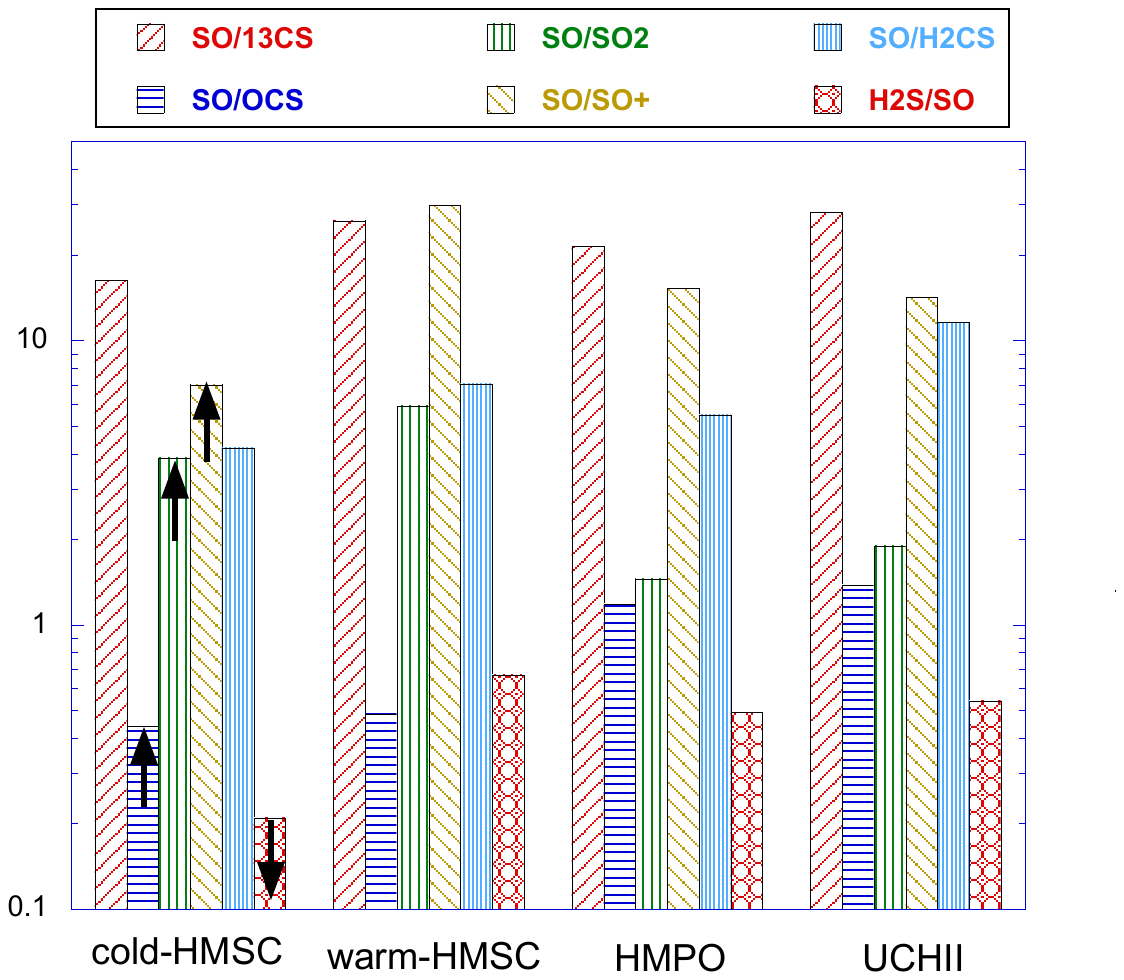}
       \includegraphics[clip,trim=3cm 1cm 3cm 1cm,width=8cm]{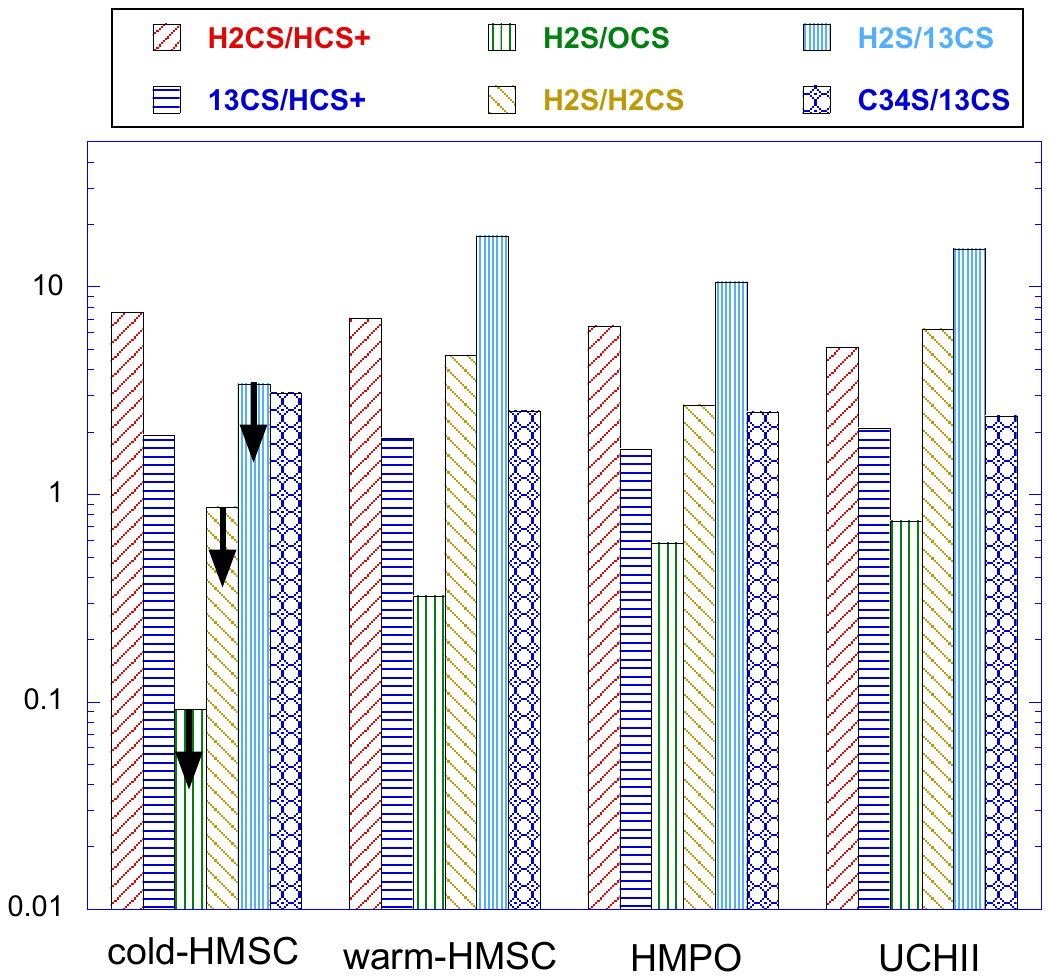}
       \includegraphics[clip,trim=3cm 1cm 3cm 1cm,width=8cm]{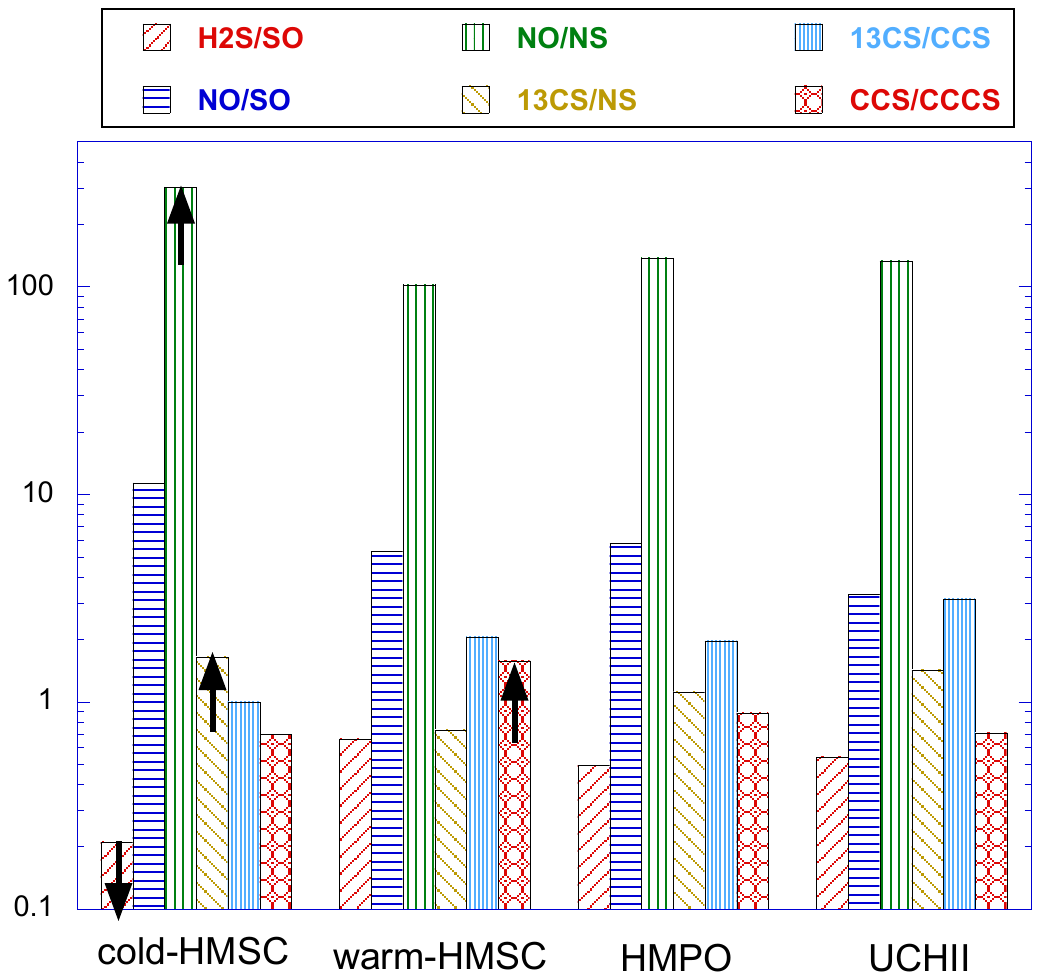}
      \caption{Column density ratios as a function of the evolutionary stage. Arrows pointing upwards show lower limits. Arrows pointing downwards show upper limits.}
         \label{fig:abundances-ratios}
\end{figure}

Previous works suggest that some abundance ratios between S-bearing species depend on the temperature of the host core. 
\citet{herpin09} studied two infrared (IR) dark and two IR bright high-mass dense cores in several sulphur-bearing molecules, and found that the SO/CS ratio increases from the IR dark to the IR bright objects, owing to an increase of the SO abundance. 
On the other hand, \citet{akel22} find that the SO/CS ratio decreases from the cold envelope of intermediate-mass protostars to warmer inner regions. 
Our observed SO/CS ratios are shown in~Fig.\ref{fig:abundances-ratios}.
The ratio is the lowest in cold HMSCs, and then slightly increases in the later stages, in agreement with the study of \citet{herpin09}.
Other ratios claimed to increase with the evolutionary stage by \citet{herpin09} are H$_2$S/OCS and SO/SO$_2$, while CS/H$_2$S and OCS/SO$_2$ are found to decrease. 
As it can be seen from Fig.~\ref{fig:abundances-ratios}, the increasing trend for H$_2$S/OCS and the decreasing trends for CS/H$_2$S and OCS/SO$_2$ are confirmed in our sources. 

The SO/SO$_2$ ratio was proposed by \citet{wakelam11} to decrease with time in massive cores.
Our observations show a decrease by a factor 3-6 from the HMSCs to the later stages (Fig.~\ref{fig:abundances-ratios}).
SO$_2$ can be produced on the surfaces of dust grains via oxygenation of SO, but also in the gas phase via (e.g.~Vastel et al.~\citeyear{vastel18}):
\begin{equation}
{\rm  SO + OH \rightarrow SO_2 + H}\;,
\end{equation}
which is barrierless and favoured at high temperatures by the enhanced presence of OH in the gas, leading to a decrease in the SO/SO$_2$ ratio.

Finally, the NO/NS ratio can be seen as a proxy to the O/S elemental abundance ratio. 
As discussed in Sect.~\ref{abundances-stage}, the atomic abundance of both S and O is expected to increase with the evolutionary stage, as
the enhancement of atomic S is due to dust grain evaporation and/or sputtering, and that of O to photodesorption and following photodissociation of water.
We note that the NO/NS ratio is $\sim 100$ and constant in the evolved groups, while it is higher than 120 in the cold HMSCs. 
Fig.~\ref{fig:abundances-stage-mean} indicates that this is mostly due to the lower abundance of NS in the cold HMSCs, as the abundance of NO is relatively constant in all evolutionary groups.
This may be due to a higher depletion of S with respect to O in the early stages.
Eq.(\ref{eq:NS}) indicates that NS is destroyed by atomic O to form NO.
However, this channel is a very minor destruction mechanism of NO and has no impact on the NO/NS ratio. 

%There is also the reaction:
%\begin{equation}
%{\rm CS + OH \rightarrow OCS + H}\;,
%\end{equation}
%which may help forming OCS at higher temperatures for the same reason.

Interestingly, the SO/$^{13}$CS ratio below 13 in all evolutionary groups (see Fig.~\ref{fig:abundances-ratios}) indicates that the SO/$^{12}$CS is smaller than 0.2 assuming the standard ${\rm ^{12}C/^{13}C=68}$ ratio,
consistent with carbon-rich environments.

Furthermore, the NO/NS ratio is $\sim 100$ and 
constant in the evolved groups, and higher than 120 in cold HMSCs. 
This points to a higher depletion of S with respect to O in this early stage, and a similar relative abundance in the later stages.
We stress, however, that care needs to be taken in the interpretation of these ratios, since the emission of the various molecules can arise from different regions inside the telescope beam.

\subsection{Sulphur and carbon isotopic ratio}
\label{isotopic-stage}

From $^{13}$CS and C$^{34}$S $J=2-1$, we have estimated the double isotopic ratio $^{34}$S/$^{32}$S$\times$$^{12}$C/$^{13}$C.
The reference Solar System values are $^{12}$C/$^{13}$C$\sim 89$, and $^{32}$S/$^{34}$S$\sim 22$ \citep{lodders03}, so the reference Solar value is $^{34}$S/$^{32}$S$\times$$^{12}$C/$^{13}$C$\sim 4$. 
The local interstellar medium (ISM) $^{12}$C/$^{13}$C ratio derived from molecular data is $\sim 68$ \citep{milam05}, and hence in this case $^{34}$S/$^{32}$S$\times$$^{12}$C/$^{13}$C would be $\sim 3$.
The plot in Figure~\ref{fig:abundances-ratios} shows the average column density ratio C$^{34}$S/$^{13}$CS as a function of the evolutionary stage. 
The ratio is $\sim 3$, consistent with the local ISM $^{34}$S/$^{32}$S$\times$$^{12}$C/$^{13}$ value. In particular, it decreases from $\sim 3$ to $\sim 2.3$ from the cold HMSCs to the warm HMSCs, and then it is almost constant in the later stages.

Fractionation processes for Sulphur are usually considered negligible in dense clouds especially in warm cores, as indicated by theoretical and observational works \citep{loison19,humire20,yan23}. 
On the other hand,
the models of \citet{colzi20} predict that the $^{12}$C/$^{13}$C ratio in CS can attain values up to a factor $\sim 3$ higher with respect to the local value in cold ($\sim 10$~K) and dense ($2\times 10^4$~\cmc) cores.
Such physical parameters are appropriate for the gaseous envelope of quiescent HMSCs, in which the measured C$^{34}$S/$^{13}$CS ratio is the highest. 
The constant trend in warm, later stages indicates that local carbon (and sulphur) fractionation processes are irrelevant at these stages in the emission regions probed by our observations.

\section{Summary and conclusions}
\label{conc}

We analysed observations of S-containing molecules obtained with the IRAM 30m telescope towards 15 star-forming cores divided equally in the evolutionary stages HMSCs, HMPOs, and UCHIIs.
The HMSCs are further divided into cold HMSCs and warm HMSCs, as already done in previous works.
We have derived column densities assuming LTE conditions of SO, SO$^+$, NS, C$^{34}$S, $^{13}$CS,
   SO$_2$, CCS, H$_2$S, HCS$^+$, OCS, 
   H$_2$CS, and CCCS. 
   We also analyse for the first time lines of the NO molecule to complement the analysis.
   For CCS, HCS$^+$, SO, OCS, and SO$_2$, we also derived the excitation temperatures. 
   The line widths and excitation temperatures indicate that  C$^{34}$S, $^{13}$CS, CCS, HCS$^+$, and NS trace preferentially quiescent and likely extended material, SO and maybe SO$^+$ both quiescent and turbulent material depending on the target, while OCS, and SO$_2$ trace warmer, more turbulent, and likely denser and more compact material. The nature of the emission of the H$_2$S, H$_2$CS, and CCCS emission is less clear.
   We also found a global increasing trend of the total gaseous sulphur going from cold HMSCs to UCHIIs. 
   This is likely due to the increasing amount of S sputtered from dust grains owing to the increasing protostellar activity with evolution. Still, the maximum total abundance that we find is $\sim 10^{-7}$ towards the UCHII G75, which is two orders of magnitude lower than the elemental S abundance.
   The abundances that show the best positive correlations with the evolutionary indicators are those of SO and SO$_2$, and in general oxygen-bearing species have higher abundances in more evolved stages than carbon-bearing ones have, perhaps due to the larger availability of atomic oxygen with evolution produced by photodissociation of water.
    We confirm that the abundance ratios SO/CS, H$_2$S/OCS, CS/H$_2$S, and OCS/SO$_2$, claimed to be possible evolutionary indicators, are found to vary with the evolution also in our study.
    Interestingly, we find that the NO/NS ratio, which can be considered a proxy of the O/S abundance ratio, is higher in the HMSCs than in the other three groups, perhaps due to the higher S-depletion with respect to O in this early stage.
   Our observational work represents an observational test-bed for theoretical studies aimed at modelling the chemistry of sulphur during the evolution of high-mass star-forming cores, and need to be complemented by higher-angular resolution observations to overcome possible problems due to different filling factors.

\begin{table*}
\label{tab:abundances-2}
\caption{Abundances with respect to H$_2$ of the molecules with three and four atoms}
\begin{tabular}{cccccccc}
\hline
\hline
 source  &    CCS           &     HCS$^+$           &     p-H$_2$S              &       OCS                &      SO$_2$             &       CCCS              &    o-H$_2$CS       \\
 \hline
\multicolumn{8}{c}{HMSCs} \\
\hline
 \hline
 00117b  & 8(3)$\times 10^{-11}$    &  3.1(1.3)$\times 10^{-11}$  &  2$\times 10^{-10}$     &  2$\times 10^{-9}$      &    5$\times 10^{-11}$    &  1.1(0.4)$\times 10^{-10}$  &  5(2)$\times 10^{-10}$  \\
   AFGLa  & 3.4(1.7)$\times 10^{-11}$     & 3.2(1.1)$\times 10^{-11}$          &  2.0(0.7)$\times 10^{-9}$           &   7.5(3.2)$\times 10^{-9}$           &  2.4(0.7)$\times 10^{-10}$           &   0.9$\times 10^{-11}$          &   8.3(3.4)$\times 10^{-10}$     \\
 05358a  & 4.0(1.2)$\times 10^{-11}$     & 4.8(1.2)$\times 10^{-11}$          &  6.1(1.7)$\times 10^{-10}$           &   6.4(2.6)$\times 10^{-10}$           &  4.3(1.7)$\times 10^{-10}$           &   3.8$\times 10^{-11}$          &   8.8(2.3)$\times 10^{-10}$     \\
 20293a  & 2.0(0.8)$\times 10^{-11}$     & 1.3(0.4)$\times 10^{-11}$          &  1.6$\times 10^{-10}$           &   5.4$\times 10^{-10}$           &  6.8$\times 10^{-11}$           &   1.4(0.5)$\times 10^{-11}$          &   1.2(0.5)$\times 10^{-10}$     \\
 22134b  & 2.5$\times 10^{-11}$     & 1.9(0.8)$\times 10^{-11}$          &  5.4$\times 10^{-11}$           &   2.0$\times 10^{-9}$           &  4.0$\times 10^{-10}$           &   6.3(2.7)$\times 10^{-11}$          &   7.9$\times 10^{-11}$     \\
 \hline
\multicolumn{8}{c}{HMPOs} \\
\hline
 00117a  & 5(2)$\times 10^{-11}$  &  3.5(1.3)$\times 10^{-11}$  &  3.5$\times 10^{-10}$     &  6.9$\times 10^{-10}$     &  2.4$\times 10^{-10}$    &  5.1(1.9)$\times 10^{-11}$  & 6(2)$\times 10^{-10}$  \\
 05358b  & 5.0(1.7)$\times 10^{-11}$     & 5.2(1.5)$\times 10^{-11}$          &  8.4(2.9)$\times 10^{-10}$           &   9.5(4.3)$\times 10^{-10}$           &  5.7(1.7)$\times 10^{-10}$           &   1.1$\times 10^{-11}$          &   1.1(0.3)$\times 10^{-9}$     \\
  18517  & 3.2(0.9)$\times 10^{-11}$     & 4.7(0.9)$\times 10^{-11}$          &  1.0(0.2)$\times 10^{-9}$           &   2.5(0.9)$\times 10^{-9}$           &  2.5(0.6)$\times 10^{-9}$           &   1.6(0.4)$\times 10^{-11}$          &   7.6(1.5)$\times 10^{-10}$     \\
  21307  & 2.7$\times 10^{-11}$     & 3.3(1.0)$\times 10^{-11}$          &  2.7$\times 10^{-10}$           &   1.1$\times 10^{-9}$           &  1.1$\times 10^{-9}$           &   4.0(1.2)$\times 10^{-11}$          &   7.9$\times 10^{-10}$     \\
  23385  & 7.1(2.4)$\times 10^{-11}$     & 1.0(0.3)$\times 10^{-10}$          &  3.9$\times 10^{-10}$           &   1.4(1.0)$\times 10^{-9}$           &  1.0(0.4)$\times 10^{-9}$           &   1.0(0.3)$\times 10^{-10}$          &   1.3(0.4)$\times 10^{-9}$     \\
  \hline
\multicolumn{8}{c}{UCHIIs} \\
\hline
     G75  & 5.6(2.0)$\times 10^{-11}$     & 1.3(0.4)$\times 10^{-10}$          &  7.4(2.1)$\times 10^{-9}$           &   9.3(5.2)$\times 10^{-9}$           &  7.2(1.9)$\times 10^{-9}$           &   2.7(0.9)$\times 10^{-11}$          &   1.8(0.5)$\times 10^{-9}$     \\
  19410  & 2.0(0.6)$\times 10^{-11}$     & 4.0(0.9)$\times 10^{-11}$          &  3.7(0.9)$\times 10^{-10}$           &   5.1(1.7)$\times 10^{-10}$           &  3.8(1.2)$\times 10^{-10}$           &   1.1(0.3)$\times 10^{-11}$          &   6.4(1.7)$\times 10^{-10}$     \\
  22134  & 7.2(2.7)$\times 10^{-11}$     & 5.5(1.6)$\times 10^{-11}$          &  1.4(0.6)$\times 10^{-9}$           &   2.4$\times 10^{-9}$           &  8.7(2.7)$\times 10^{-10}$           &   1.4(0.5)$\times 10^{-10}$          &   1.1(0.4)$\times 10^{-9}$     \\
  23033  & 3.9(1.3)$\times 10^{-11}$     & 3.9(1.0)$\times 10^{-11}$          &  3.2(1.6)$\times 10^{-10}$           &   3.8(1.3)$\times 10^{-10}$           &  2.9(0.8)$\times 10^{-10}$           &   1.1$\times 10^{-11}$          &   1.1(0.3)$\times 10^{-9}$     \\
   NGC7538  & 2.4(0.8)$\times 10^{-11}$     & 4.5(1.1)$\times 10^{-11}$          &  5.3(1.7)$\times 10^{-10}$           &   5.8(2.7)$\times 10^{-10}$           &  1.1(0.3)$\times 10^{-9}$           &   5.3$\times 10^{-12}$          &   7.5(1.9)$\times 10^{-10}$     \\
   \hline
\end{tabular}
\tablefoot{The numbers in brackets are the uncertainties calculated according to the propagation of errors.}
\end{table*}

\begin{acknowledgements}
We thank the anonymous Referee for their careful reading of the paper and their constructive comments.
F.F. is grateful to the IRAM staff for their precious help during observations.
L.C. acknowledges financial support through the Spanish
grant PID2019-105552RB-C41 funded by MCIN/AEI/10.13039/501100011033.
This work is based on observations carried out under project number 116-16 with the IRAM 30m telescope. IRAM is supported by INSU/CNRS (France), MPG (Germany) and IGN (Spain).
\end{acknowledgements}


\begin{thebibliography}{}

\bibitem[Adams(2010)]{adams10}
Adams, F.C.~2010, ARA\&A, 48, 47

\bibitem[el Akel et al.(2022)]{akel22}
el Akel, M., Kristensen, L.E., Le Gal, R., van der Walt, S.J., Pitts, R.L., Dulieu, F.~2023, A\&A, 659, A100

\bibitem[Altwegg et al.(2019)]{altwegg19}
Altwegg, K., Balsiger, H., Fuselier, S.A.~2019, ARA\&A, 57, 113 

\bibitem[Anderson et al.(2013)]{anderson13}
Anderson, D.E., Bergin, E.A., Maret, S., Wakelam, V. 2013, ApJ, 779, 141

\bibitem[Asplund(2009)]{asplund09} 
Asplund, M., Grevesse, N., Sauval, A.J., Scott, P.~2009, ARA\&A, 47, 481

\bibitem[Beuther et al.(2009)]{beuther09}
Beuther, H., Zhang, Q., Bergin, E.A., Sridharan, T.K.~2009, AJ, 137, 406

\bibitem[Boogert et al.(2015)]{boogert15}
Boogert A.C.A., Gerakines P.A., Whittet D.C. B.~2015, ARA\&A, 53, 541

\bibitem[Calmonte et al.(2016)]{calmonte16}
Calmonte, U., Altwegg, K.; Balsiger, H., et al.~2016, MNRAS, 462, 253

\bibitem[Carpenter(2000)]{carpenter00}
Carpenter, J.M.~2000, AJ, 120, 3139

\bibitem[Cazaux et al.(2022)]{cazaux22}
Cazaux, S., Carrascosa, H., Mu\~{n}oz Caro, G.M., Caselli, P., Fuente, A., Navarro-Almaida, D., Rivi\'ere-Marichalar, P.~2022, A\&A, 657, A100

\bibitem[Cernicharo et al.(2021)]{cernicharo21}
Cernicharo, J., Cabezas, C., Ag\'undez, M.~2021, A\&A, 648, L3

\bibitem[Codella et al.(2018)]{codella18}
Codella, C., Viti, S., Lefloch, B., et al.~2018, MNRAS, 474, 5694

\bibitem[Coletta et al.(2020)]{coletta20}
Coletta, A., Fontani, F., Rivilla, V.M., Mininni, C., Colzi, L., S\'anchez-Monge, \'A., Beltr\'an, M.T.~2020, A\&A, 641, 54

\bibitem[Colzi et al.(2018)]{colzi18}
Colzi, L., Fontani, F., Caselli, P., Ceccarelli, C., Hily-Blant, P., Bizzocchi, L.~2018, A\&A, 609, 129

\bibitem[Colzi et al.(2019)]{colzi19}
Colzi, L.,  Fontani, F., Caselli, P., Leurini, S., Bizzocchi, L., Quaia, G.~2019, MNRAS, 485, 5543

\bibitem[Colzi et al.(2020)]{colzi20}
Colzi, L., Sipil\"{a}, O., Roueff, E., Caselli, P., Fontani, F.~2020, A\&A, 640, 51

\bibitem[Drdla et al.(1989)]{drdla89}
Drdla, K., Knapp, G.R. \& van Dishoeck, E.F.~1989, ApJ 345, 815

\bibitem[Endres et al.(2016)]{endres16}
Endres, P., Schlemmer, S., Schilke, P., Stutzki, J., M\"{u}ller, H.S.P.~2016, J.Mol.Spec., 327, 95

\bibitem[Esplugues et al.(2014)]{esplugues14}
Esplugues, G.B., Viti, S., Goicoechea, J.R., Cernicharo, J.~2014, A\&A, 567, 95

\bibitem[Esplugues et al.(2022)]{esplugues22}
Esplugues, G., Fuente, A., Navarro-Almaida, D., et al.~2022, A\&A, 662, A52

\bibitem[Esteban et al. (2004)]{esteban04}
Esteban, C., Peimbert, M., Garc\'ia-Rojas, J., Ruiz, M.T., Peimbert, A., Rodr\'iguez, M.~2004, MNRAS, 355, 229

%\bibitem[Dom{\'\i}nguez-Guzm{\'a}n et al. 2022]{esteban22}
%\bibitem[Dom{\'\i}nguez-Guzm{\'an} et al. (2022)]{dominguez22}
%Dom{\'\i}nguez-Guzm{\'an}, G., Rodr{\'\i}guez, M., Garc{\'\i}a-Rojas, J.,  Esteban, C., Toribio San Cipriano, L.~2022, MNRAS, 517, 4497

\bibitem[Evans et al.(2009)]{evans09}
Evans, N.J.II, Dunham, M.M., J\o{}rgensen, J. K., et al. 2009, ApJS, 181, 321

\bibitem[Fontani et al.(2004)]{fontani04}
Fontani, F., Cesaroni, R., Testi, L., et al.~2004, A\&A, 414, 299 

\bibitem[Fontani et al.(2007)]{fontani07}
Fontani, F., Pascucci, I., Caselli, P., Wyrowski, F., Cesaroni, R., Walmsley, C.M.~2007, A\&A, 470, 639

\bibitem[Fontani et al.(2011)]{fontani11}
Fontani, F., Palau, Aina, Caselli, P., et al.~2011, A\&A, 529, L7

\bibitem[Fontani et al.(2014)]{fontani14}
Fontani, F., Sakai, T., Furuya, K., Sakai, N., Aikawa, Y., Yamamoto, S.~2014, MNRAS, 440, 448

\bibitem[Fontani et al.(2015a)]{fontani15a}
Fontani, F., Caselli, P., Palau, Aina, Bizzocchi, L., Ceccarelli, C.~2015a, ApJ, 808, L46

\bibitem[Fontani et al.(2015b)]{fontani15b}
Fontani, F., Busquet, G., Palau, Aina, et al.~2015b, A\&A, 575, 87

\bibitem[Fontani et al.(2021)]{fontani21}
Fontani, F., Barnes, A.T., Caselli, P., et al.~2021, MNRAS, 503, 4320

\bibitem[Fontani et al.(2018)]{fontani18}
Fontani, F., Vagnoli, A., Padovani, M., Colzi, L., Caselli, P., Rivilla, V.M.~2018, MNRAS, 481, L79

\bibitem[Fontani et al.(2021)]{fontani21}
Fontani, F., Colzi, L., Redaelli, E., Sipil\"{a}, O., Caselli, P.~2021, A\&A, 651, 94

\bibitem[Fuente et al.(2016)]{fuente16}
Fuente, A., Cernicharo, J., Roueff, E., et al.~2016, A\&A, 593, 94

\bibitem[Fuente et al.(2019)]{fuente19}
Fuente, A., Navarro, D.G., Caselli, P., et al.~2019, A\&A, 624, A105 

\bibitem[Fuente et al.(2023)]{fuente23}
Fuente, A., Rivi\`ere-Marichalar, P., Beitia-Antero, L., et al.~2023, A\&A, 670, A114

\bibitem[Goicoechea et al.(2006)]{goicoechea06}
Goicoechea, J.R., Pety, J., Gerin, M., Teyssier, D., Roueff, E., Hily-Blant, P., Baek, S. 2006, A\&A 456, 565

\bibitem[Hatchell et al.(1998)]{hatchell98}
Hatchell, J., Thompson, M.A., Millar, T.J., MacDonald, G.H.~1998, A\&A 338, 713

\bibitem[Herpin et al.(2009)]{herpin09}
Herpin, F., Marseille, M., Wakelam, V., Bontemps, S., Lis, D.C.~2009, A\&A, 504, 853

\bibitem[Heyl et al. (2022)]{heyl22}
{Heyl}, J.,  {Sellentin}, E.,  {Holdship}, J. and {Viti}, S.~2022, MNRAS 517, 38 

\bibitem[Hily-Blant et al.(2022)]{hily-blant22}
Hily-Blant, P., Pineau des For\^{e}ts, G., Faure, A., Lique, F.~2022, A\&A, 658, A168

\bibitem[Holdship et al.(2019)]{holdship19}
Holdship, J., Jim\'enez-Serra, I., Viti, S., et al.~2019, ApJ, 878, 64

\bibitem[Howk et al.(2006)]{howk06}
Howk, J.C., Sembach, K.R., Savage, B.D.~2006, ApJ, 637, 333

\bibitem[Humire et al.(2020)]{humire20}
Humire, P.K., Thiel, V., Henkel, C., et al.~2020, A\&A, 642, A222

\bibitem[Jenkins(2009)]{jenkins09}
Jenkins, E.B.~2009, ApJ, 700, 1299

\bibitem[Jim\'enez-Escobar \& Mu\~{n}oz-Caro(2011)]{jem11}
Jim\'enez-Escobar A., Mu\~{n}oz Caro G.M.~2011, A\&A, 536, A91

\bibitem[Kurtz et al.(2000)]{kurtz00}
Kurtz, S., Cesaroni, R., Churchwell, E., Hofner, P., Walmsley, C.M.~2000, in Protostars and Planets IV (Book - Tucson: University of Arizona Press; eds Mannings, V., Boss, A.P., Russell, S. S.), p. 299-326

\bibitem[Kutner \& Ulich(1981)]{keu81} 
Kutner, M.L., \& Ulich, B.L. 1981, ApJ, 250, 341

\bibitem[Laas \& Caselli(2019)]{lec19}
Laas, J. \& Caselli, P.~2019, A\&A, 624, A118

\bibitem[Lodders(2003)]{lodders03}
Lodders, K.~2003, ApJ, 591, 1220L

\bibitem[Loison et al.(2019)]{loison19}
Loison, J.-C., Wakelam, V., Gratier, P., et al.~2019, MNRAS, 485, 5777

\bibitem[Mart\'in et al.(2019)]{martin19}
Mart\'in, S., Mart\'in-Pintado, J., Blanco-S\'anchez, C., et al.~2019, A\&A, 631, 159

\bibitem[McClure et al.(2023)]{mcclure23}
McClure, M.K., Rocha, W.R.M., Pontoppidan, K.M., et al.~2023, Nature Astronomy, 7, 431

\bibitem[McGuire(2018)]{mcguire18}
McGuire, B.A.~2018, ApJS, 239, 17

\bibitem[Milam et al.(2005)]{milam05}
Milam, S.N., Savage, C., Brewster, M.A., Ziurys, L.M.~2005, ApJ, 634, 1126

\bibitem[Millar et al. (1986)]{millar:86}
{{Millar}, T.~J., {Adams}, N.~G., {Smith}, D., {Lindinger}, W. and {Villinger}, H.}~1986, MNRAS 221, 673

\bibitem[Mininni et al.(2018)]{mininni18}
Mininni, C., Fontani, F., Rivilla, V.M., Beltr\'an, M.T., Caselli, P., Vasyunin, A.~2018, MNRAS, 476, L39

\bibitem[Mininni et al.(2021)]{mininni21}
Mininni, C., Fontani, F., S\'anchez-Monge, A., Rivilla, V.M., Beltr\'an, M.T.~2021, A\&A, 653, 87

\bibitem[Navarro-Almaida et al.(2020)]{navarro20}
Navarro-Almaida, D., Le Gal, R., Fuente, A.,~et al.~2020, A\&A, 637, A39

\bibitem[Oppenheimer \& Dalgarno(1974)]{oed74}
Oppenheimer, M. \& Dalgarno, A.~1974, ApJ, 187, 231

\bibitem[Pickett et al.(1998)]{pickett98}
Pickett, H.M., Poynter, R.L., Cohen, E.A., et al. 1998, J. Quant. Spectr. Rad. Transf., 60, 883

\bibitem[Pineau des for\^{e}ts et al.(1986)]{pineau86}
Pineau des For\^{e}ts, G., Roueff, E. \& Flower, D.R., 1986, MNRAS 223, 743

\bibitem[Pfalzner(2013)]{pfalzner13}
Pfalzner, S.~2013, A\&A, 549, A82

\bibitem[Rivilla et al.(2017)]{rivilla17}
Rivilla, V.M., Beltr\'an, M.T., Cesaroni, R., Fontani, F., Codella, C., Zhang, Q.~2017, A\&A, 598, 59

\bibitem[Rivilla et al.(2020a)]{rivilla20}
Rivilla, V.M., Colzi, L., Fontani, et al.~2020a, MNRAS, 496, 1990

\bibitem[Rivilla et al.(2020b)]{rivilla20b}
Rivilla, V.M., Drozdovskaya, M. N.; Altwegg, K., et al.~2020b, MNRAS, 492, 1180

\bibitem[Shingledecker et al.(2020)]{shingledecker20}
Shingledecker, C.N., Lamberts, T., Laas, J.C., et al.~2020, ApJ, 888, 52

\bibitem[Tabone et al.(2021)]{tabone21}
Tabone, B., van Hemert, M.C., van Dishoeck, E.F., Black, J.H.~2021, A\&A, 650, A192

\bibitem[Taniguchi et al.(2018)]{taniguchi18}
Taniguchi, K., Saito, M., Sridharan, T.K., Minamidani, T.~2018, ApJ, 854, 133

\bibitem[Turner(1995)]{turner95}
Turner, B.~1995, ApJ, 455, 556

\bibitem[van Dishoeck et al.(2021)]{vandishoeck21}
van Dishoeck, E.F., Kristensen, L.E., Mottram, J.C., et al.~2021, A\&A, 648, A24

\bibitem[Vastel et al.(2018)]{vastel18}
Vastel, C., Qu\'enard, D., Le Gal, R., et al.~2018, MNRAS, 478, 5514 

\bibitem[Vasyunina et al.(2011)]{vasyunina11}
Vasyunina, T., Linz, H., Henning, Th., Zinchenko, I., Beuther, H., Voronkov, M.~2011, A\&A, 527, 88

\bibitem[Vidal et al.(2017)]{vidal17}
Vidal, T.H.G., Loison, J.-C., Jaziri, A.Y., Ruaud, M., Gratier, P., Wakelam, V.~2017, MNRAS, 469, 435

\bibitem[Wakelam et al.(2011)]{wakelam11}
Wakelam, V., Hersant, F., Herpin, F.~2011, A\&A, 529, A112

\bibitem[Wang et al.(2016)]{wang16}
Wang, Y., Audard, M., Fontani, F., et al.~2016, A\&A, 587, A69

\bibitem[Woods et al.(2015)]{woods15}
Woods, P.M., Occhiogrosso, A., Viti, S., Kaňuchová, Z., Palumbo, M.E., Price, S.D.

\bibitem[Yan et al.(2023)]{yan23}
Yan, Y.T., Henkel, C., Kobayashi, C., et al.~2023, A\&A, 670, A98

\bibitem[Zanchet et al.(2013)]{zanchet13}
Zanchet, A., Ag\'undez, M., Herrero, V.J., et al.~2013, AJ, 146, 125

%\bibitem[Zhang et al.(2002)]{zhang02}
%Zhang, Q., Hunter, T.R., Sridharan, T.K., \& Ho, P.T.P. 2002, ApJ, 566, 982
   
\end{thebibliography}

\newpage

\begin{appendix}

\section{Spectra}
\label{app:spectra}

In this appendix, we show the spectra of all molecular species analysed. For those having more than three transitions detected. we show only the brightest ones.

   \begin{figure*}
   \centering
   \includegraphics[width=15cm]{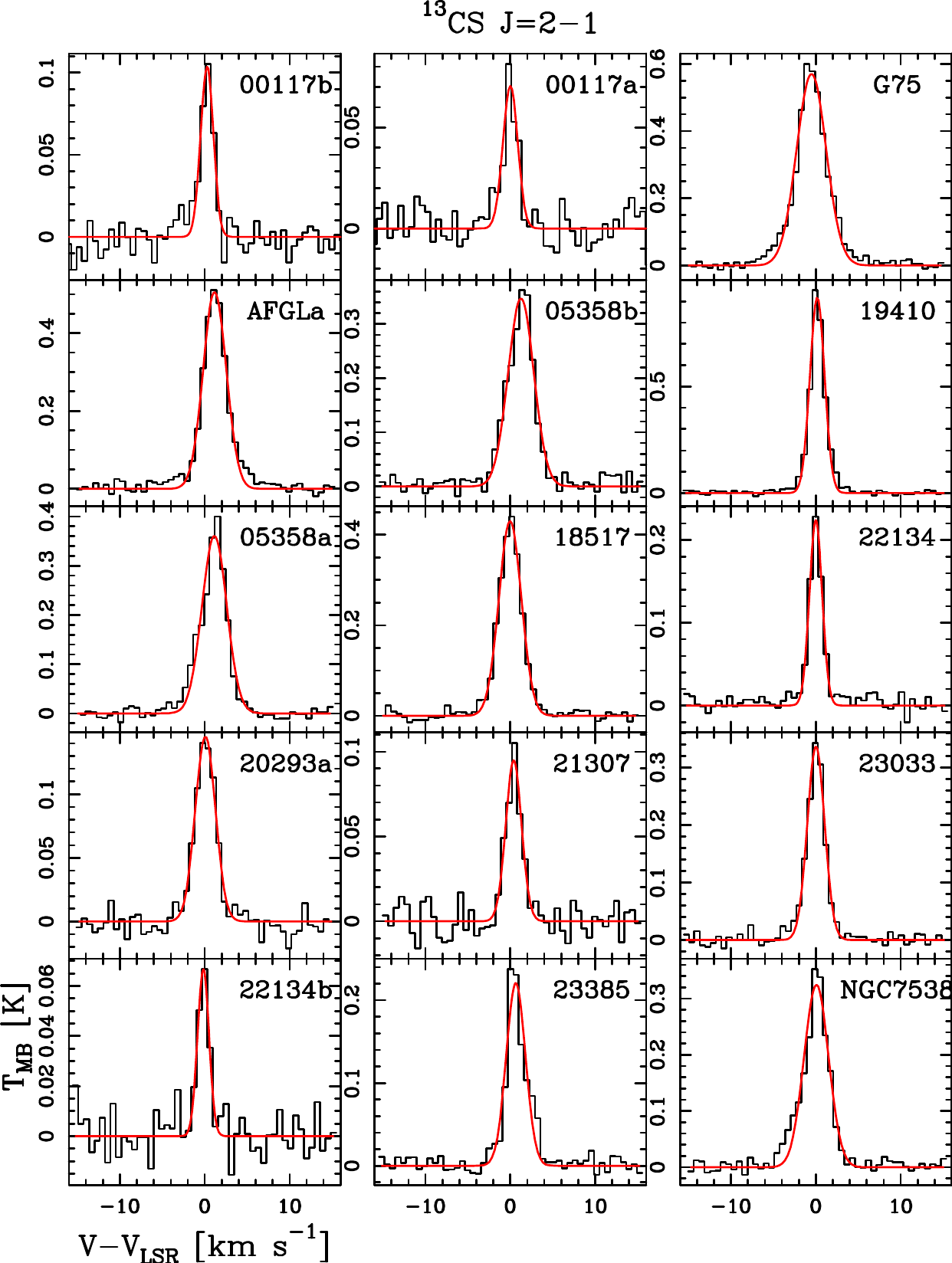}
      \caption{Spectra of the $^{13}$CS $J=2-1$ lines. 
      On the x-axis we show the velocity shift from the systemic velocity $V_{\rm LSR}$ listed in Table~\ref{tab:sources}.
      In each frame, the red curve represents the best Gaussian fit performed with {\sc madcuba} (see Sect.~\ref{reduction}). The first column contains the HMSCs, the second the HMPOs, and the third the UCHIIs.}
         \label{fig:spec-13cs}
   \end{figure*}

    \begin{figure*}
   \centering
   \includegraphics[width=15cm]{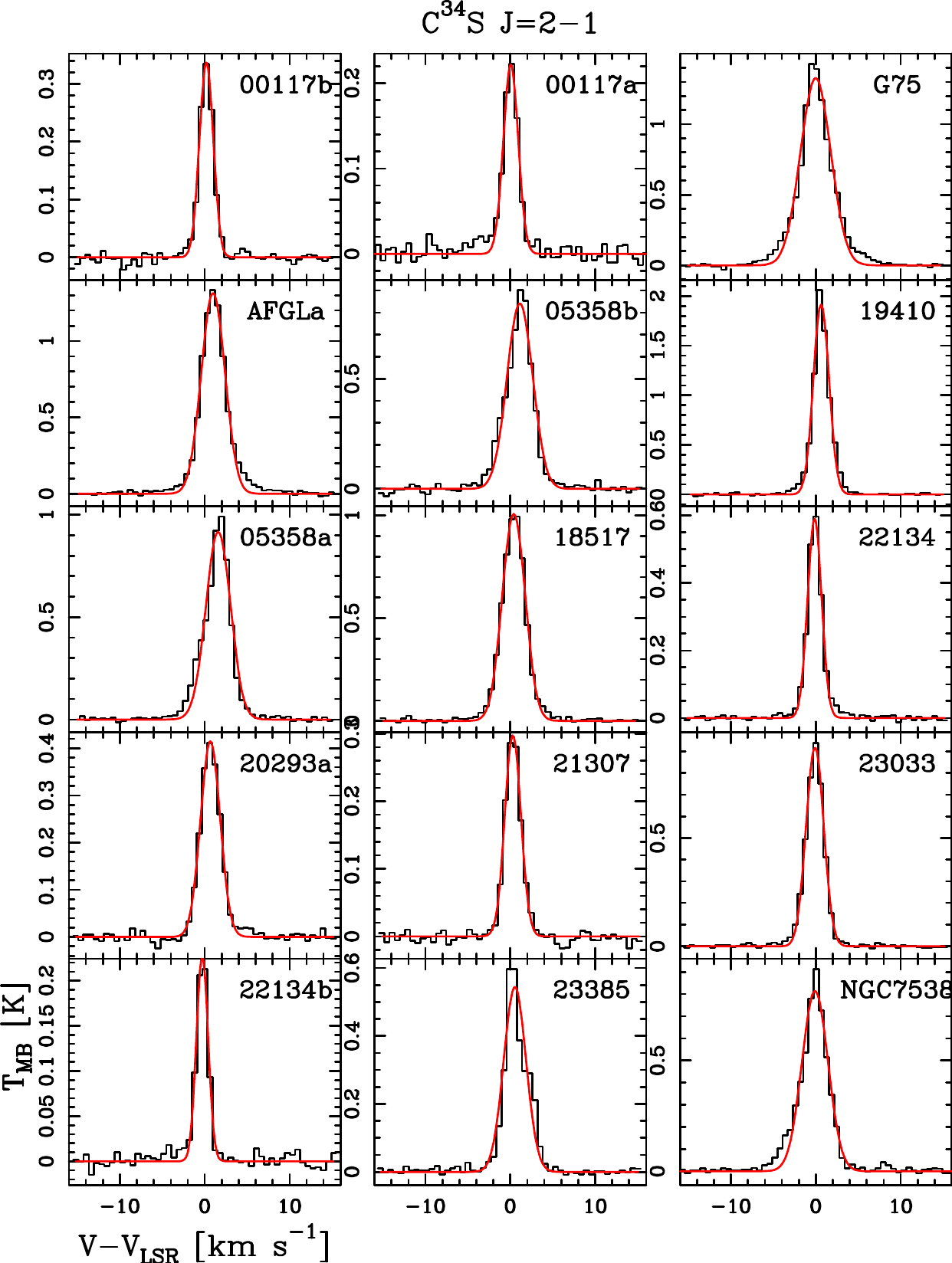}
      \caption{Same as Fig.~\ref{fig:spec-13cs} for the C$^{34}$S $J=2-1$ lines.}
         \label{fig:spec-c34s}
   \end{figure*}

     \begin{figure*}
   \centering
   \includegraphics[width=15cm]{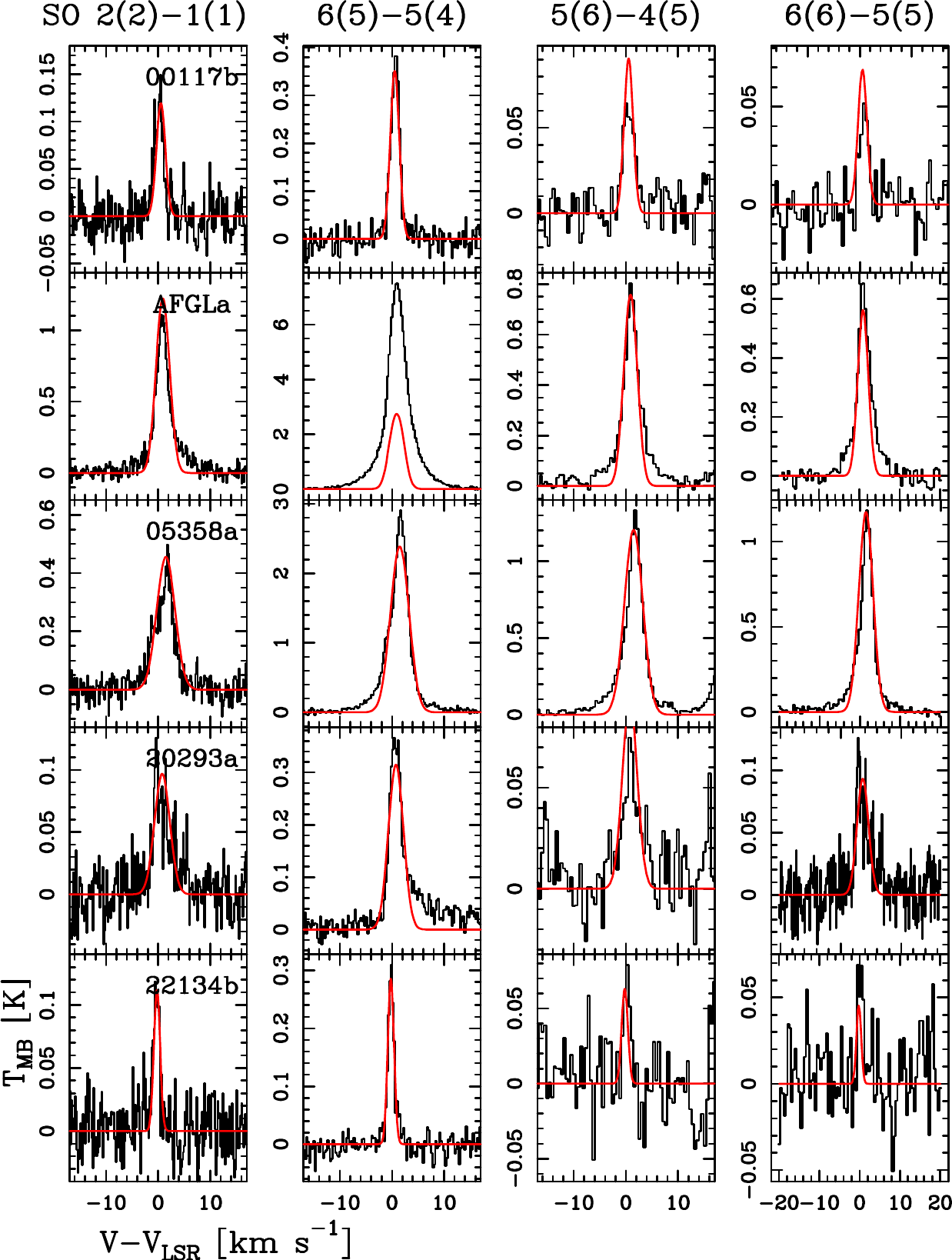}
      \caption{Spectra of SO $J(K)=2(1)-1(1)$, $6(5)-5(4)$, $5(6)-4(5)$, and $6(6)-5(5)$ observed towards the HMSCs. The red curves indicate the best-fit Gaussians performed with \sc{madcuba}.}
         \label{fig:spec-so-hmsc}
   \end{figure*}

        \begin{figure*}
   \centering
   \includegraphics[width=15cm]{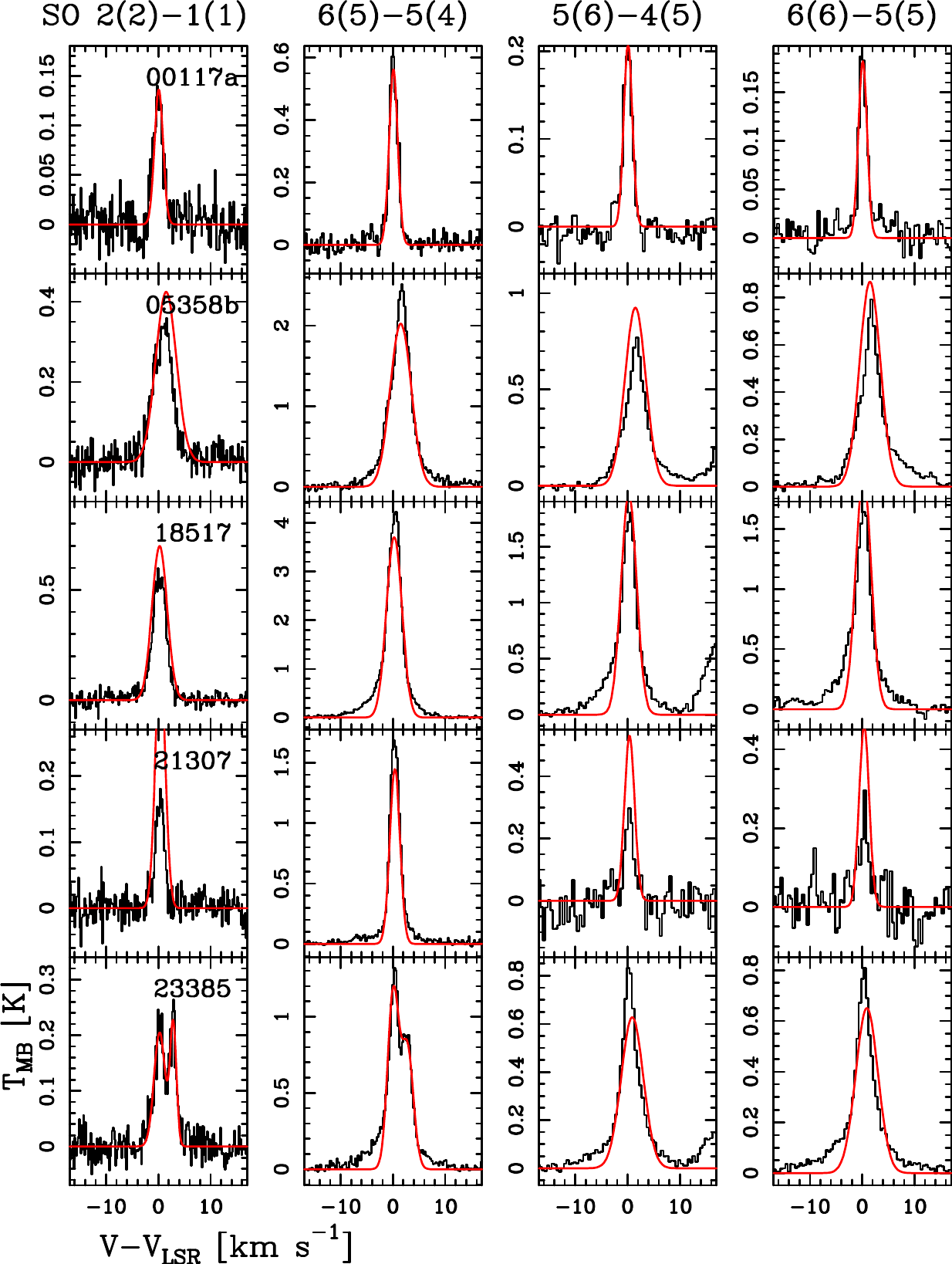}
      \caption{Same as Fig.~\ref{fig:spec-so-hmsc} for the HMPOs.}
         \label{fig:spec-so-hmpo}
   \end{figure*}

        \begin{figure*}
   \centering
   \includegraphics[width=15cm]{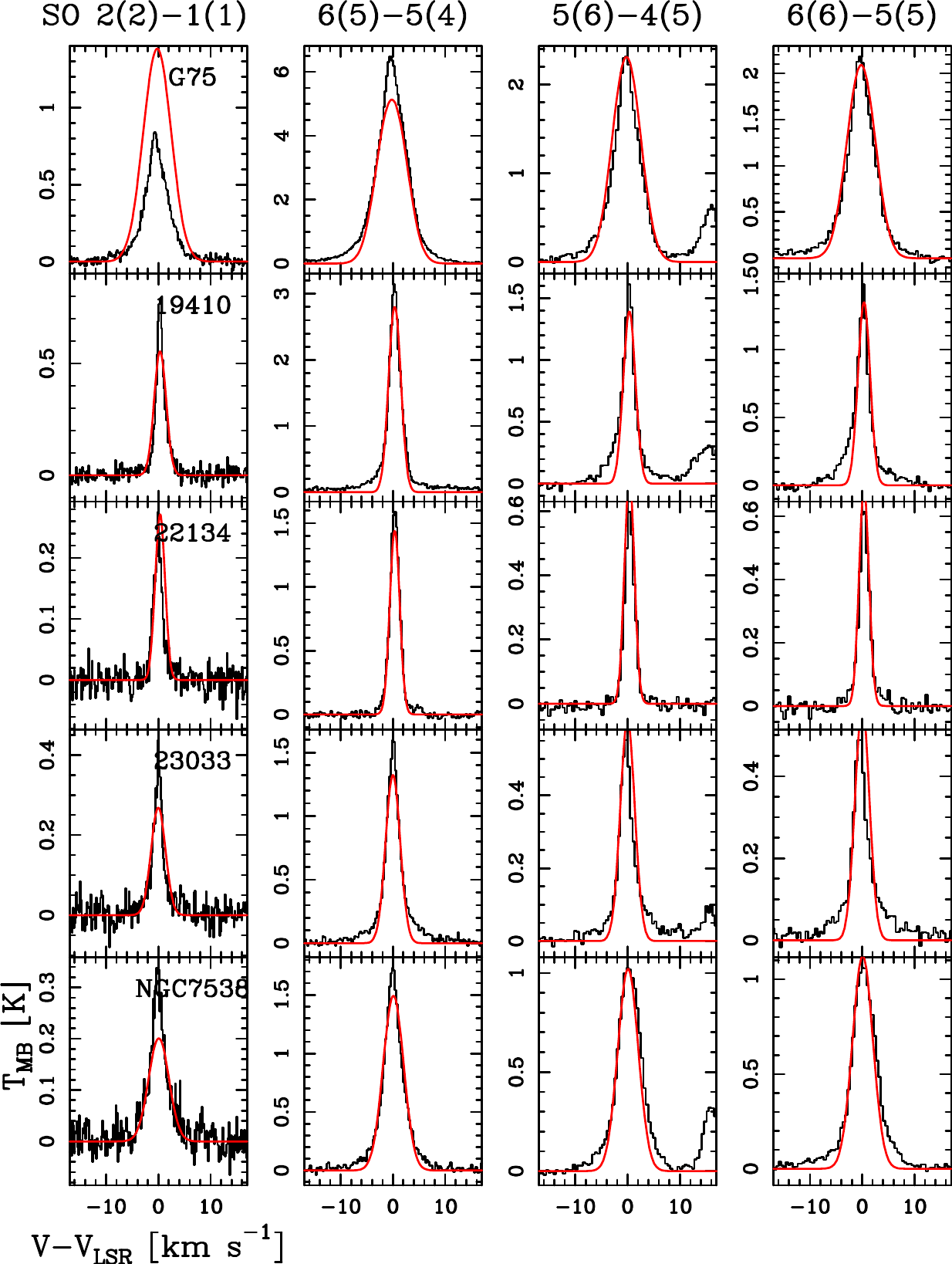}
      \caption{Same as Fig.~\ref{fig:spec-so-hmsc} for the UCHIIs.}
         \label{fig:spec-so-uchii}
   \end{figure*}

       \begin{figure*}
   \centering
   \includegraphics[width=15cm]{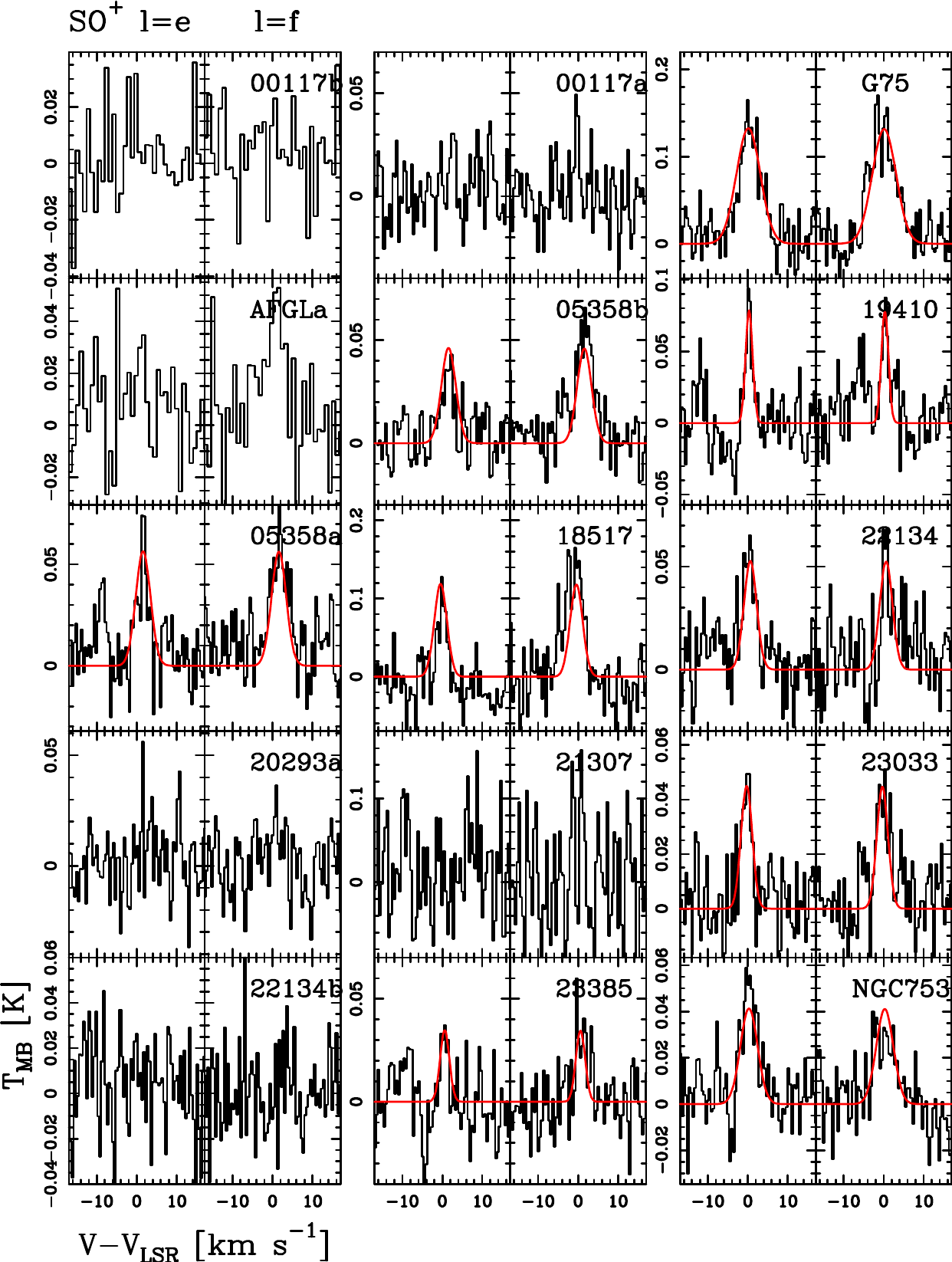}
      \caption{Same as Fig.~\ref{fig:spec-13cs} for the SO$^+$ $J=11/2-9/2$ lines.}
         \label{fig:spec-sop}
   \end{figure*}

       \begin{figure*}
   \centering
   \includegraphics[width=15cm]{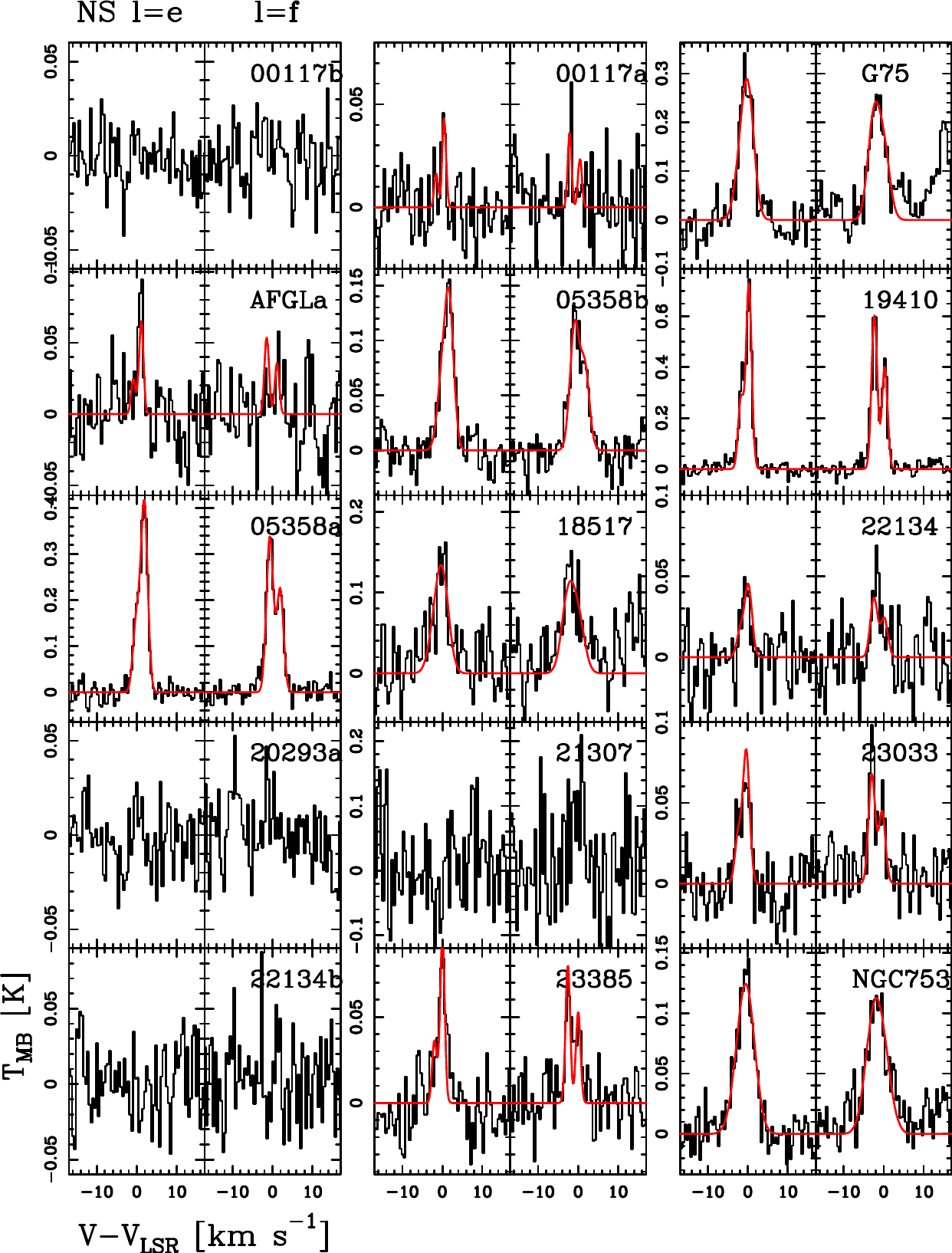}
      \caption{Same as Fig.~\ref{fig:spec-13cs} for the NS $J=11/2-9/2, F=13/2-11/2$ lines.}
         \label{fig:spec-ns}
   \end{figure*}

 \begin{figure*}
   \centering
   \includegraphics[width=15cm]{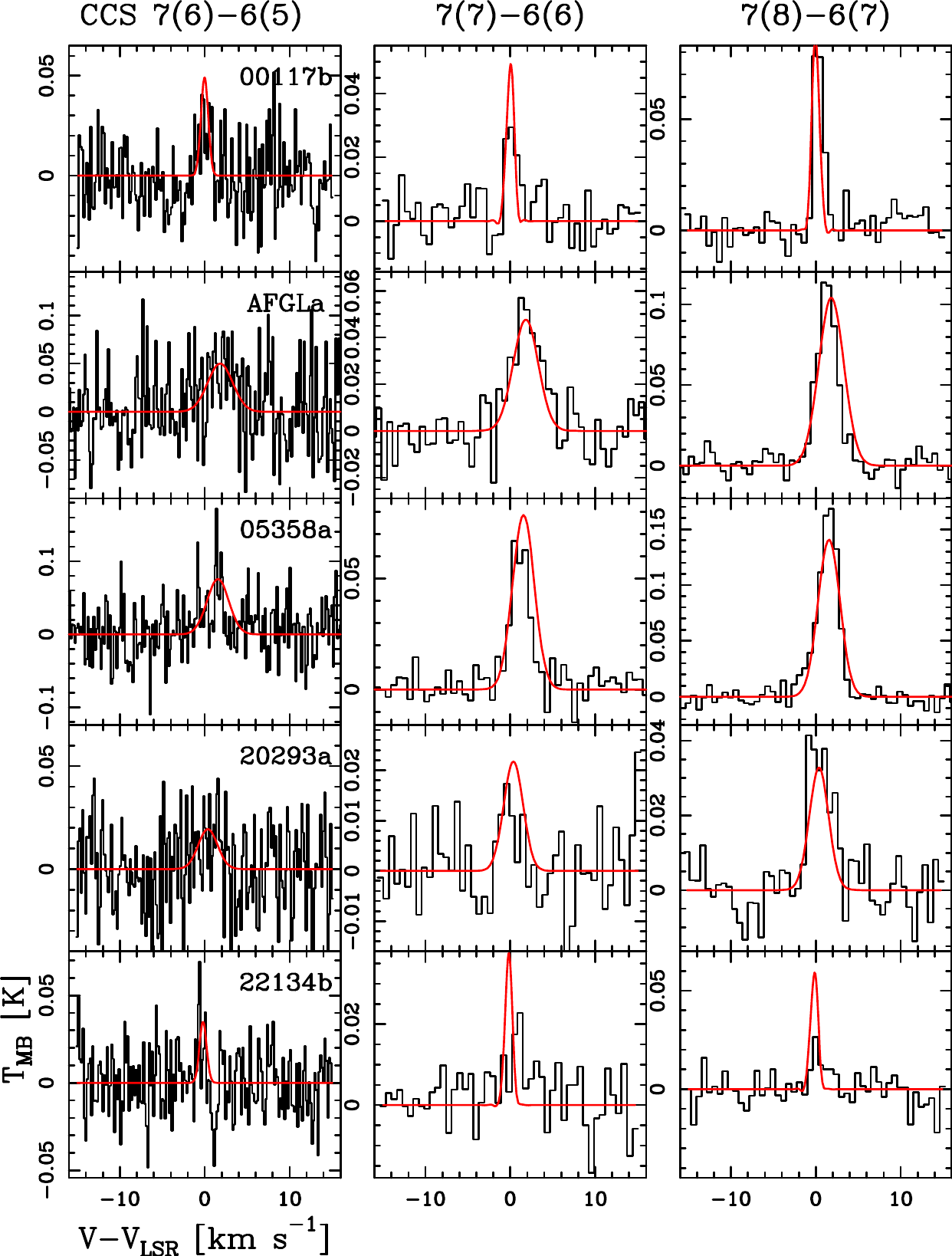}
      \caption{Spectra of CCS $N(J)=7(6)-6(5)$, $7(7)-6(6)$, and $7(8)-6(7)$ for the HMSCs. The red curves indicate the best fit Gaussian curves obtained with \sc{madcuba}}
         \label{fig:spec-ccs-hmsc}
   \end{figure*}

    \begin{figure*}
   \centering
   \includegraphics[width=15cm]{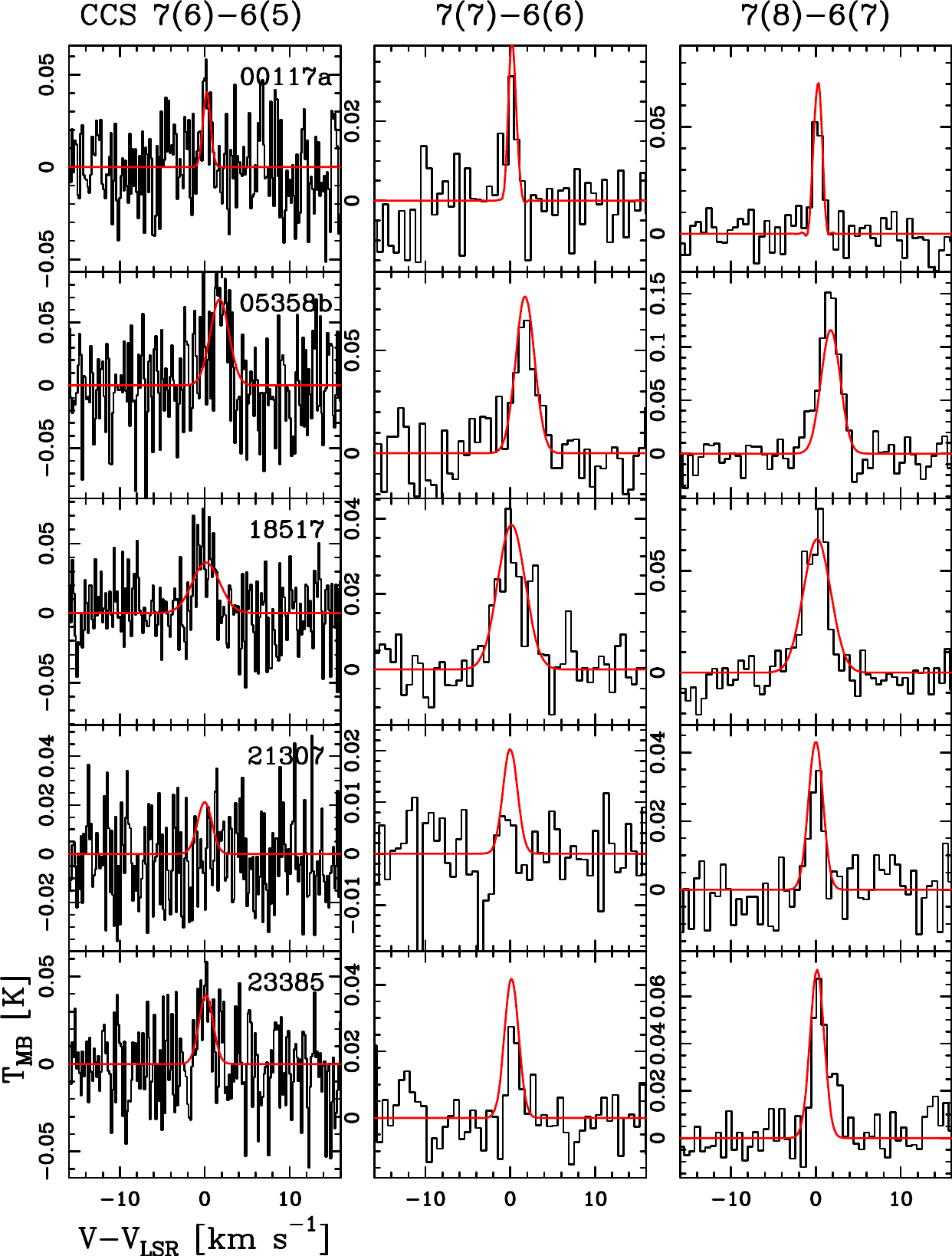}
      \caption{Same as Fig.~\ref{fig:spec-ccs-hmsc} for the HMPOs lines.}
         \label{fig:spec-ccs-hmpo}
   \end{figure*}

    \begin{figure*}
   \centering
   \includegraphics[width=15cm]{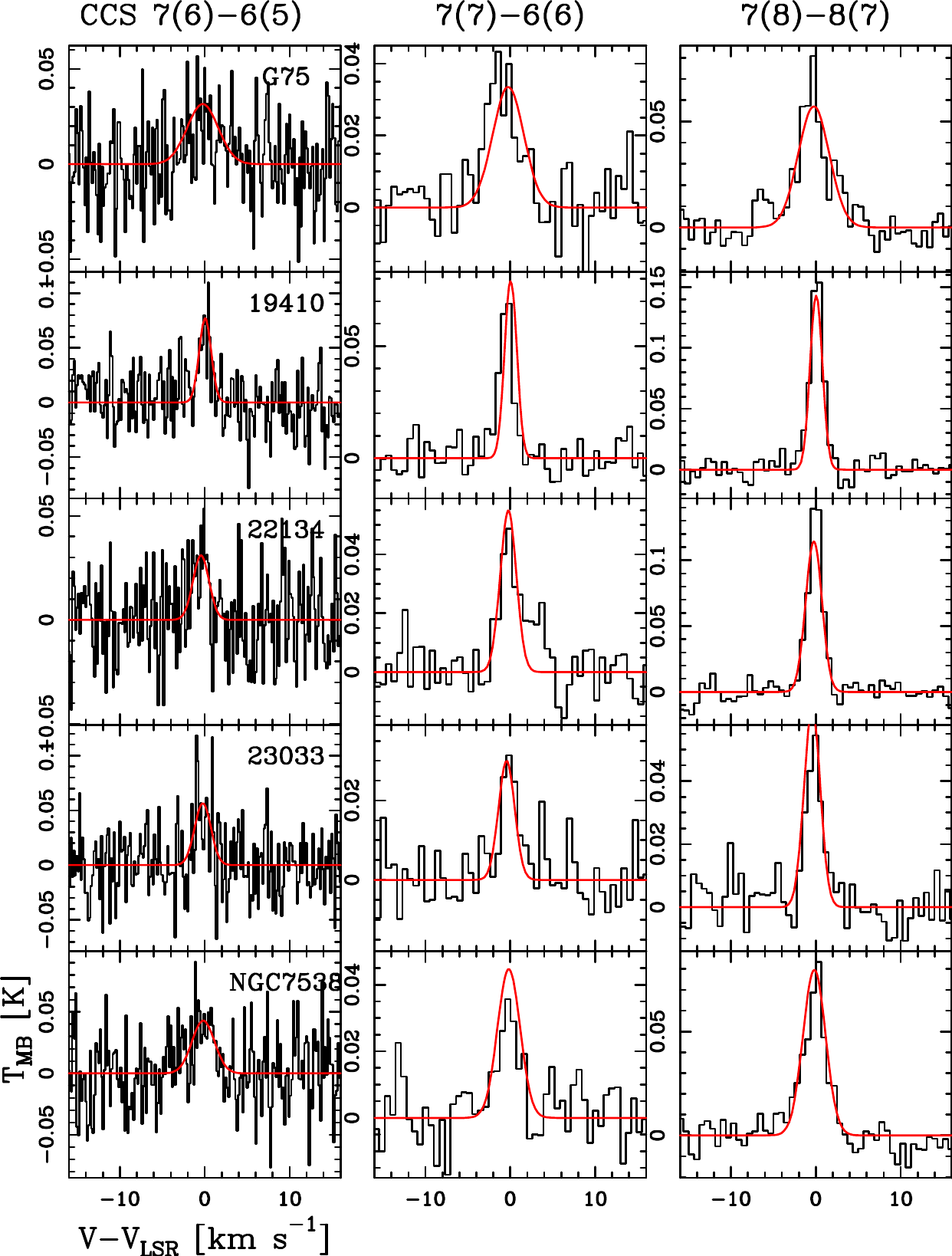}
      \caption{Same as Fig.~\ref{fig:spec-ccs-hmsc} the UCHIIs.}
         \label{fig:spec-ccs-uchii}
   \end{figure*}

 \begin{figure*}
   \centering
   \includegraphics[width=15cm]{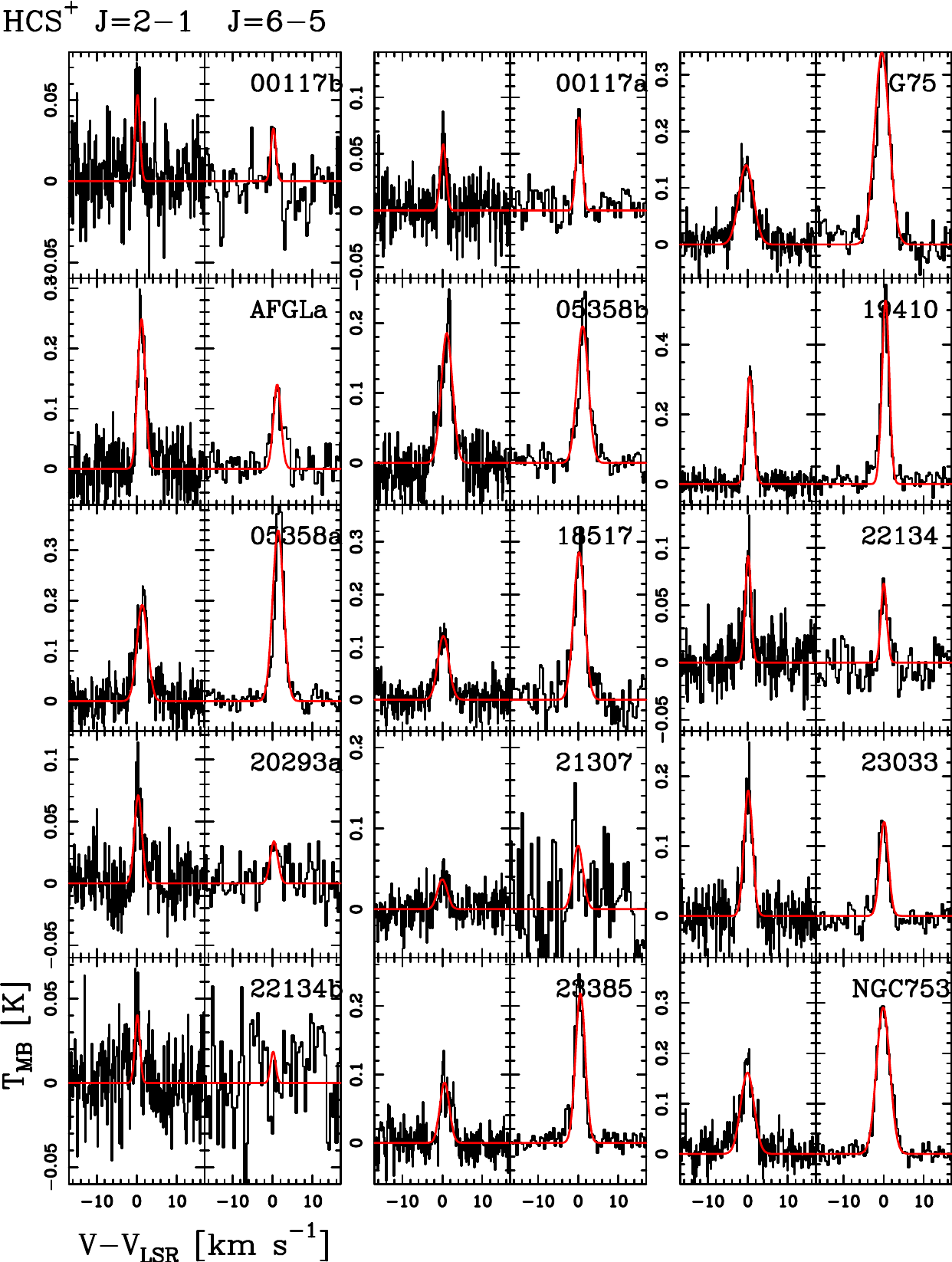}
      \caption{Same as Fig.~\ref{fig:spec-13cs} for the HCS$^+$ $J=2-1$ and $J=6-5$ lines.}
         \label{fig:spec-hcsp}
   \end{figure*}

        \begin{figure*}
   \centering
   \includegraphics[width=15cm]{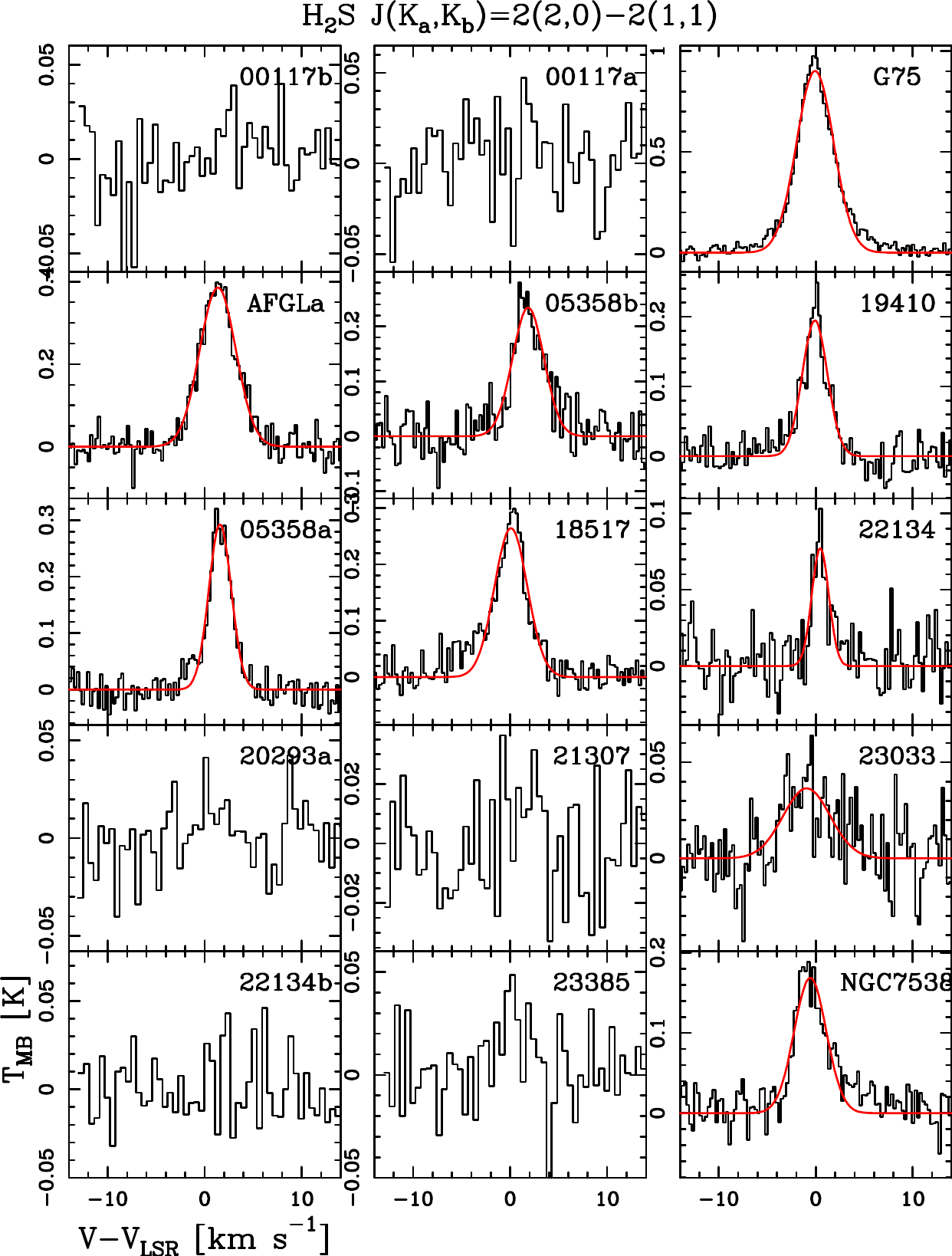}
      \caption{Same as Fig.~\ref{fig:spec-13cs} for the H$_{2}$S $J(K_{\rm a},K_{\rm b}=2(2,0)-2(1,1)$ lines.}
         \label{fig:spec-h2s}
   \end{figure*}

       \begin{figure*}
   \centering
   \includegraphics[width=15cm]{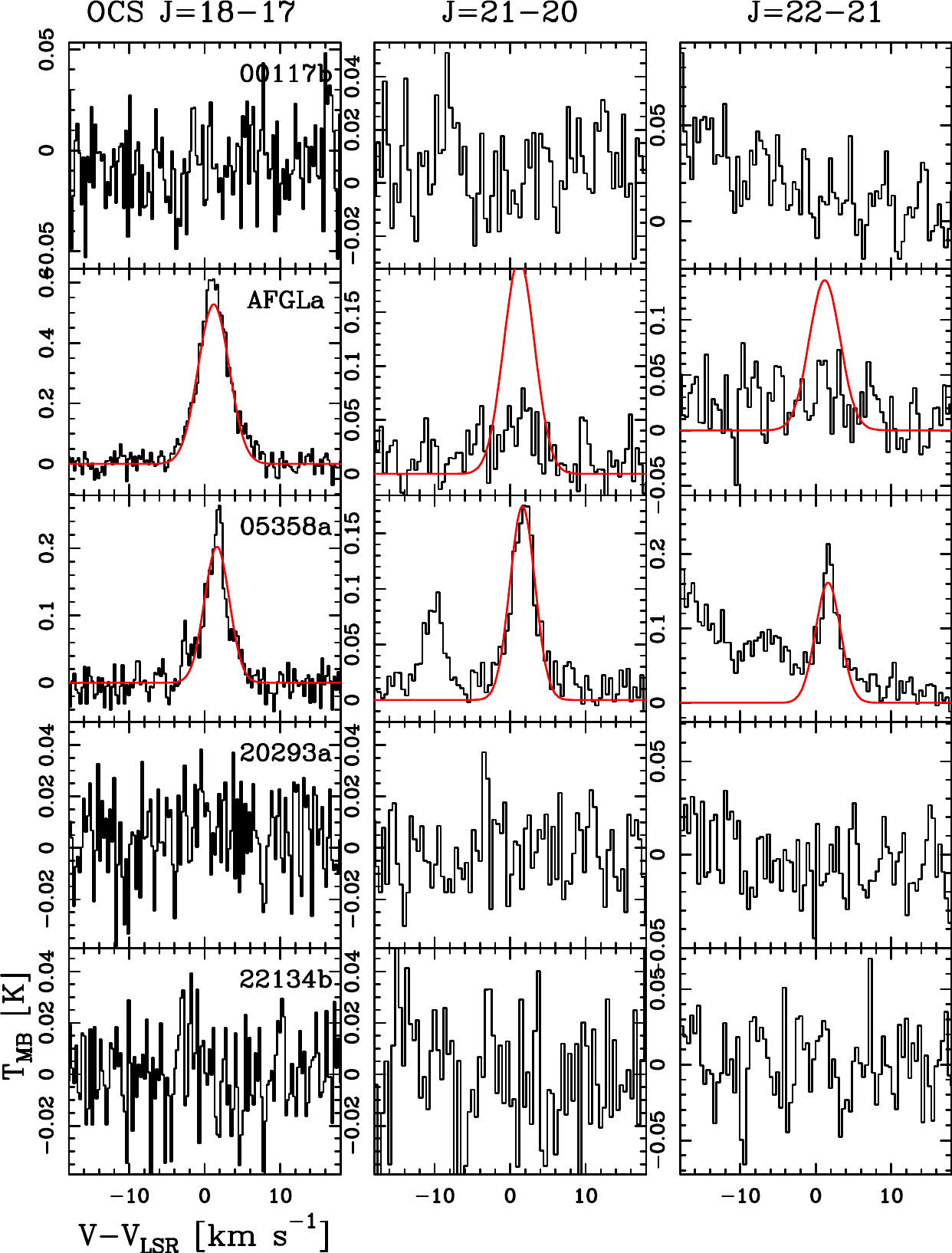}
      \caption{Spectra obtained towards the HMSCs of the OCS $J=18-17$, $J=21-20$ and $J=22-21$ lines. The red curves indicate the best fit obtained with \sc{MADCUBA}.}
         \label{fig:spec-ocs-hmsc}
   \end{figure*}

      \begin{figure*}
   \centering
   \includegraphics[width=15cm]{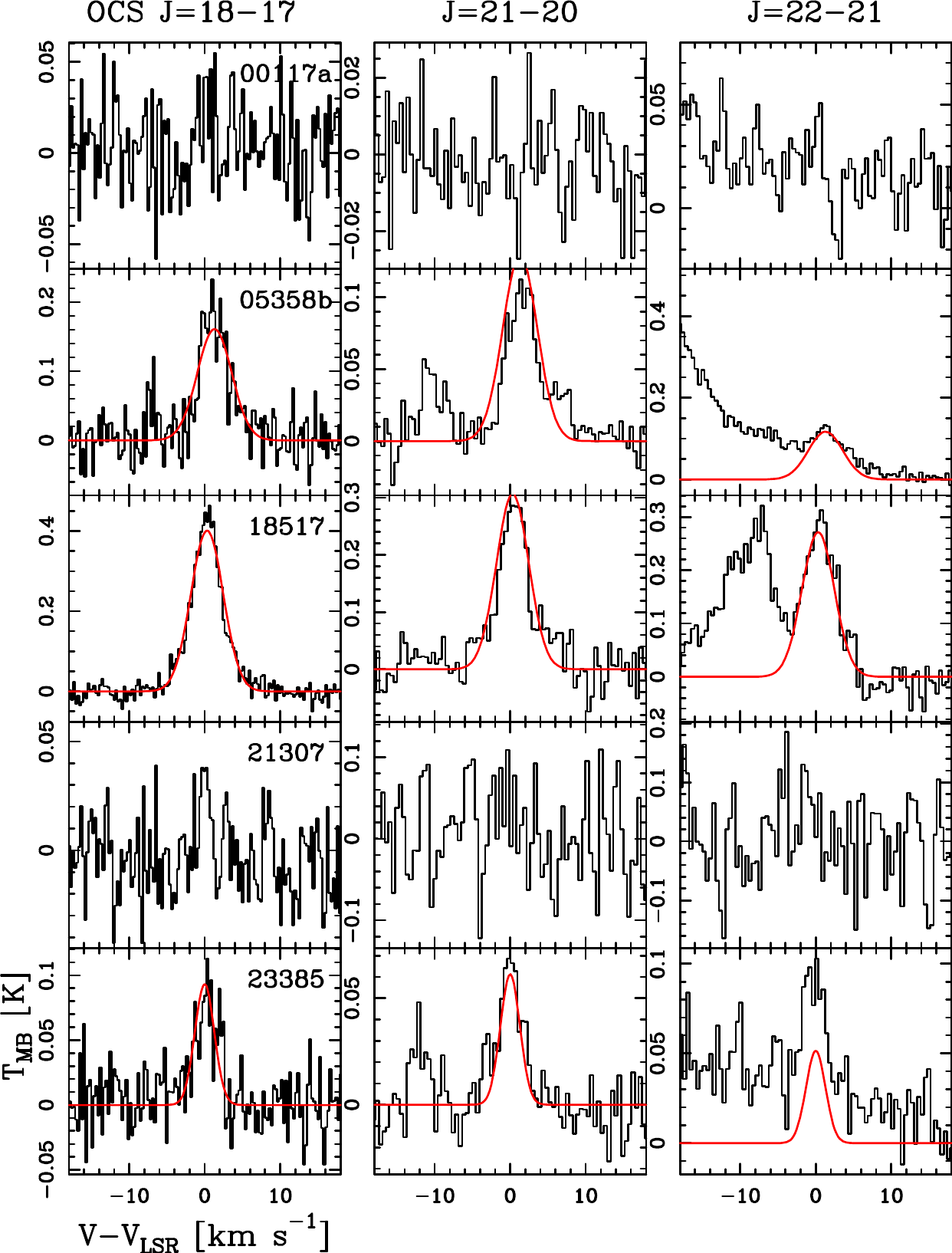}
      \caption{Same as Fig.~\ref{fig:spec-ocs-hmsc} for the HMPOs.}
         \label{fig:spec-ocs-hmpo}
   \end{figure*}

      \begin{figure*}
   \centering
   \includegraphics[width=15cm]{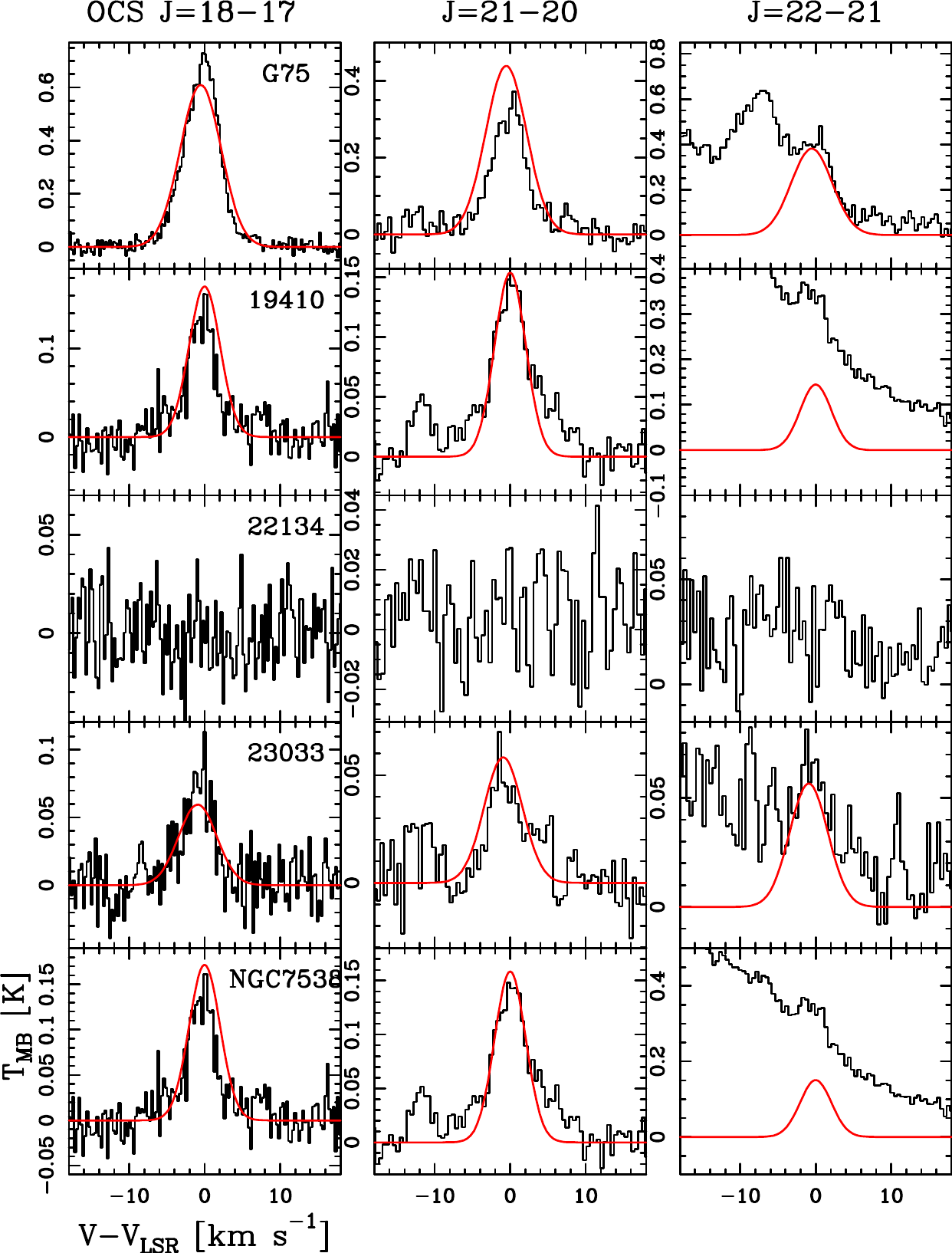}
      \caption{Same as Fig.~\ref{fig:spec-ocs-hmsc} for the UCHIIs.}
         \label{fig:spec-ocs-uchii}
   \end{figure*}

      \begin{figure*}
   \centering
   \includegraphics[width=15cm]{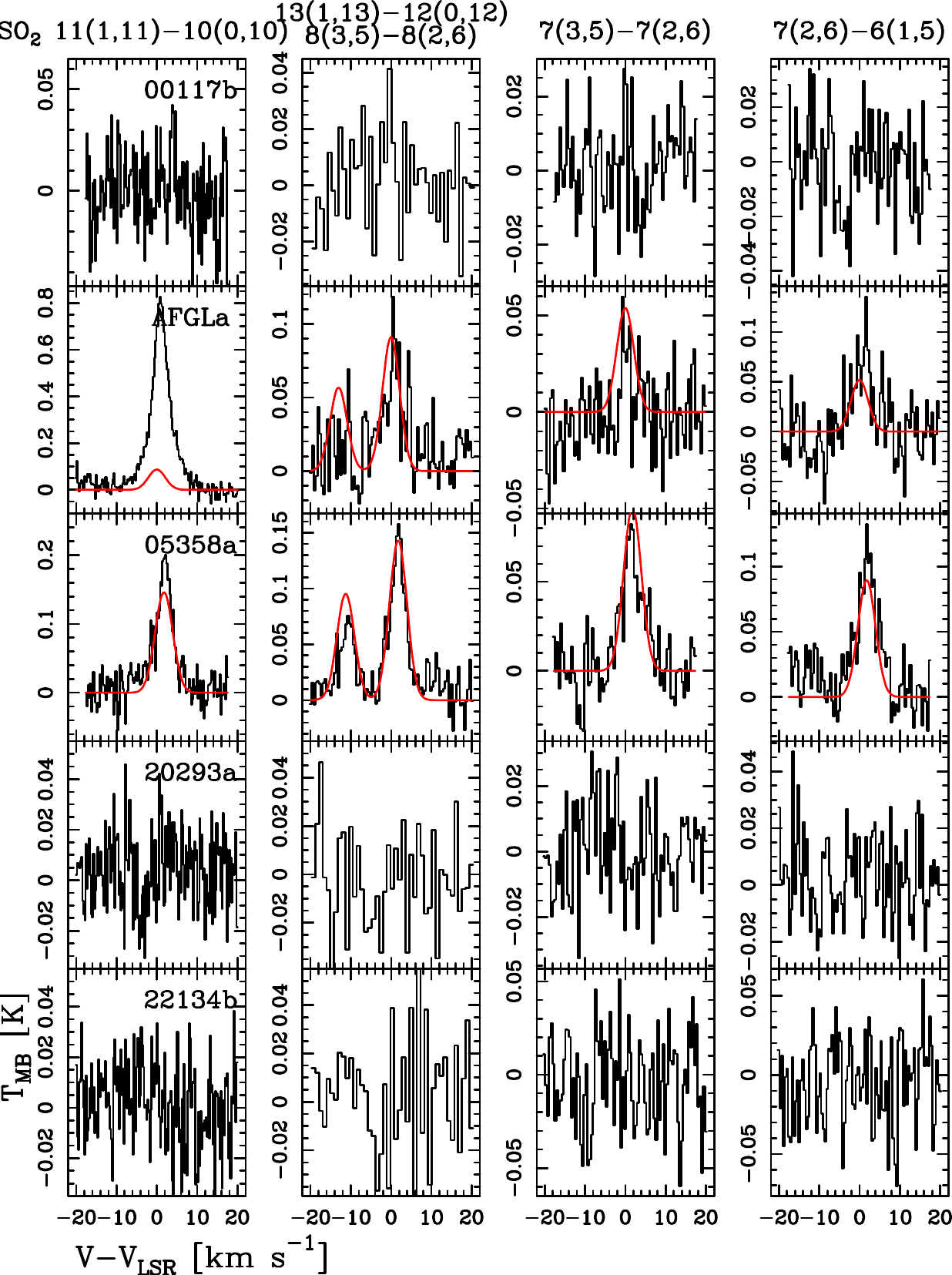}
      \caption{Spectra obtained towards the HMSCs of the SO$_2$ $J(K_a,K_b)=11(1,11)-10(0,10)$, $13(1,13)-12(0,12)$, $8(3,5)-8(2,6)$, $7(3,5)-7(2,6)$, and $7(2,6)-6(1,5)$ lines. The red curves indicate the best fit obtained with \sc{MADCUBA}.}
         \label{fig:spec-so2-hmsc}
   \end{figure*}

     \begin{figure*}
   \centering
   \includegraphics[width=15cm]{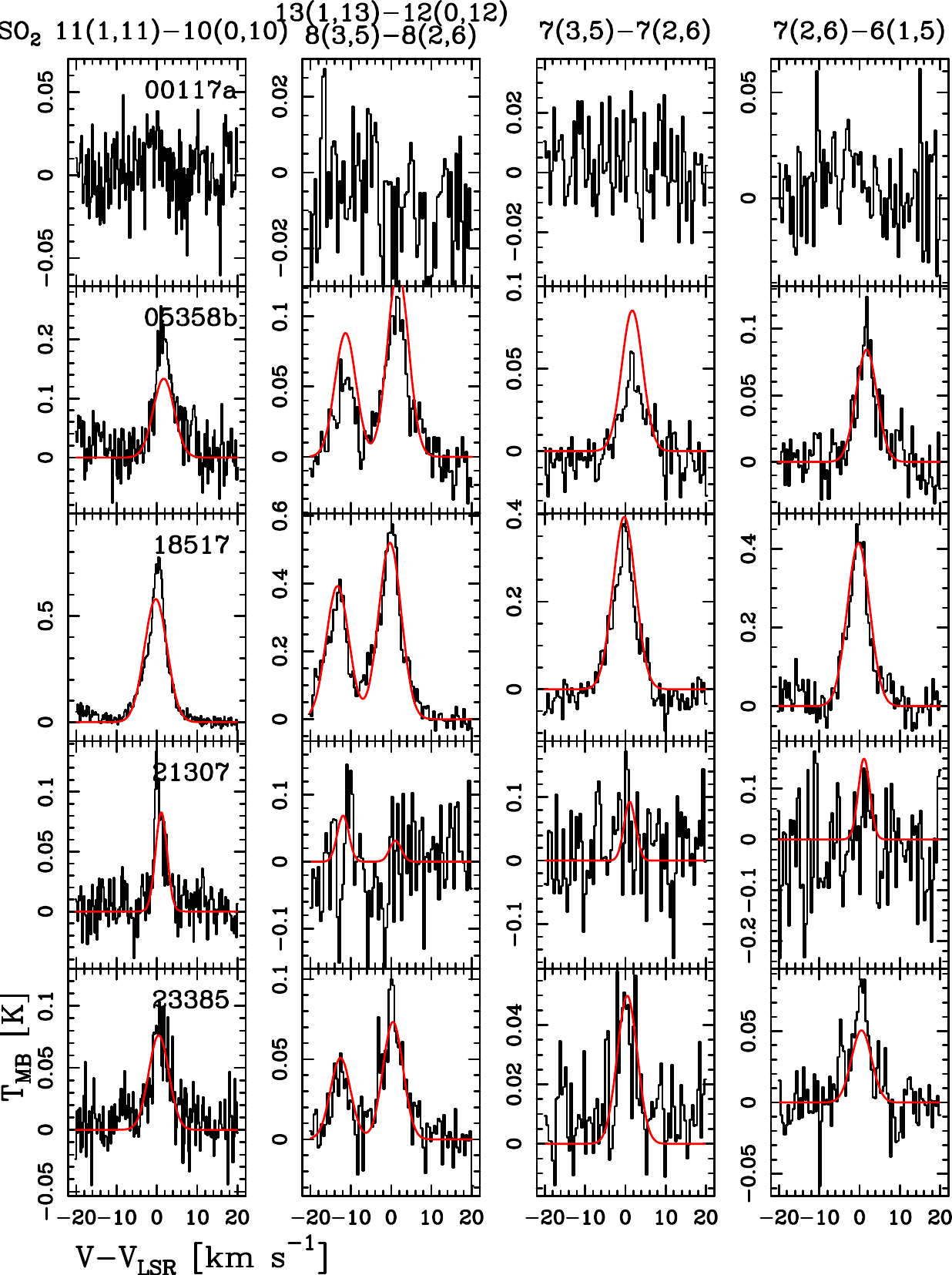}
      \caption{Same as Fig.~\ref{fig:spec-so2-hmsc} for HMPOs.}
         \label{fig:spec-so2-hmpo}
   \end{figure*}

     \begin{figure*}
   \centering
   \includegraphics[width=15cm]{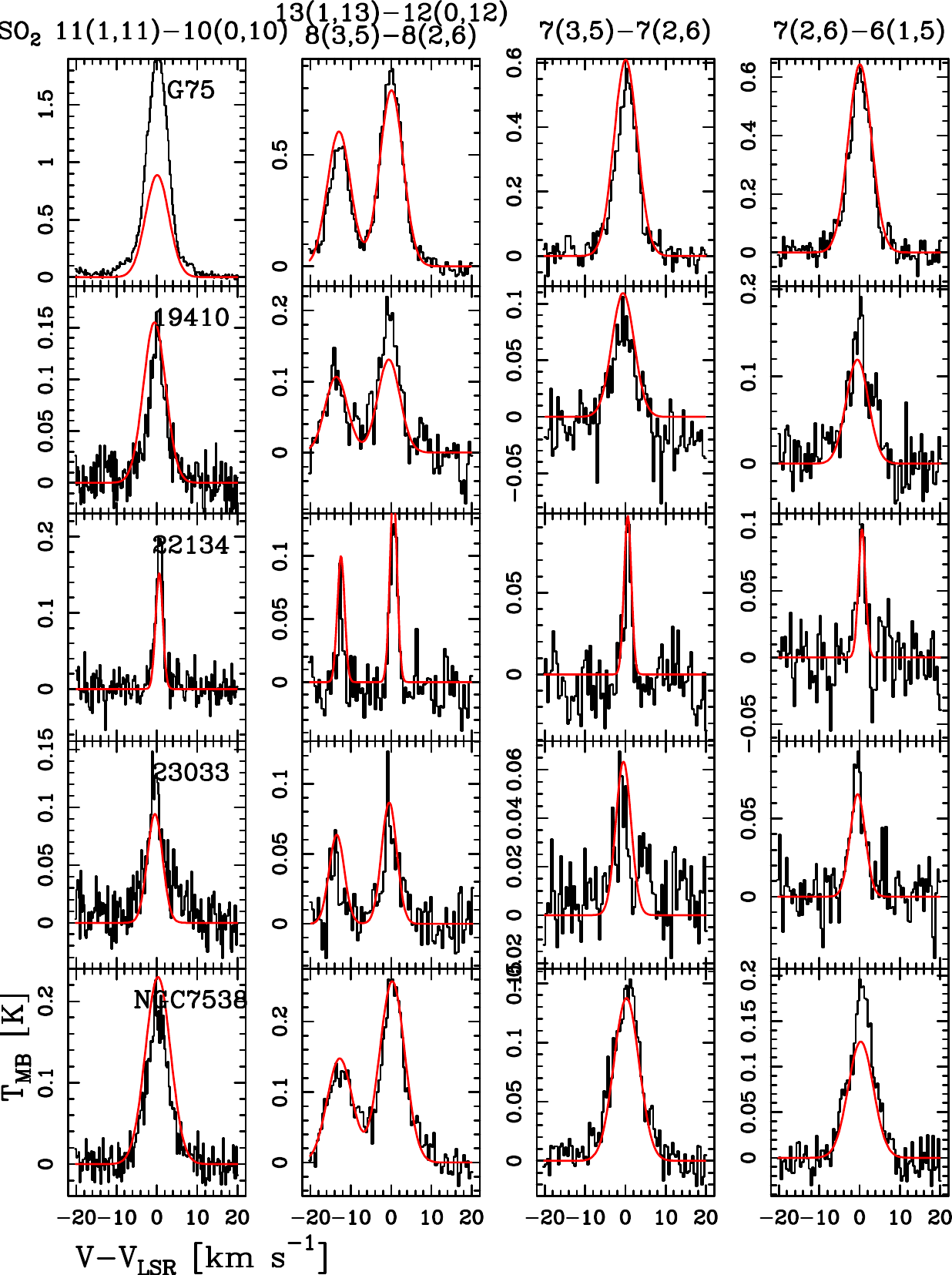}
      \caption{Same as Fig.~\ref{fig:spec-so2-hmsc} for UCHIIs.}
         \label{fig:spec-so2-uchii}
   \end{figure*}

       \begin{figure*}
   \centering
   \includegraphics[width=15cm]{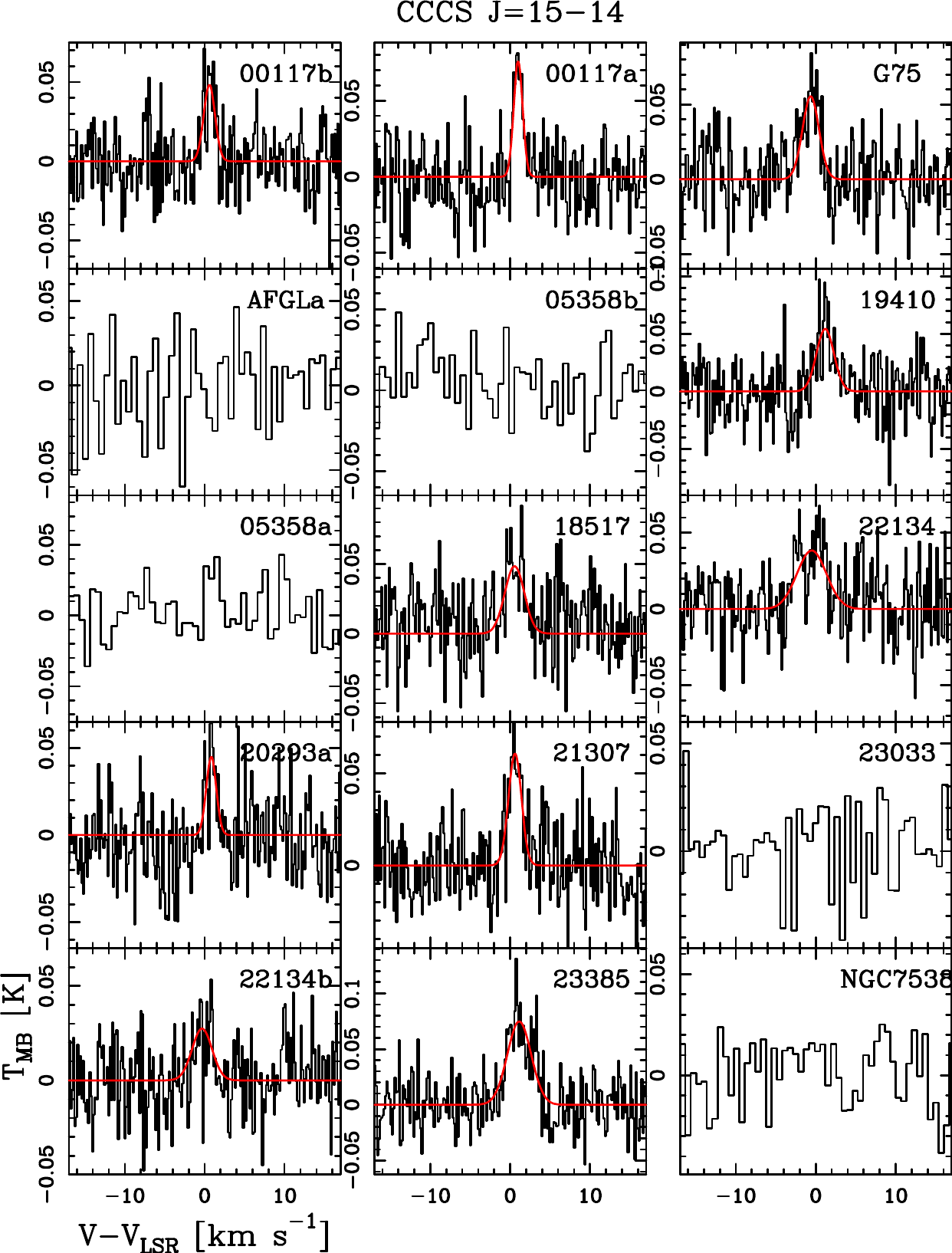}
      \caption{Same as Fig.~\ref{fig:spec-13cs} for the CCCS $J=15-14$ lines.}
         \label{fig:spec-cccs}
   \end{figure*}
      
          \begin{figure*}
   \centering
   \includegraphics[width=15cm]{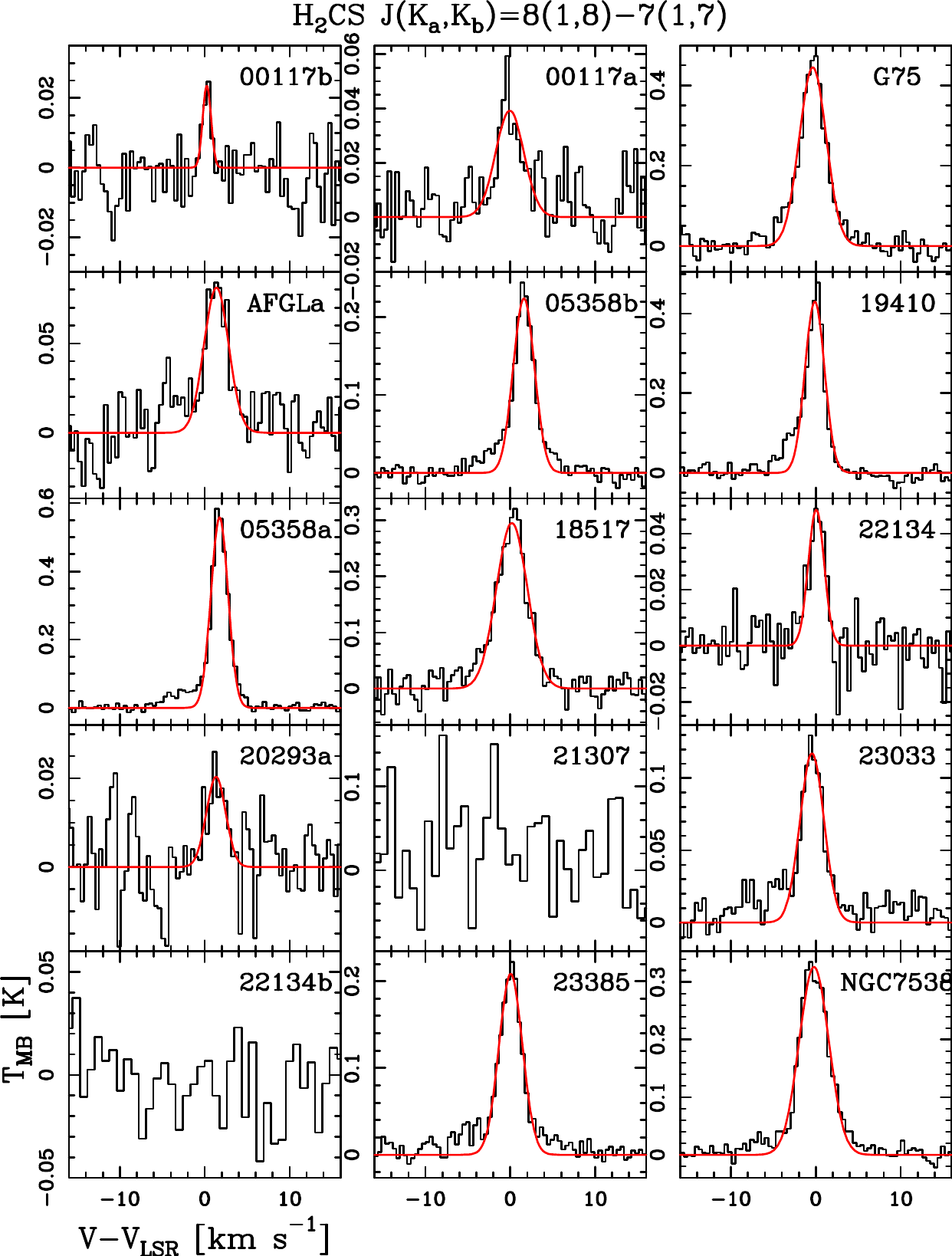}
      \caption{Same as Fig.~\ref{fig:spec-13cs} for the H$_{2}$CS $J(K_{\rm a}),K_{\rm b}=8(1,8)-7(1,7)$ lines.}
         \label{fig:spec-h2cs}
   \end{figure*}

          \begin{figure*}
   \centering
   \includegraphics[width=15cm]{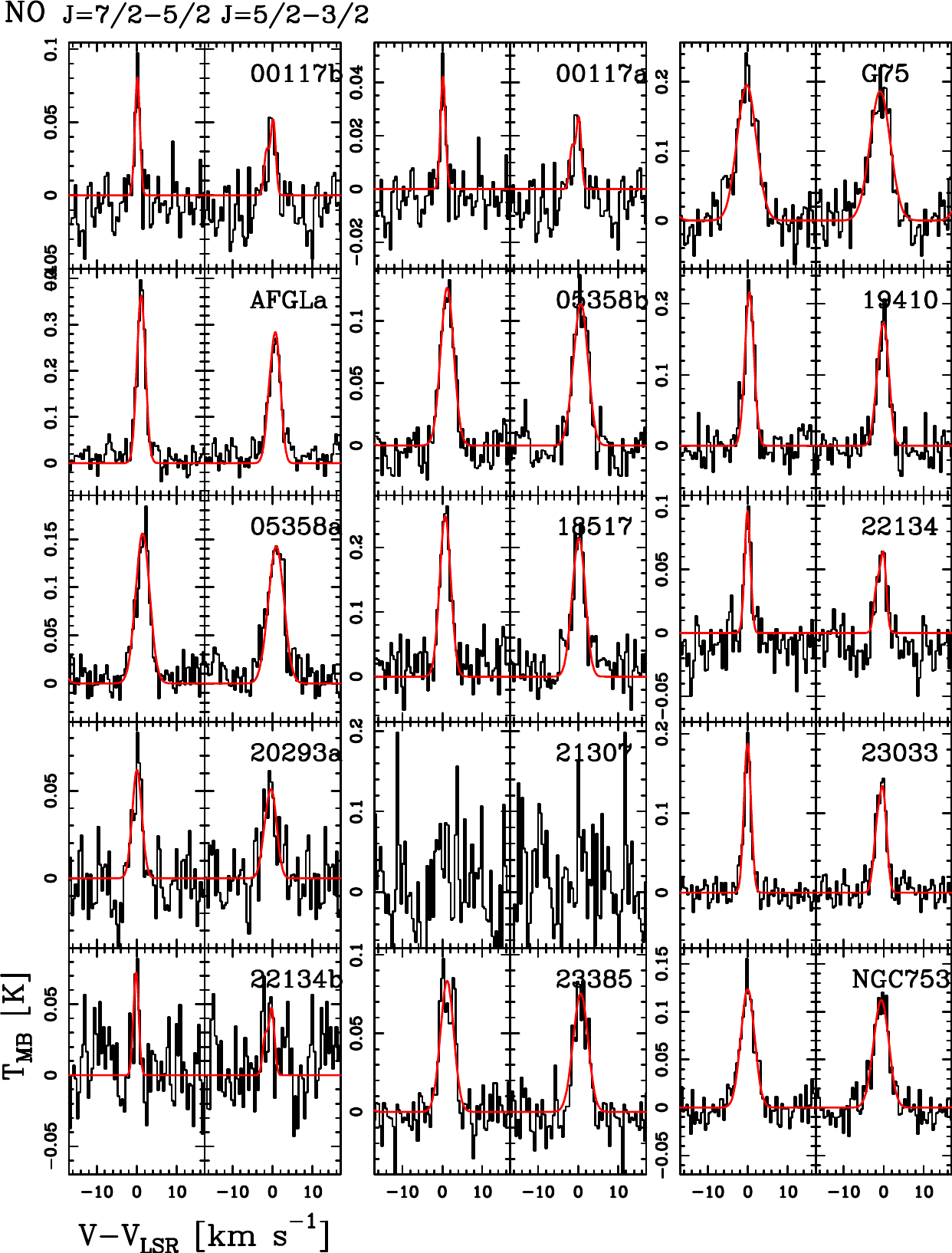}
      \caption{Same as Fig.~\ref{fig:spec-13cs} for the NO lines listed in Table~\ref{tab:lines-2}.}
         \label{fig:spec-no}
   \end{figure*}

\newpage

\FloatBarrier
\section{Best-fit parameters}
\label{app:tables}

In this appendix, we include the best-fit parameters of the species analysed obtained with {\sc madcuba} (Sect.~\ref{reduction}): centroid velocities, line widths at half maximum, excitation temperatures, and total column densities.

\begin{table*}
\label{tab:velocities}
    \caption{Best fit centroid velocities in \kms\ for the molecules with two atoms analysed. Numbers in brackets give the uncertainties.}
    \begin{tabular}{ccccccc}
\hline
\hline
sources  &  $^{13}$CS     &   C$^{34}$S     &        SO     &     SO$^+$    &          NS     & NO  \\                              
\hline
\multicolumn{7}{c}{HMSCs} \\
\hline        
00117b   &   --36.02(0.08) &  --36.07(0.02) &  --35.77(0.02) &   --        &   --       & --36.1(0.12)  \\
AFGLa    &   --2.71(0.05) &  --2.87(0.04) &  --2.04(0.05)    &   --        &  --2.79(0.14)  & --2.79(0.04)  \\
05358a   &   --16.46(0.08) &  --16.48(0.08) &  --16.17(0.08) &  --16.0(0.2) &  --15.70(0.02)  &--16.18(0.08) \\ 
20293a   &   6.39(0.05) &  7.00(0.02)     &  7.08(0.04)     &   --         &   --       &   6.37(0.16)   \\
22134b   &   --18.46(0.08) &  --18.55(0.02) &  18.53(0.02)   &   --        &   --       &--18.46(0.13) \\
\hline
\multicolumn{7}{c}{HMPOs} \\
\hline                                                       
00117a   &   --36.02(0.08) &  --36.20(0.03) &  --36.21(0.02) &   --        &  --35.83(0.12  & --36.14(0.09)  \\
05358b   &   --16.33(0.07) &  --16.43(0.06) &  --16.12(0.03)  &  --16.1(0.2)   &  --15.91(0.06  & --16.10(0.08)  \\
18517    &   43.70(0.03) &  44.14(0.02)   &  43.94(0.02)     &  43.2(0.3)    &  43.82(0.14)   &   44.44(0.08)  \\
21307    &   --46.29(0.06) &  --46.35(0.02) &  --46.33(0.03) &   --        &   --       &      --    \\
23385    &   --49.8(0.1)  &  --49.9(0.1)  &  --49.54(0.03)  &  --49.9(0.3)   &  --50.3(0.1)   & --49.24(0.15)   \\
\hline
\multicolumn{7}{c}{UCHIIs} \\
\hline     
G75       &   --0.30(0.07) &  0.22(0.07)   &   --0.01(0.5)   &  0.4(0.2)      &  0.30(0.09)    & --0.04(0.11)   \\         
19410    &   22.58(0.03) &  23.07(0.04)   &  22.75(0.02)    &  22.7(0.1)     &  22.67(0.02)   &   22.80(0.06)   \\
22134    &   --18.29(0.03) &  --18.41(0.02) &  --17.97(0.02) &  --17.8(0.2)  &  --18.1(0.2)   & --18.40(0.10)   \\
23033    &   --52.96(0.03) &  --53.06(0.02) &  --53.02(0.02) &  --53.3(0.2)  &  --53.4(0.1)   & --53.09(0.05)  \\
NGC7538     &   --56.93(0.08) &  --57.06(0.07) &  --56.96(0.02) &  --56.6(0.2)  &  --57.11(0.09) & --57.04(0.08)   \\ 
\hline
    \end{tabular}
    \label{tab:velocities}
\end{table*}

\begin{table*}
\label{tab:velocities2}
    \caption{Same as Table~\ref{tab:velocities} for the remaining molecular species}
    \begin{tabular}{cccccccc}
\hline
\hline
sources  &           CCS     &     HCS$^+$ &        H$_2$S    &        OCS    &         SO$_2$   &         CCCS  &        H$_2$CS \\
\hline
\multicolumn{8}{c}{HMSCs} \\
\hline
00117b   &   --36.16(0.09)   &   --36.13(0.09) &   --   &    --   &    --   &  --35.7(0.1)  &  --36.04(0.09)  \\ 
AFGLa    &   --2.1(0.2)      &   --2.72(0.03) &  --2.54(0.05) &   --2.72(0.04) &   --2.7(0.1)  &   --   &  --2.52(0.17)  \\ 
05358a   &   --16.0(0.1)     &   --16.33(0.06) &  --16.02(0.05) &   --16.0(0.1)  &   --15.7(0.1)  &   --   &  --15.84(0.04)  \\ 
20293a   &   7.1(0.3)        &   6.67(0.06) &   --   &    --   &    --   &  7.14(0.07) &  7.6(0.3) \\ 
22134b   &    --           &   --18.37(0.06) &   --   &    --   &    --   &  --18.6(0.02) &   --  \\ 
\hline
\multicolumn{8}{c}{HMPOs} \\
\hline
00117a   &   --36.1(0.1)     &   --36.03(0.07) &   --   &    --   &    --   &  --35.29(0.04) &  --36.3(0.2) \\ 
05358b   &   --15.9(0.1)     &   --16.62(0.06) &  --15.7(0.1)  &   --16.3(0.1)  &   --16.0(0.1)  &   --   &  --15.99(0.06)  \\ 
18517    &   43.9(0.2)       &   43.90(0.05) &  43.78(0.08) &   43.98(0.09) &   43.4(0.1)  &  44.3(0.1)  &  43.90(0.08)  \\ 
21307    &    --         &   --46.40(0.12) &   --   &    --   &    --   &  --46.09(0.07) &   --  \\ 
23385    &   --50.34(0.09)   &   --49.65(0.07) &   --   &   --50.3(0.2)  &   --50.1(0.1)  &  --49.4(0.1)  &  --50.44(0.07)  \\ 
\hline
\multicolumn{8}{c}{UCHIIs} \\
\hline
G75       &   --0.03(0.02)    &   --0.37(0.05) &  0.13(0.04) &   --0.17(0.09) &   0.40(0.08) &  --0.41(0.09) &  --0.15(0.06)  \\    
19410    &   22.46(0.05)     &   22.95(0.02) &  22.31(0.05) &    21.8(0.1)  &   22.20(0.08) &  23.6(0.1)  &  22.29(0.07)  \\ 
22134    &   --18.7(0.13)    &   --18.27(0.05) &  --17.8(0.1)  &    --   &  --17.82(0.06) & --18.8(0.2)  & --18.2(0.2) \\ 
23033    &   --53.23(0.09)   &   --52.87(0.03) &  --53.9(0.4)  &   --53.8(0.2)  &   --53.6(0.1)  &   --   & --53.44(0.09)  \\ 
NGC7538     &   --57.2(0.13)    &   --57.08(0.05) &  --57.6(0.1)  &   --57.4(0.5)  &   --56.7(0.1)  &   --   & --57.20(0.06)  \\ 
\hline
\end{tabular}
\label{tab:velocities2}
\end{table*}

\begin{table*}
\label{tab:linewidths}
    \caption{Best fit line widths at half maximum in \kms\ for the molecules with two atoms analysed. Numbers in brackets give the uncertainty.}
    \begin{tabular}{ccccccc}
\hline
\hline
source  &   $^{13}$CS  &     C$^{34}$S  &      SO       &     SO$^+$    &     NS        & NO  \\ 
\hline
\multicolumn{7}{c}{HMSCs} \\
\hline
00117b  &    1.8(0.2)  &     1.89(0.04)  &     2.02(0.05)  &     --  &   --       & 1.5(0.3)  \\ 
AFGLa  &    3.2(0.1)  &     3.19(0.09)  &     4.4(0.1)    &     --  &   1.3(0.3)  & 2.6(0.1) \\ 
05358a  &    3.4(0.2)  &     3.4(0.2)    &     4.1(0.2)    &     4.3(0.4) &   2.01(0.05) & 4.2(0.2)  \\ 
20293a  &    2.8(0.1)  &     2.70(0.06)  &     3.46(0.09)  &     --  &   --   & 2.9(0.4)   \\ 
22134b  &    1.6(0.2)  &     1.61(0.06)  &     1.45(0.05)  &     --  &   --  & 1.6(0.3)  \\ 
\hline
\multicolumn{7}{c}{HMPOs} \\
\hline              
00117a  &    1.8(0.2)  &     2.02(0.08)  &     1.94(0.03)  &     2.02  &   1.2(0.3)  & 2.0(0.2)  \\ 
05358b  &    3.6(0.2)  &     3.50(0.15)  &     4.61(0.07)  &     5.3(0.6) &   2.73(0.16) & 3.3(0.1)  \\
18517   &   3.11(0.06)  &    3.12(0.05) &     3.34(0.05)  &     4.5(0.7) &   2.9(0.4)    & 3.3(0.2)  \\ 
21307   &    2.2(0.2)  &    2.20(0.05) &     2.25(0.06)  &     --        &   --    &  --  \\ 
23385   &    2.7(0.3)  &     3.1(0.3)  &     4.94(0.08)  &     2.7(0.7)  &   2.1(0.2)    & 4.2(0.4)  \\ 
\hline
\multicolumn{7}{c}{HMSCs} \\
\hline
G75      &    4.1(0.2)  &    4.12(0.15) &     6.24(0.10)  &     7.0(0.5 ) &  3.5(0.3 )   & 5.7(0.3) \\
19410   &   2.01(0.08)  &    2.11(0.08) &     2.65(0.03)  &     2.0(0.3) &   1.65(0.04)  & 2.6(0.2)  \\ 
22134   &   1.77(0.07)  &    1.91(0.06) &     2.21(0.04)  &     4.0(0.4) &   2.3(0.5)    & 1.7(0.2)   \\ 
23033   &   2.40(0.08)  &    2.40(0.05) &     3.09(0.05)  &     3.8(0.4) &    2.0(0.2)   & 2.1(0.1)   \\ 
NGC7538    &    3.4(0.2)  &     3.6(0.2)   &     4.39(0.04)  &     5.3(0.6) &   5.0(0.2)    & 4.0(0.2)   \\ 
\hline
\end{tabular}
\tablefoot{The values without uncertainty are the FWHMs assumed to compute the column density upper limits.}
\label{tab:linewidths}
\end{table*}

\begin{table*}
\label{tab:linewidths2}
    \caption{Same as Table~\ref{tab:linewidths} for the remaining molecular species}
    \begin{tabular}{cccccccc}
\hline
\hline
source &     CCS     &  HCS$^+$    &   p-H$_2$S   &    OCS  &   SO$_2$  &   CCCS  & o-H$_2$CS \\
\hline
\multicolumn{8}{c}{HMSCs} \\
\hline
00117b &   1.5(0.2)  &    1.6(0.2) &    2.02  &   2.02  &  2.02  &   1.7(0.2) &   1.13(0.22)  \\ 
AFGLa  &   3.5(0.5)  &    2.28(0.08)  &    4.40(0.12)  &   4.69(0.08)  &  4.7(0.3) &   2.28  &   3.4(0.4) \\ 
05358a  &   3.0(0.2) &     3.41(0.14)  &     2.7(0.1)  &   3.89(0.2) &  5.2(0.3) &   3.41  &   2.33(0.09)  \\
20293a &   2.9(0.6)  &    2.12(0.15)  &    3.46  &   3.46  &  3.46  &   1.5(0.2) &   2.7(0.6) \\ 
22134b &   1.61  &    2.18(0.30)  &    1.45  &      1.45  &  1.45  &   2.9(0.5) &   1.61  \\
\hline
\multicolumn{8}{c}{HMPOs} \\
\hline
00117a &   1.0(0.3)  &    1.7(0.2) &    1.94  &   1.94  &  1.94  &   1.3(0.1) &   3.9(0.6) \\ 
05358b  &   2.7(0.3) &     3.74(0.14)  &    3.8(0.3) &   5.4(0.3) &  6.3(0.2) &   3.74  &   2.92(0.15)  \\ 
18517  &   3.8(0.6)  &    3.27(0.11)  &    4.0(0.2) &   4.7(0.2) &  5.9(0.2) &   2.9(0.3) &   4.27(0.18)  \\ 
21307  &   0.0(0.0)  &    2.9(0.3) &    2.25  &   2.25  &  3.37  &   1.9(0.2) &   0. (0. ) \\ 
23385  &   2.0(0.2)  &    3.10(0.17)  &    4.94  &   4.0(0.4) &  5.9(0.3) &   3.5(0.2) &   3.23(0.16)  \\ 
\hline
\multicolumn{8}{c}{UCHIIs} \\
\hline
G75     &  4.3 (0.4 ) &    4.52(0.12) &   4.48(0.09) &  6.3 (0.2 ) & 6.8 (0.2 ) &  2.4 (0.2 ) &  3.73(0.14) \\
19410  &   1.7(0.1)  &    2.09(0.04)  &    3.1(0.1) &    5.5(0.3)  &  6.4(0.3) &   2.6(0.2) &   2.69(0.16)  \\ 
22134  &   2.2(0.3)  &    1.89(0.12)  &    2.0(0.3) &   0. (0. ) &  2.5(0.1) &   4.3(0.4) &   2.2(0.4) \\ 
23033  &   2.3(0.2)   &   2.49(0.08)  &    6.0(1.0) &   4.5(0.4) &  4.4(0.2) &   2.49  &   3.3(0.2) \\ 
NGC7538   &   3.1(0.3)   &   3.59(0.18)  &    3.9(0.2) &   5.3(1.1) &  5.4(0.2) &   3.59  &   4.10(0.13)  \\ 
\hline
\end{tabular}
\tablefoot{The values without uncertainty are the FWHMs assumed to compute the column density upper limits}
\label{tab:linewidths2}
\end{table*}

\begin{table*}
\label{tab:temperatures}
\caption{Excitation temperatures (in K) derived with {\sc madcuba}. }
    \begin{tabular}{cccccc}
\hline
\hline
source   &      SO       &     CCS  &     HCS$^+$    &     OCS     &   SO$_2$    \\    
\hline
\multicolumn{6}{c}{HMSCs} \\
\hline
00117b   &    11.9(0.4)  &    12(3)  &    13(3)  &     --    &    --    \\  
AFGLa    &    11(1)      &    18(3)  &    12.8(0.7)  &   --\tablefootmark{a}   &  73\tablefootmark{b}   \\  
05358a   &    24.2(1.0)  &    17(3)  &    22.0(1.0)  &   76(3)  &  68(3)   \\  
20293a   &    12.6(0.4)  &    12(3)    &   12.1(0.9)  &   --    &   --    \\  
22134b   &    10.7(0.4)  &    12(4)   &    12(2)     &     --    &    --    \\  
\hline
\multicolumn{6}{c}{HMPOs} \\
\hline
00117a   &    16.7(0.4)  &    13(3)  &    19(2)  &     21    &    --     \\  
05358b   &    20.9(0.5)  &    26(4)  &    16.6(0.6)  &   66(5)  &  63(5)  \\  
18517    &    25.8(0.6)  &    21(3)  &    26.4(0.9)  &   60(3)  &  48(4)   \\  
21307    &    16.5(0.9)  &    16(4)    &    25(2)  &     21    &    17(2)   \\  
23385    &    33.0(0.6)  &    22(3)  &    27.8(1.0)  &   48(2)  &  56(5)   \\  
\hline
\multicolumn{6}{c}{UCHIIs} \\
\hline
G75       &    18.4(0.9)  &    29(5)   &    27.3(0.5)  &   54(4)  &  47(5)   \\  
19410    &    23.4(0.4)  &    25(2)  &    21.5(0.3)  &   85(8)  &  42(5)     \\  
22134    &    23.9(0.6)  &    25(5)  &    14.3(1.2)  &   --    &  62(6)      \\  
23033    &    21.3(0.4)  &    13(2)  &    14.4(0.6)  &   105(8)  &  50(4)   \\  
NGC7538     &    49.7(1.0)  &    21(3)  &    22.3(0.5)  &   90(6)  &  93(7)   \\
\hline 
\end{tabular}
\tablefoot{The values without uncertainty are the temperatures assumed to derive the upper limits on the total column densities.
\tablefoottext{a}{the $J=18-17$ line is too strong with respect to the LTE predictions. It is unlikely that the discrepancy in intensity between this line and the LTE prediction is due to contamination with nearby lines because no reliable candidates have been found around it;}
\tablefoottext{b}{\Tex\ needs to be fixed to 73~K to obtain a reasonable fit, but the $J(K_a,K_b)=11(1,11)-10(0,10)$ intensity is largely underestimated probably due to contamination with methyl formate $J(K_a,K_b)=11(5,6)-11(3,9)$ at 221.969 GHz.}
}
\label{tab:temperatures}
\end{table*}

\begin{table*}
\label{tab:column1}
\caption{Column densities in \cmq\ of the analysed diatomic molecules}
    \begin{tabular}{ccccccc}
\hline
\hline
      source   &     $^{13}$CS              &          C$^{34}$S  &       SO                    &      SO$^+$               &      NS                        &    NO   \\
\hline
\multicolumn{7}{c}{HMSCs} \\
\hline                                         
      00117b &   1.0(0.1)$\times 10^{12}$    &   3.16(0.06)$\times 10^{12}$   &  1.7(0.1)$\times 10^{13}$     &  3.2$\times 10^{12}$      &   5.0$\times 10^{11}$          &  1.9(0.2)$\times 10^{14}$  \\ 
       AFGLa &  8.9(0.3)$\times 10^{12}$     &   2.34(0.06$\times 10^{13}$    &  3.0(0.2)$\times 10^{14}$     &  1.3$\times 10^{13}$       &  2.5(0.8)$\times 10^{12}$     &  1.48(0.05)$\times 10^{15}$ \\ 
      05358a &  7.6(0.3)$\times 10^{12}$     &   1.82(0.08)$\times 10^{13}$   &  1.4(0.2)$\times 10^{14}$     &  7.1(0.4)$\times 10^{12}$  &   1.95(0.02)$\times 10^{13}$  &  8.5(0.3)$\times 10^{14}$  \\ 
      20293a &  2.24(0.08)$\times 10^{12}$   &   5.9(0.1)$\times 10^{12}$     &  2.2(0.1)$\times 10^{13}$     &  5.0$\times 10^{12}$       &  5.0$\times 10^{11}$          &  2.3(0.3)$\times 10^{14}$  \\ 
      22134b &  5.9(0.6)$\times 10^{11}$     &   1.95(0.06)$\times 10^{12}$   &  1.20(0.06)$\times 10^{13}$    &  3.5$\times 10^{11}$      &  7.9$\times 10^{11}$          &  1.5(0.3)$\times 10^{14}$  \\ 
\hline
\multicolumn{7}{c}{HMPOs} \\
\hline 
      00117a &  1.1(0.1)$\times 10^{12}$      &   2.45(0.08)$\times 10^{12}$   &  1.78(0.06)$\times 10^{13}$    &  4.3$\times 10^{12}$      &   9(2)$\times 10^{11}$        &  1.8(0.2)$\times 10^{14}$  \\ 
      05358b &  7.1(0.3)$\times 10^{12}$     &   1.62(0.06)$\times 10^{13}$   &  1.38(0.04)$\times 10^{14}$    &  7.2(0.5)$\times 10^{12}$  &  6.8(0.2)$\times 10^{12}$    &  6.3(0.3)$\times 10^{14}$  \\ 
       18517 &  8.7(0.1)$\times 10^{12}$     &   1.95(0.03)$\times 10^{13}$   &  1.82(0.05)$\times 10^{14}$    &  1.4(0.1)$\times 10^{13}$  &  7.8(0.3)$\times 10^{12}$    &  1.07(0.05)$\times 10^{15}$ \\ 
       21307 &  1.29(0.08)$\times 10^{12}$   &   3.80(0.06)$\times 10^{12}$   &  5.4(0.2)$\times 10^{13}$     &  5.5$\times 10^{12}$        &  1.6$\times 10^{12}$         &  2(1)$\times 10^{14}$    \\ 
       23385 &  3.8(0.3)$\times 10^{12}$     &   1.02(0.08)$\times 10^{13}$   &  7.1(0.2)$\times 10^{13}$     &  2.8(0.3)$\times 10^{12}$    &  2.5(0.1)$\times 10^{12}$   &  4.3(0.3)$\times 10^{14}$ \\ 
\hline
\multicolumn{7}{c}{UCHIIs} \\
\hline 
        G75 &  1.55(0.05)$\times 10^{13}$   &   3.5(0.1)$\times 10^{13}$    &  6.2(0.1)$\times 10^{14}$     &  2.3(0.1)$\times 10^{13}$     &  1.02(0.03)$\times 10^{13}$ &  1.45(0.06)$\times 10^{15}$ \\ 
       19410 &  1.15(0.04)$\times 10^{13}$   &   2.46(0.08)$\times 10^{13}$   &  1.10(0.03)$\times 10^{14}$   &  4.8(0.4)$\times 10^{12}$    &  1.41(0.03)$\times 10^{13}$ &  7.6(0.3)$\times 10^{14}$  \\ 
       22134 &  2.19(0.08)$\times 10^{12}$   &   6.0(0.2)$\times 10^{12}$    &  4.4(0.1)$\times 10^{13}$     &  1.32(0.04)$\times 10^{13}$  &  1.1(0.3)$\times 10^{12}$   &  2.2(0.2)$\times 10^{14}$  \\ 
       23033 &  4.5(0.1)$\times 10^{12}$     &   1.20(0.02)$\times 10^{13}$  &  6.2(0.3)$\times 10^{13}$     &  5.1(0.4)$\times 10^{12}$    &  1.9(0.3)$\times 10^{12}$   &  5.1(0.2)$\times 10^{14}$  \\ 
        NGC7538 &  6.8(0.3)1$\times 10^{12}$    &   1.70(0.06)$\times 10^{13}$  &  1.02(0.02)$\times 10^{14}$   &  5.5(0.3)$\times 10^{12}$    &  5.6(0.2)$\times 10^{12}$   &  6.6(0.3)$\times 10^{14}$  \\ 
\hline
\end{tabular}
\label{tab:column1}
\end{table*}

\begin{table*}
\label{tab:column2}
\caption{Column densities in \cmq\ of the molecules with three and four atoms}
    \begin{tabular}{cccccccc}
\hline
\hline
      source  &    CCS                    &     HCS$^+$                &     p-H$_2$S                &       OCS                  &      SO$_2$               &       CCCS              &    o-H$_2$CS       \\
\hline
\multicolumn{8}{c}{HMSCs} \\
\hline 
      00117b  &  1.3(0.6)$\times 10^{12}$  &  5.0(0.5)$\times 10^{11}$  &  3.2$\times 10^{12}$        &  3.2$\times 10^{13}$       &  7.9$\times 10^{11}$      &  1.7(0.2)$\times 10^{12}$  & 8(1)$\times 10^{12}$   \\ 
       AFGLa  &  3.8(0.8)$\times 10^{12}$  &  3.6(0.1)$\times 10^{12}$  &  2.29(0.06)$\times 10^{14}$ &  9(1)$\times 10^{14}$      &  2.8$\times 10^{13}$      &   1.0$\times 10^{12}$     & 9(1)$\times 10^{13}$    \\ 
      05358a  &  4.3(0.3)$\times 10^{12}$  &  5.1(0.2)$\times 10^{12}$  &  6.5(0.3)$\times 10^{13}$   &  7(1)$\times 10^{13}$      &  4.6(0.8)$\times 10^{13}$  &  4.0$\times 10^{12}$     & 9.3(0.3)$\times 10^{13}$  \\ 
      20293a  &  1.4(0.3)$\times 10^{12}$  &  9.8(0.6)$\times 10^{11}$  &  1.2$\times 10^{13}$        &  4.0$\times 10^{13}$     &  5.0$\times 10^{12}$        &  1.0(0.1)$\times 10^{12}$  &  9(2)$\times 10^{12}$    \\ 
      22134b  &  5.0(0.9)$\times 10^{11}$  &  3.7(0.4)$\times 10^{11}$  &  1.1$\times 10^{12}$        &  4.0$\times 10^{13}$     &  7.9$\times 10^{12}$        &  1.3(0.2)$\times 10^{12}$   & 1.6$\times 10^{12}$      \\
\hline
\multicolumn{8}{c}{HMPOs} \\
\hline 
      00117a  &  1.1(0.2)$\times 10^{12}$  &  7.9(0.6)$\times 10^{11}$  &  7.9$\times 10^{12}$        &  1.6$\times 10^{13}$     &  5.5$\times 10^{12}$        &  1.18(0.08)$\times 10^{12}$  & 1.4(0.2)$\times 10^{13}$  \\ 
      05358b  &  4.2(0.4)$\times 10^{12}$  &  4.4(0.2)$\times 10^{12}$  &  7.1(0.6)$\times 10^{13}$   &  8(2)$\times 10^{13}$     &  4.8(0.2)$\times 10^{13}$   &  9.3$\times 10^{11}$       &  9.6(0.4)$\times 10^{13}$  \\ 
       18517  &  2.5(0.3)$\times 10^{12}$  &  3.72(0.08)$\times 10^{12}$  &  8.1(0.5)$\times 10^{13}$ &  2.0(0.4)$\times 10^{14}$ &  2.0(0.2)$\times 10^{14}$   &  1.3(0.1)$\times 10^{12}$   & 6.0(0.2)$\times 10^{13}$  \\ 
       21307  &  7.9(0.9)$\times 10^{11}$  &  9.6(0.6)$\times 10^{11}$  &  7.9$\times 10^{12}$        &  3.2$\times 10^{13}$      &  3.3$\times 10^{13}$       &  1.15(0.08)$\times 10^{12}$  & 2.3$\times 10^{13}$     \\ 
       23385  &  1.7(0.2)$\times 10^{12}$  &  2.5(0.1)$\times 10^{12}$  &  9.3$\times 10^{12}$        &  3(2)$\times 10^{13}$     &  2.5(0.3)$\times 10^{13}$   &  2.4(0.1)$\times 10^{12}$   & 3.2(0.2)$\times 10^{13}$ \\ 
\hline
\multicolumn{8}{c}{UCHIIs} \\
\hline 
          G75  &  2.5(0.3)$\times 10^{12}$  &  5.9(0.1)$\times 10^{12}$  &  3.2(0.1)$\times 10^{14}$   &  4(1)$\times 10^{14}$     &  3.16(0.04)$\times 10^{14}$  &  1.2(0.1)$\times 10^{12}$   & 7.8(0.3)$\times 10^{13}$ \\ 
       19410  &  2.8(0.2)$\times 10^{12}$  &  5.50(0.08)$\times 10^{12}$  &  5.0(0.2)$\times 10^{13}$  &  6.9(0.8)$\times 10^{13}$  &  5.1(0.5)$\times 10^{13}$  &  1.5(0.1)$\times 10^{12}$   & 8.7(0.4)$\times 10^{13}$ \\ 
       22134  &  1.5(0.2)$\times 10^{12}$  &  1.15(0.06)$\times 10^{12}$  &  3.0(0.5)$\times 10^{13}$  &  5.0$\times 10^{13}$     &   1.8(0.1)$\times 10^{13}$   &  3.0(0.3)$\times 10^{12}$   & 2.4(0.4)$\times 10^{13}$ \\ 
       23033  &  3.0(0.3)$\times 10^{12}$  &  3.02(0.08)$\times 10^{12}$  &  2.5(0.6)$\times 10^{13}$  &  3.0(0.3)$\times 10^{13}$  &  2.2(0.1)$\times 10^{13}$  &  8.9$\times 10^{11}$       &  8.7(0.5)$\times 10^{13}$ \\ 
        NGC7538  &  2.9(0.3)$\times 10^{12}$  &  5.4(0.2)$\times 10^{12}$  &   6.3(0.6)$\times 10^{13}$   &  7(2)$\times 10^{13}$     &  1.3(0.1)$\times 10^{14}$   &   6.3$\times 10^{11}$      &  8.9(0.2)$\times 10^{13}$ \\ 
\hline
\end{tabular}
\label{tab:column2}
\end{table*}

\newpage

\FloatBarrier
\section{Plots of fractional abundances versus all evolutionary parameters}
\label{app:plots-abb}

In this appendix, we show plots of the fractional abundances of all molecular species analysed as a function of the physical parameters expected to vary with evolution: dust temperature (\Td), kinetic temperature (\Tk, already shown in Fig.~\ref{fig:evol-Tk}), source bolometric luminosity ($L$), and lumonisity-to-mass ratio ($M/L$).

\FloatBarrier
\begin{figure*}
       \centering
   \includegraphics[width=15cm]{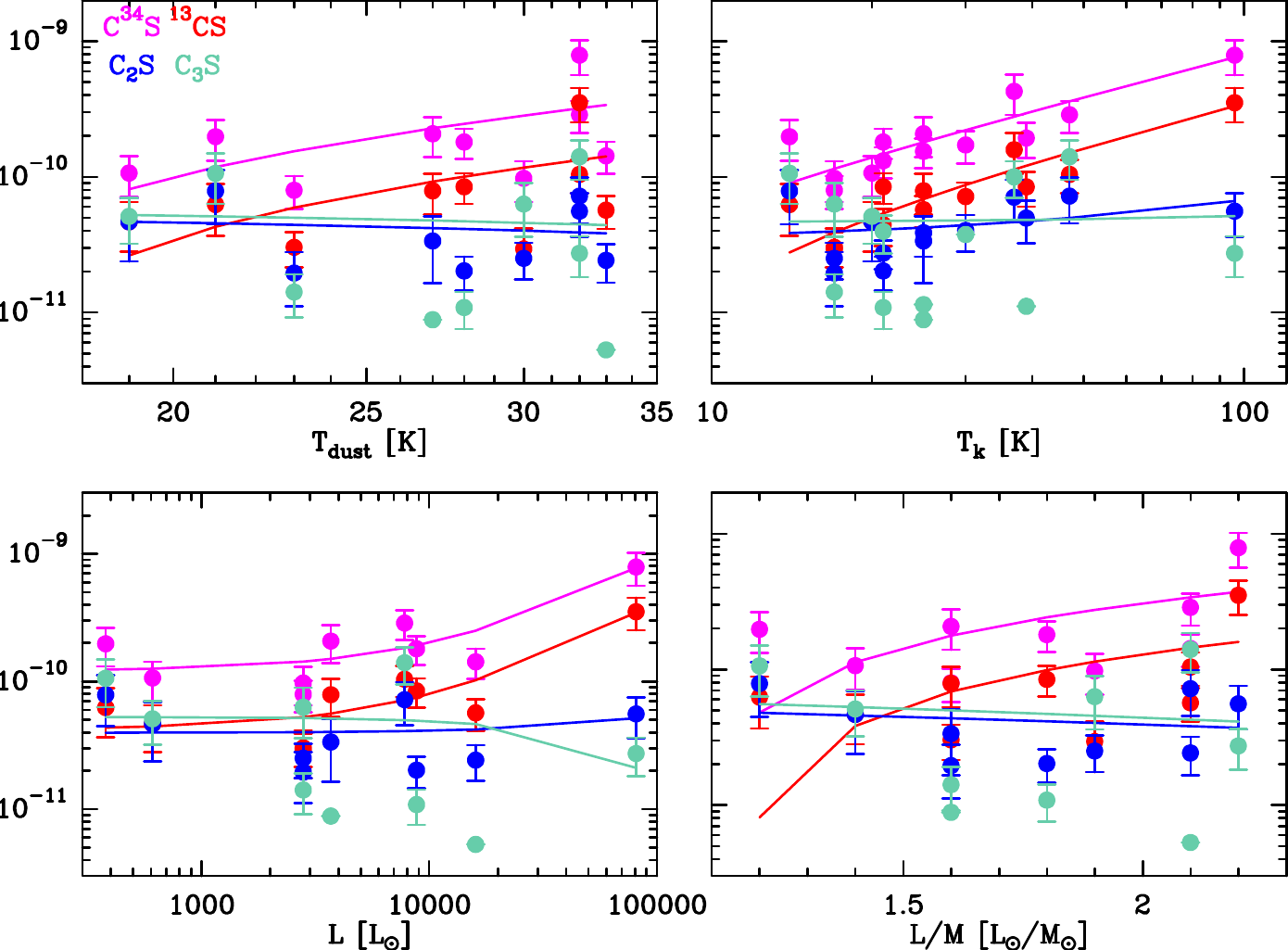}
      \caption{Fractional abundances as a function of evolutionary parameters.
      {\it Top-left panel:} measured abundances of molecules containing only sulphur and carbon as a function of the dust temperature. $T_{\rm d}$ was derived by fitting the Spectral Energy Distribution (SED) \citep{mininni21}. The symbols without uncertainty are upper limits. The curves represent linear fits to the data including upper limits;
      {\it Top-right panel:} same abundances as in top-left panel as a function of the gas kinetic temperature derived from ammonia \citep{fontani11};
      {\it Bottom-left panel:} same abundances as a function of the source bolometric luminosity derived from the SED \citep{mininni21};
      {\it Bottom-right panel:} same abundances as a function of the luminosity-to-mass ratio \citep{mininni21}.
      }
         \label{fig:abundances-parameters-C}
\end{figure*}

\begin{figure*}
       \centering
   \includegraphics[width=15cm]{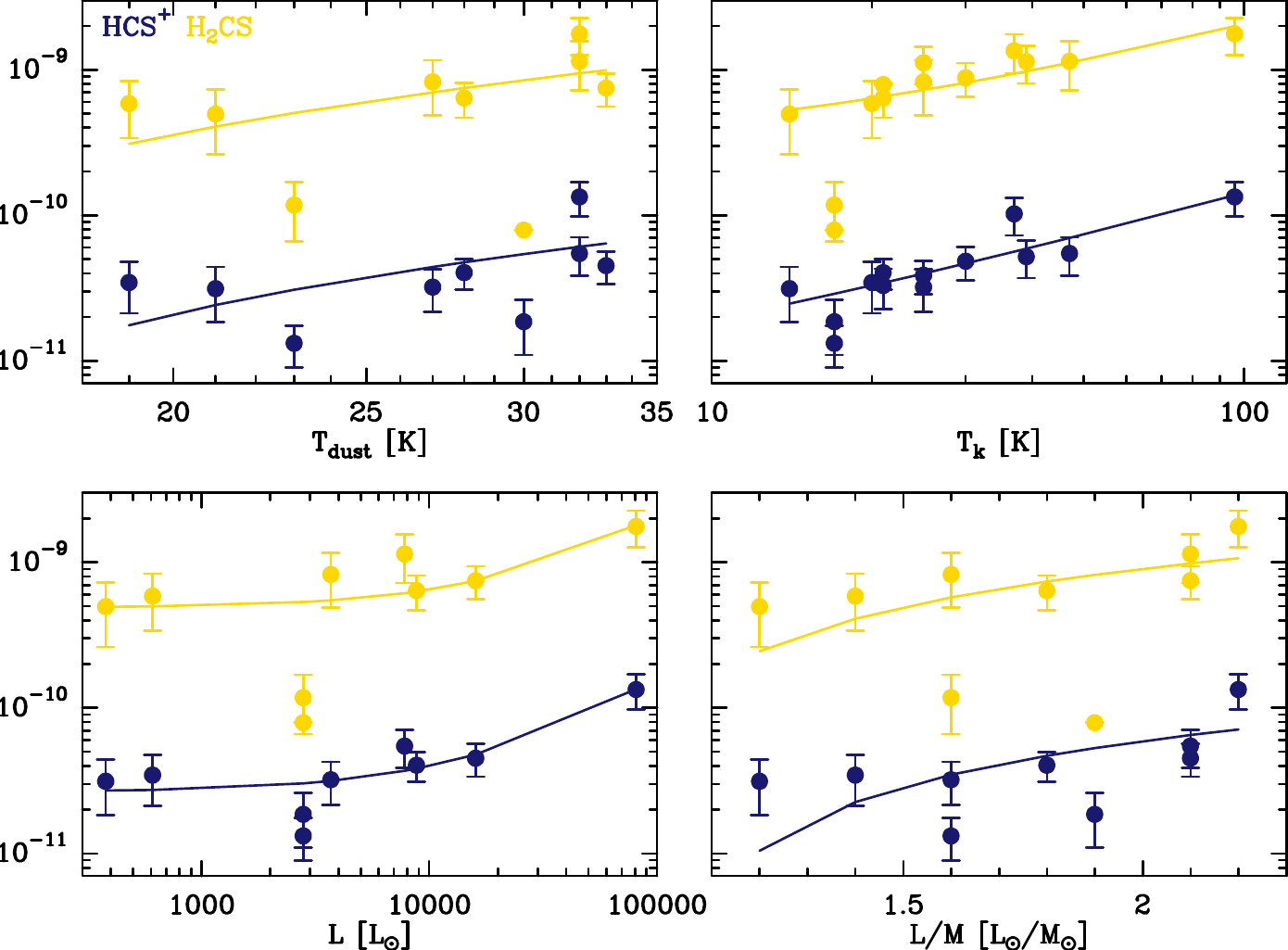}
      \caption{Same as Fig.~\ref{fig:abundances-parameters-C} for molecules containing only sulphur, carbon and hydrogen.
      }
         \label{fig:abundances-parameters-CH}
\end{figure*}

\begin{figure*}
       \centering
   \includegraphics[width=15cm]{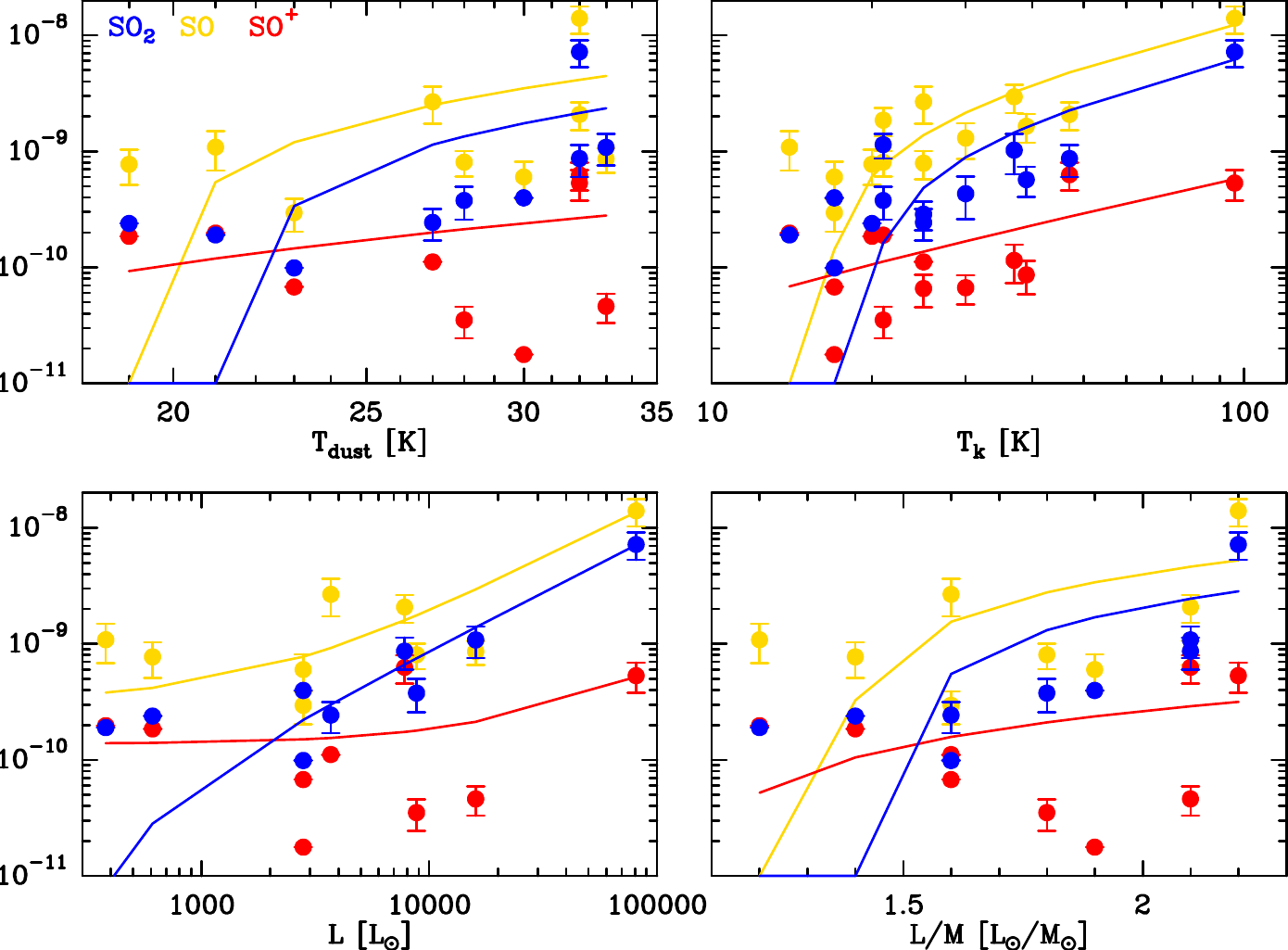}
      \caption{Same as Fig.~\ref{fig:abundances-parameters-C} for molecules containing only sulphur and oxygen.
      }
         \label{fig:abundances-parameters-O}
\end{figure*}

\begin{figure*}
       \centering
   \includegraphics[width=15cm]{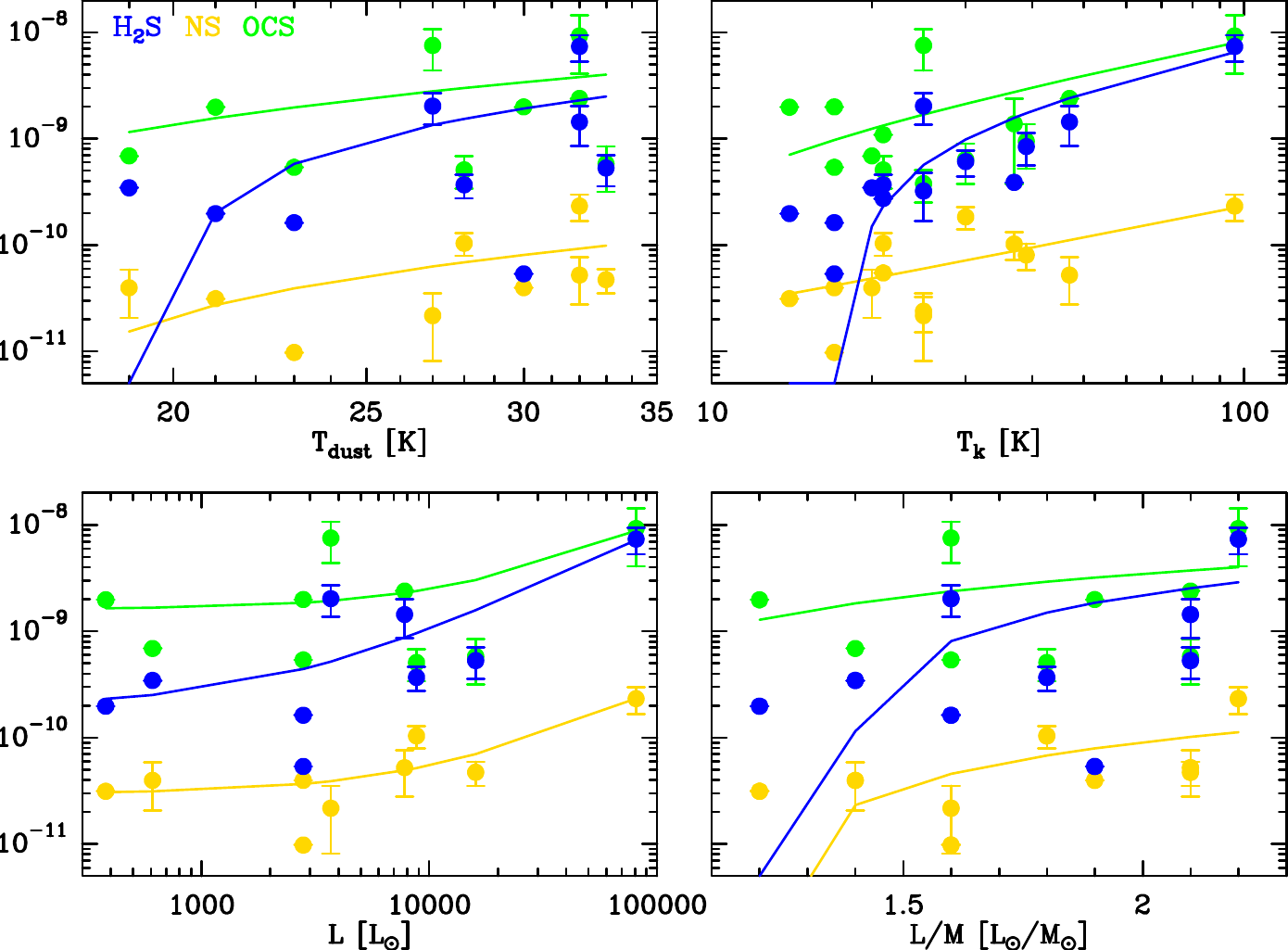}
      \caption{Same as Fig.~\ref{fig:abundances-parameters-C} for all other molecules.
      }
         \label{fig:abundances-parameters-others}
\end{figure*}

\end{appendix}

\end{document}